\pdfoutput=1
\documentclass[aps,prd,superscriptaddress,notitlepage,11pt,final,longbibliography,nofootinbib]{revtex4-1}
\usepackage{graphicx}
\usepackage{amsfonts}
\usepackage{amsmath}
\usepackage{amsmath,amssymb,mathrsfs}
\usepackage[colorlinks=true, a4paper=true, pdfstartview=FitV, linkcolor=blue, citecolor=blue, urlcolor=blue]{hyperref}

\usepackage{subfigure}
\usepackage{epsfig}
\usepackage[usenames,dvipsnames]{xcolor}
\usepackage{bbm}
\usepackage{color}
\usepackage{amsmath}
\usepackage{natbib}
\usepackage{ulem}

\graphicspath{{./fig/}}

\newcommand{\be}{\begin{equation}}
\newcommand{\ee}{\end{equation}}
\newcommand{\bea}{\begin{eqnarray}}
\newcommand{\eea}{\end{eqnarray}}

\newcommand{\eqn}{\begin{eqnarray}}
\newcommand{\eqnx}{\end{eqnarray}}

\numberwithin{equation}{section}

\begin{document}

\title{Compact Q-balls and Q-shells within a $CP^N$  Skyrme-Faddeev type model}
\author{E. L. Cola\c co~}
\email{eduardo.luizcolaco@gmail.com}
\affiliation{Departamento de F\'isica, Universidade Federal de Santa Catarina, Campus Trindade, 88040-900, Florian\'opolis-SC, Brazil}
\author{P. Klimas~}
\email{pawel.klimas@ufsc.br}
\affiliation{Departamento de F\'isica, Universidade Federal de Santa Catarina, Campus Trindade, 88040-900, Florian\'opolis-SC, Brazil}

\author{L. R. Livramento~}
\email{livramento@usp.br}
\affiliation{Instituto de F\'\i sica de S\~ao Carlos,\\
Universidade de S\~ao Paulo, 13560-970, S\~ao Carlos-SP, Brazil\\}
\affiliation{Institute of Theoretical Physics,  Jagiellonian University,
Lojasiewicza 11, Krak\'{o}w, Poland}

\author{N. Sawado~}
\email{sawadoph@rs.tus.ac.jp}
\affiliation{Department of Physics and Astronomy, Tokyo University of Science, Noda, Chiba 278-8510, Japan}

\author{S. Yanai}
\email{yanai@toyota-ct.ac.jp}
\affiliation{Department of Natural Sciences, National Institute of Technology, Toyota College, Toyota, Aichi 471-8525, Japan}

\begin{abstract}

\end{abstract}

\begin{abstract}
While $CP^N$ models with analytic potentials are known to support finite-energy compact Q-ball and Q-shell solutions, their behavior in more complex Lagrangian frameworks remains a subject of active research. This work explores these non-topological structures within an extended Skyrme-Faddeev-type model that incorporates quartic derivative terms. In this context, harmonic time dependence and the presence of quartic terms constitute two independent stabilization mechanisms that allow the configurations to circumvent Derrick's scaling argument. We investigate the necessary conditions for the existence of these solutions and analyze the influence of quartic terms on the properties of the resulting compactons, specifically examining the $E(Q)$ relationship between energy and Noether charge. Our findings provide valuable insights into the stability and characteristics of compact boson stars within $CP^N$ models featuring higher-order derivative terms.
\end{abstract}

\maketitle

\section{Introduction}
Scalar fields play a crucial role in theoretical physics. They are extensively employed for modeling physical matter across a diverse range of contexts, from fundamental fields such as the Higgs field \cite{PhysRevLett.13.508} to effective descriptions like the order parameter field in the Landau-Ginzburg model \cite{Ginzburg:1950sr}. Depending on the inherent complexity of the physical phenomenon under investigation, scalar fields can be represented by straightforward real-valued functions or by more intricate structures, such as multiple complex-valued fields defined on complex projective spaces ($CP^N$ models) \cite{zakrzewski1988low}.

This paper examines a scalar field theory characterized by a $CP^N$ target space. The model under consideration represents an extension of the conventional Skyrme-Faddeev (SF) model, specifically through the inclusion of additional quartic terms and a potential function. Our central objective is the identification of stationary, finite-energy solutions, which are commonly referred to as Q-balls. Under specific selections for the potential, these Q-balls are realized as {\it compactons}, meaning they are solutions characterized by compact support.

The Skyrme-Faddeev (SF) model, initially proposed by L. D. Faddeev in 1975 \cite{Faddeev:1975tz, Faddeev:1995au}, is a modified sigma model defined in three spatial dimensions. This model is characterized by a topological charge associated with the Hopf map $S^3\rightarrow S^2$. Faddeev introduced this particular model with the goal of identifying finite-energy knot soliton solutions. This theoretical concept was subsequently validated by numerical studies \cite{Battye:1998pe, Sutcliffe:2007ui}, which successfully demonstrated the existence of stable toroidal solitons and knot structures.

In contrast, the analytical solutions for this model remain largely unexamined. The crucial relevance of the model within emerging physical contexts is the main reason behind continuous efforts to find these solutions using analytical methods. Specifically, a conjecture based on the Cho-Faddeev-Niemi decomposition postulates that the SF model can be obtained as a low-energy, non-perturbative effective model for $SU(2)$ Yang-Mills theory \cite{Faddeev:1998yz}. Furthermore, results established by Gies \cite{Gies:2001hk} indicate that the SF model can be derived from an analysis of the Wilsonian effective action for $SU(2)$ Yang-Mills theory. This analysis not only reproduces the original SF Lagrangian but also generates {\it novel terms} that are quartic in derivatives. The inclusion of these terms within the Lagrangian yields a modified SF model. Notably, the presence of a quartic term was instrumental in the construction of the exact time-dependent vortex solution in $3+1$ dimensions by Ferreira \cite{Ferreira:2008nn}, a solution which exists for a specific combination of coupling constants.

The model can also be extended through the generalization of its target space. For instance, the SF model (without additional non-Skyrme type quartic terms) defined on the coset space $SU(N+1)/U(1)^N$ was proposed in \cite{Faddeev:1999cj}. This specific case was identified as the "maximum case" in the classification presented in \cite{Kondo:2008xa}. An alternative generalization, designated the "minimum case" by the same authors, utilizes the coset space $SU(N+1)/SU(N)\otimes U(1)$, which corresponds to complex projective $CP^N$ space. A modified SF model incorporating the $CP^N$ target space was subsequently proposed in \cite{Ferreira:2010jb}. This particular formulation introduces two additional coupling constants that multiply two distinct types of quartic terms, which can be included for $N\ge 2$. When these coupling constants satisfy a specific constraint, the model is shown to support vortex solutions within its holomorphic (or anti-holomorphic) sector. Further analysis of these vortex solutions was conducted in \cite{Ferreira:2011ja}. It was later observed in \cite{Ferreira:2011mz} that the condition imposed on the coupling constants can be relaxed if the model is augmented by a potential (non-derivative) term. The authors of \cite{Ferreira:2011mz} successfully discovered exact and numerical vortex solutions for the model with a $CP^1$ target space. The concept of vortex solutions in the simultaneous presence of both quartic terms and a potential was also investigated in \cite{Sawado:2013yza}. This principle retains its validity when the model's target space is extended to $CP^N$ space. According to the findings in \cite{Amari:2015sva}, vortex solutions exist for a certain class of potentials, even when the coupling constants do not satisfy the usual condition initially introduced in \cite{Ferreira:2010jb}.

A crucial requirement for physical relevance is that the solutions must possess finite total energy. However, all known vortex solutions of the extended Skyrme-Faddeev (ESF) model extend infinitely along the $z$-axis. Consequently, they exhibit infinite total energy, even though their energy per unit length remains finite. For realistic physical systems, finite-energy solutions in $3+1$ dimensions are intuitively expected to be spatially confined. Analogous to vortices observed in liquid helium, which necessarily terminate at the boundaries of their container, these solutions should not extend indefinitely in any spatial direction.

The construction of finite-energy solutions within higher-dimensional $CP^N$ target spaces necessitates rigorous theoretical refinements. In the $CP^1$ framework, the existence of topological solutions is facilitated by the non-trivial homotopy class $\pi_3(CP^1) \cong \mathbb{Z}$. Conversely, for $N \ge 2$, higher-dimensional target spaces do not support such configurations because the third homotopy group is trivial, $\pi_3(CP^N) = 0$. In the absence of conserved topological charges, the analysis is restricted to non-topological solutions \cite{LEE1992251}, which require auxiliary stabilization mechanisms to ensure localized energy density. For static configurations, Derrick's theorem \cite{Derrick:1964ww} imposes strict constraints on the dimensionality of stable solutions, derived from the scaling behavior of the energy functional with respect to the soliton size. To circumvent the limitations imposed by Derrick's theorem, two primary strategies are generally employed. First, the Lagrangian may be augmented with terms containing higher powers of derivatives; these terms exhibit scaling properties opposite to those of the standard quadratic kinetic terms, thereby allowing for a global energy minimum at a finite scale parameter. Alternatively, one may consider stationary solitons characterized by a harmonic time dependence. Such solutions, extensively investigated in \cite{Lee:1991ax, Rosen:1968mfz, Werle:1977cc}, are known as Q-balls. These configurations possess a temporal dependence of the form $e^{i\omega t}$, where the frequency $\omega$ facilitates the stabilization of the non-topological charge. Furthermore, non-topological solitons may possess conserved Noether charges, which provide an additional mechanism for addressing the issue of stability. Our previous investigations regarding compact Q-balls were restricted to complex field models containing only quadratic kinetic terms; consequently, harmonic temporal dependence represented the sole method for evading the constraints of Derrick's theorem. In the present work, we extend this analysis to a model incorporating quartic terms. This inclusion introduces a secondary stabilization mechanism, allowing for a more comprehensive exploration of the interplay between higher-order derivatives and stationary field configurations.

The existence of non-topological solutions within the ESF model, analogous to its non-integrable sector's vortices, necessitates the inclusion of a potential term. The precise functional form of this potential in the vicinity of its minimum directly governs the field's asymptotic behavior at spatial infinity. However, models featuring sharp, V-shaped potentials at their minimum, when combined with standard quadratic derivative terms in the Lagrangian, exhibit an unconventional field behavior near that minimum. In these cases, the field approaches its vacuum quadratically--characterized by a parabolic approach--thereby reaching the minimum within a finite spatial distance. Solutions possessing this characteristic are non-zero exclusively within a compact support, leading to their designation as compactons.

The signum-Gordon model \cite{Arodz:2005gz, Arodz:2005bc} provides the simplest supporting example of compacton solutions. This model is essentially a Klein-Gordon model in which the typical quadratic potential, $\phi^2$, is substituted by the sharp potential, $|\phi|$. A version of this model employing a complex-valued scalar field is known to support compact Q-balls \cite{Arodz:2008jk, Arodz:2009ye, Arodz:2008nm}. Importantly, compactons are classified as solutions in a weak sense \cite{richtmyer1978, evans1998}, implying they satisfy the integrated Euler-Lagrange equations. Interest in compact solutions has increased substantially in recent years, with compactons emerging in various domains of physics, including condensed matter physics \cite{PhysRevE.60.6218}, nuclear physics \cite{Adam:2010fg}, and cosmology \cite{Hartmann:2013kna}. More recently, oscillons \cite{Arodz:2007jh}, another class of compact solutions, were discovered within the BPS submodel, which satisfies the Bogomol'nyi--Prasad--Sommerfield bound, of the Skyrme model \cite{Adam:2017srx, Klimas:2018woi}. These oscillons are expected to play a crucial role in the radiation spectrum observed during skyrmion scattering. Analogously, we expect that time-dependent non-topological solutions will appear in the radiation spectrum of the ESF model when a potential term is included.

This paper examines compact non-topological solutions within the ESF model augmented with a potential term. Initial research into these solutions, detailed in \cite{Klimas:2017eft}, established the existence of compact Q-balls and Q-shells in the $CP^N$ model (specifically for odd $N$) when a particular V-shaped potential is introduced. These solutions are characterized by a non-trivial radial function in spherical coordinates and exhibit time dependence via the factor $e^{i\omega t}$. Their angular component is described by spherical harmonics.

The remainder of this paper is organized as follows. Section \ref{sec2} details the model and its parameterization. Section \ref{sec3} is dedicated to the discussion of Noether charges. Subsequently, in Section \ref{sec4}, we present and analyze the numerical compact solutions. The final section is reserved for summarizing the key findings and discussing the implications of the results.

\section{The extended Skyrme-Faddeev model with potential}
\label{sec2}

\subsection{The Lagrangian and Euler-Lagrange equations}
The Extended Skyrme-Faddeev (ESF) model augmented by a potential, defined on a $CP^N$ target space, is conveniently formulated using the principal variable $X$ \cite{Ferreira:1998zx}. The Lagrangian density is expressed by derivative terms containing $L_{\mu}\equiv X^{-1}\partial_{\mu}X$, as well as a non-derivative potential term $V(X)$. We examine a model whose Lagrangian density, $\mathcal{L}$, comprises five terms: a potential term, $\mathcal{L}_0$; a standard quadratic term, $\mathcal{L}_2$; a (quartic) Skyrme term, $\mathcal{L}_4^{(S)}$; and two additional quartic terms, $\mathcal{L}_{4}^{(1)}$ and $\mathcal{L}_{4}^{(2)}$. The total Lagrangian density is given by:
\begin{equation}
{\cal L}={\cal L}_2+{\cal L}_0+{\cal L}_4^{(S)}+{\cal L}_{4}^{(1)}+ {\cal L}_{4}^{(2)}.\label{lagr1}
\end{equation}
These individual Lagrangian terms are explicitly defined as:
\begin{align*}
{\cal L}_0&=-\mu^2V(X)\\
{\cal L}_2&=-\frac{M^2}{2}\;{\rm Tr}(L_{\mu}L^{\mu})\\
{\cal L}_4^{(S)}&=\frac{1}{e^2}\;{\rm Tr}([L_{\mu},L_\nu][L^{\mu},L^\nu])\\
{\cal L}_4^{(1)}&=\frac{\beta}{2}\;\big({\rm Tr}(L_{\mu}L^{\mu})\big)^2\\
{\cal L}_4^{(2)}&=\gamma\;{\rm Tr}(L_{\mu}L_{\nu}){\rm Tr}(L^{\mu}L^{\nu}).
\end{align*}

The term $\mathcal{L}_2$ constitutes the standard $CP^N$ Lagrangian. When combined with a properly chosen potential term $\mathcal{L}_0$, these two terms collectively support compact Q-ball and Q-shell solutions \cite{Klimas:2017eft}. While the additional quartic terms, $\mathcal{L}_4^{(1)}$ and $\mathcal{L}_4^{(2)}$, have been investigated in other physical contexts, their application to Q-ball solutions represents a novel approach. This paper, therefore, primarily focuses on analyzing the impact of these quartic terms on the morphology of the resulting Q-balls. 

It is important to note that for the $N=1$ case, the terms $\mathcal{L}_4^{(1)}$ and $\mathcal{L}_4^{(2)}$ are functionally equivalent. This $N=1$ scenario, previously discussed in \cite{Ferreira:2008nn}, included only a single additional quartic term, which was defined using a unit complex three-component vector $\vec{n}$.

The coupling constant $M^2$ has the dimension of $L^{-2}$, where $L$ denotes the dimension of length. Similarly, $\mu^2$ has the dimension of $L^{-4}$. Conversely, the constants $e^{-2}$, $\beta$, and $\gamma$ are dimensionless. In the context of the Lagrangian, the parameters are required to satisfy the constraints $M^2>0$, $e^2<0$, and $\mu^2>0$. The specific values of the dimensionless constants $\beta$ and $\gamma$ will be discussed subsequently, with particular attention to ensuring the positive definiteness of the energy density.

The principal variable $X$, as introduced in \cite{EICHENHERR1980528, Ferreira:1984bi}, is employed to parameterize symmetric spaces, with the complex projective space $CP^N$ serving as a canonical example. We consider the $(N+1)$-dimensional defining representation wherein the $SU(N+1)$-valued element $g$ is parameterized by a set of $N$ complex fields $u_i$ according to the following matrix form:
\begin{align}
g&\equiv\frac{1}{\sqrt{1+u^{\dagger} \cdot u}}\left(\begin{array}{cc}
\Delta&iu\\
iu^{\dagger}&1
\end{array}\right)\label{parametryzacja}
\end{align}
where the $N \times N$ block $\Delta$ is defined as:
\begin{align}
\Delta_{ij}&\equiv \sqrt{1+u^{\dagger} \cdot u}\,\delta_{ij}-\frac{u_iu_j^*}{1+\sqrt{1+u^{\dagger} \cdot u}}.
\end{align}
The principal variable $X$ is then expressed in terms of the complex fields $u_i$ as the square of the element $g$:
\begin{align}
X(g)&=g^2=
\left(\begin{array}{cc}
{\mathbb{I}}_{N\times N} & 0 \nonumber \\
0 & -1 \nonumber 
\end{array}\right)
+
\frac{2}{1+\sqrt{1+u^{\dagger} \cdot u}}\left(\begin{array}{cc}
-u\otimes u^\dagger & iu \nonumber \\
iu^\dagger & 1  
\end{array}\right).
\end{align}

The Lagrangian given in Equation \eqref{lagr1} simplifies to the following expression:
\be
{\cal L}=-\frac{1}{2}\left[M^2\eta_{\mu\nu}+C_{\mu\nu}\right]\tau^{\nu\mu}-\mu^2V.\label{lagr2}
\ee
The potential $V$ adopted for this investigation, defined in \cite{Klimas:2017eft}, is given by:
\be
V=\frac{1}{2}\sqrt{{\rm Tr}(1-X)}=\sqrt{\frac{u^{\dagger}\cdot u}{1+u^{\dagger}\cdot u}}.\label{potencjal}
\ee
In Equation \eqref{lagr2}, $\eta_{\mu\nu}$ is the Minkowski metric, defined as $\eta_{\mu\nu}={\rm diag}(1,-1,-1,-1)$. The derivative-coupling tensor $\tau_{\nu\mu}$ represents the kinetic and gradient term and is given by:
\be
\tau_{\nu\mu}:=-4\frac{\partial_{\mu}u^{\dagger}\cdot\Delta^2\cdot\partial_{\nu}u}{(1+u^{\dagger}\cdot u)^2},\label{tau}
\ee
where the matrix $\Delta^2$ is defined as:
\[
\Delta^2_{ij}=(1+u^{\dagger} \cdot u)\;\delta_{ij}-u_iu^*_j.
\]
The coefficients $C_{\mu\nu}$ are defined by:
\begin{align}
C_{\mu\nu}:=&M^2\eta_{\mu\nu}-\frac{4}{e^2}\Big[(\beta e^2-1)\tau^{\alpha}_{\,\,\,\,\alpha}\eta_{\mu\nu}+(\gamma e^2-1)\tau_{\mu\nu}+(\gamma e^2+2)\tau_{\nu\mu}\Big].
\end{align}

Only the $M^2$--proportional term in the Lagrangian stems directly from quadratic terms. The remaining contributions, however, originate from quartic interactions. The influence of these quartic terms can be interpreted as a perturbation--provided their magnitude is sufficiently small--acting upon the principal solution determined by the combination of the quadratic term and the potential $V$.

To facilitate the analysis of this model, we will reformulate it using dimensionless coordinates. This procedure requires the introduction of a characteristic length scale, denoted as $r_0$. Although various definitions based on the coupling constants are possible, a specific choice must be made. The selection of the length scale has varied across prior studies; for instance, in our previous investigation of the $CP^N$--type model with a potential \cite{Klimas:2017eft}, we defined the scale simply as $r_0=M^{-1}$. Conversely, in the study of vortices within the extended Skyrme-Faddeev model \cite{Amari:2015sva}, the length scale was set as $r_0:=\frac{2}{M}\frac{1}{\sqrt{-e^{2}}}$, where $e^2<0$. The latter definition presents a formal inconvenience when considering the limit $e^{-2}\to 0$, as the length scale $r_0$ would vanish unless a corresponding adjustment is made to the parameter $M$. Given that a quadratic term proportional to $M^2$ is explicitly retained in the Lagrangian for the current analysis, we adopt the more straightforward definition: $r_0:=M^{-1}$.

We proceed by expressing the Lagrangian density of the model using the dimensionless counterparts of the Derivative-Coupling Tensor, $\tau_{\mu\nu}$, and the coefficient tensor, $C_{\mu\nu}$. Let $\{\xi^{\mu}\}_{\mu=0,\ldots,3}$ represent a set of dimensionless curvilinear coordinates, defined by the transformation $\xi^{\mu}=\xi^{\mu}(\tilde{x}^{\alpha})$, where $\tilde{x}^{\mu} = x^{\mu}/r_0$ are the dimensionless Cartesian coordinates. The partial derivatives transform according to the chain rule, incorporating the characteristic length scale $r_0$:
\be
\frac{\partial}{\partial x^{\mu}}=r_0^{-1}\frac{\partial}{\partial \tilde x^{\mu}}=r_0^{-1}\frac{\partial\xi^{\alpha}}{\partial \tilde x^{\mu}}\frac{\partial}{\partial  \xi^{\alpha}}.
\ee
The components of the physical flat metric $\eta_{\mu\nu}$ (in the original Cartesian coordinates $x$) are related to the components of the metric tensor in the curvilinear system, $g_{\alpha\beta}(\xi)$, by the standard tensor transformation rule:
\be
\eta_{\mu\nu}=\frac{\partial\xi^{\alpha}}{\partial \tilde x^{\mu}}\frac{\partial\xi^{\beta}}{\partial \tilde x^{\nu}}g_{\alpha\beta}(\xi).\nonumber
\ee
Consequently, the derivative--coupling tensor $\tau_{\mu\nu}(x)$, defined in Equation \eqref{tau}, transforms to its dimensionless counterpart, $\tau_{\alpha\beta}(\xi)$, which contains derivatives with respect to the curvilinear coordinates $\xi^{\alpha}$: 
\be
\tau_{\mu\nu}(x)=r_0^{-2}\frac{\partial\xi^{\alpha}}{\partial \tilde x^{\mu}}\frac{\partial\xi^{\beta}}{\partial \tilde x^{\nu}}\tau_{\alpha\beta}(\xi).\nonumber
\ee
As a result of these transformations, the original tensor $C_{\mu\nu}(x)$ can be expressed as a linear combination of its dimensionless counterpart, $\widetilde{C}_{\alpha\beta}(\xi)$, scaled by $M^2$:
\begin{align*}
C_{\mu\nu}(x)=M^2\frac{\partial\xi^{\alpha}}{\partial \tilde x^{\mu}}\frac{\partial\xi^{\beta}}{\partial \tilde x^{\nu}}\tilde C_{\alpha\beta}(\xi).
\end{align*}
The dimensionless coefficient tensor $\widetilde{C}_{\mu\nu}(\xi)$ is defined in terms of the curvilinear metric $g_{\mu\nu}(\xi)$ and the dimensionless tensor $\tau_{\mu\nu}(\xi)$ as:
\begin{align}
\widetilde C_{\mu\nu}(\xi):=g_{\mu\nu}(\xi)&-\frac{\kappa}{4}\Big[(\beta e^2-1)\tau^{\alpha}_{\,\,\,\,\alpha}(\xi)g_{\mu\nu}(\xi)\nonumber\\&+(\gamma e^2-1)\tau_{\mu\nu}(\xi)+(\gamma e^2+2)\tau_{\nu\mu}(\xi)\Big].\label{Cxi}
\end{align}

The dimensionless combination of coupling constants and the characteristic length scale $r_0$ is defined as:
\be
\kappa:=\frac{16}{M^2 r_0^2e^2}.
\ee
The parameter $\kappa$ is crucial for indicating the decoupling of the quartic terms, which originate from the Lagrangian components proportional to $e^{-2}$, $\beta$, and $\gamma$. It should be noted that $\kappa$ is constrained to be non-positive ($\kappa\le 0$). If the Skyrme term vanishes (i.e., $e^{-2}\to 0$), the parameter $\kappa$ approaches zero, provided the length scale parameter $r_0$ remains finite. Crucially, the products $\kappa \beta e^2$ and $\kappa \gamma e^2$ tend to zero as $\kappa\to 0$ if and only if both $\beta$ and $\gamma$ simultaneously approach zero ($\beta\to 0$ and $\gamma\to 0$). Should the condition $\beta\to 0$ and $\gamma\to 0$ be satisfied concurrently, the full model simplifies considerably, reducing solely to the $\mathbb{C}P^N$ model incorporating only the potential term.

The primary objective of this paper is to investigate the effect of quartic terms on Q-ball solutions. To achieve this, it is essential to demonstrate the ability to decouple these quartic terms by performing the simultaneous limit where $e^{-2}\rightarrow 0$, $\beta\rightarrow 0$, and $\gamma\rightarrow 0$. This specific limiting procedure should effectively recover the known Q-ball solutions previously obtained in \cite{Klimas:2017eft}.

The Lagrangian defined in Equation \eqref{lagr2} can be expressed in terms of dimensionless quantities as ${\cal L}=r_0^{-4}\widetilde{\cal L}$, where the dimensionless Lagrangian density $\widetilde{\cal L}$ takes the form:
\be
\widetilde{\cal L}=-\frac{1}{2}\left[g_{\mu\nu}(\xi)+\widetilde C_{\mu\nu}(\xi)\right]\tau^{\nu\mu}(\xi)-\tilde\mu^2V.
\ee
Here, $\tilde{\mu}^2$ is a dimensionless constant defined as $\tilde{\mu}^2:=\frac{r_0^2}{M^2}\mu^2$. The Euler-Lagrange equations of motion (EOMs) derived from this dimensionless Lagrangian are given by:
\begin{align}
&\frac{1}{\sqrt{-g}}\partial_{\alpha}\Big(\sqrt{-g}g^{\mu\alpha}g^{\nu\beta}\widetilde C_{\mu\nu}\partial_{\beta}u_i\Big)-\frac{g^{\mu\alpha}g^{\nu\beta}\widetilde C_{\mu\nu}}{1+u^{\dagger}\cdot u}\Big[(u^{\dagger}\cdot \partial_{\alpha}u)\partial_{\beta}u_i+(u^{\dagger}\cdot \partial_{\beta}u)\partial_{\alpha}u_i\Big]+\nonumber\\
&+\frac{\tilde\mu^2}{4}\Gamma_i=0\label{eom1}
\end{align}
where $g$ is the determinant of the curvilinear metric tensor, $g:=\det(g_{\mu\nu})<0$. The term $\Gamma_i$ is derived from the variation of the potential $V$ with respect to the complex field $u_i^*$ and is defined as:
\begin{align*}
\Gamma_i&\equiv(1+u^{\dagger}\cdot u)\sum_{k=1}^N\left[(\delta_{ik}+u_iu^*_k)\frac{\delta V}{\delta u_k^*}\right]\\
&=\frac{u_i}{2}\sqrt{\frac{1+u^{\dagger}\cdot u}{u^{\dagger}\cdot u}}.
\end{align*}
It should be noted that the form of the potential $V$, and consequently the expression for $\Gamma_i$, can be modified in several qualitatively equivalent ways while preserving the compact nature of the localized solutions (Q-balls). 

\subsection{Radial reduction}
Following the methodology established in \cite{Klimas:2017eft}, we consider an odd number of scalar fields, specifically $N=2l+1$. For convenience, the field index $i=1, 2, \ldots, N$ is replaced by the corresponding angular momentum index $m=-l, \ldots, l$. We introduce spherical coordinates $(r, \theta, \phi)$ and apply the following ansatz for the field components $u_m(t, r, \theta, \phi)$:
\be
u_m(t,r,\theta,\phi):=\sqrt{\frac{4\pi}{2l+1}}f(r)Y_{lm}(\theta,\phi)e^{i\omega t}.\label{ansatz}
\ee
In this definition, $Y_{lm}(\theta, \phi)$ represents the spherical harmonics, $f(r)$ is a non-negative radial profile function ($f(r) \ge 0$), and $\omega$ is the oscillation frequency. This specific ansatz enables the reduction of both the Lagrangian and the field equations to a radial form.

The non-vanishing components of the derivative--coupling tensor, $\tau_{\mu\nu}(\xi)$, in the spherical coordinate system are found to be:
\begin{align}
\tau_{tt}(\xi)&=-\frac{4\,\omega^2\,f^2}{(1+f^2)^2},
&\tau_{rr}(\xi)&=-\frac{4\,f'^2}{(1+f^2)^2},\nonumber\\
\tau_{tr}(\xi)&=-\frac{4i\,\omega\,ff'}{(1+f^2)^2},
&\tau_{rt}(\xi)&=-\tau_{tr}(\xi),\nonumber\\
\tau_{\theta\theta}(\xi)&=-4\,\frac{l(l+1)}{2}\frac{f^2}{1+f^2},
&\tau_{\phi\phi}(\xi)&=\sin^2\theta\,\tau_{\theta\theta}(\xi).\nonumber
\end{align}
Note that $f'$ denotes the derivative of the radial function with respect to $r$ ($f' = \partial f / \partial r$).

The components of the dimensionless coefficient tensor $\widetilde{C}_{\mu\nu}(\xi)$ in the spherical coordinate system are determined to be:
\begin{align}
\widetilde C_{tt}&=1+\kappa F_1(r),\nonumber\\ 
\widetilde C_{rr}&=-1+\kappa F_2(r),\nonumber\\
\widetilde C_{tr}&=-\widetilde C_{rt}=-3i\omega\,\kappa  F_3(r),\nonumber\\
\widetilde C_{\theta\theta}&=r^2[-1+\kappa F_4(r)],\nonumber\\
\widetilde C_{\phi\phi}&=\sin^2\theta \,\tilde C_{\theta\theta},\nonumber
\end{align}
where the functions $F_1(r)$, $F_2(r)$, $F_3(r)$, and $F_4(r)$ encapsulate the radial dependence arising from the coupling terms. These functions are explicitly defined as:
\begin{align}
F_1(r)&=(\beta e^2+2\gamma e^2)\frac{\omega^2f^2}{(1+f^2)^2}-(\beta e^2-1)\left[\frac{f'^2}{(1+f^2)^2}+\frac{l(l+1)}{r^2}\frac{f^2}{1+f^2}\right],\nonumber\\
F_2(r)&=(\beta e^2+2\gamma e^2)\frac{f'^2}{(1+f^2)^2}-(\beta e^2-1)\left[\frac{\omega^2 f^2}{(1+f^2)^2}-\frac{l(l+1)}{r^2}\frac{f^2}{1+f^2}\right],\nonumber\\
F_3(r)&=\frac{ff'}{(1+f^2)^2},\nonumber\\
F_4(r)&=(\beta e^2+\gamma e^2-{\textstyle\frac{1}{2}})\frac{l(l+1)}{r^2}\frac{f^2}{1+f^2}-(\beta e^2-1)\frac{\omega^2f^2-f'^2}{(1+f^2)^2}.\nonumber
\end{align}

The radial equation of motion (Equation \eqref{eom1}) can be reduced to the following ordinary differential equation for the radial profile function $f(r)$:
\begin{align}
&(1-\kappa F_2)\left(f''+\frac{2}{r}f'-\frac{2ff'^2}{1+f^2}\right)-\kappa F_2'f'\nonumber\\
&-3\kappa\omega^2\left(F_3'+\frac{2}{r}F_3\right)f-(1-\kappa F_4)\frac{l(l+1)}{r^2}f\nonumber\\&+\omega^2(1+\kappa F_1)\frac{(1-f^2)f}{1+f^2}-\frac{\tilde\mu^2}{8}{\rm sgn}(f)\sqrt{1+f^2}=0\label{radialeq}
\end{align}
where the prime symbol ($\prime$) denotes the derivative with respect to the radial coordinate $r$, and the functions $F_1, F_2, F_3, F_4$ (and their derivatives $F_2', F_3'$) are defined based on the coupling constants and the profile $f(r)$.

\subsection{Energy density}
A crucial prerequisite for constructing physically meaningful solutions is ensuring the positive definiteness of the energy density. Given that the Lagrangian for this model comprises multiple terms connected by various coupling constants, it is essential to guarantee a non-negative energy density across all configurations before proceeding to the derivation of solutions for specific parameter values. Consequently, the focus of the following section is a detailed analysis of the model's energy density.

The dimensionless Hamiltonian density, $\widetilde{\cal H} := r_0^4{\cal H}$, is conventionally defined via the Legendre transformation:
\be
\widetilde {\cal H}=\sum_{m=-l}^l\left(\frac{\delta \widetilde{\cal L}}{\delta(\partial_t u_m)}\partial_tu_m+\frac{\delta \widetilde {\cal L}}{\delta(\partial_t u^*_m)}\partial_tu^*_m\right)-\widetilde {\cal L}.
\ee
Given the structure of the Lagrangian density, where its variation with respect to the derivative-coupling-tensor is $\delta \widetilde{\cal L}=-\frac{1}{2}\tilde{C}_{\mu\nu}\delta \tau^{\nu\mu}$, the Hamiltonian density can be formally expressed using the components involving time derivatives as:
\be
\widetilde{\cal H}=-\widetilde C^{\mu t}\tau_{t\mu}-\widetilde C^{t\mu}\tau_{\mu t}-\widetilde{\cal L}.\label{H2}
\ee

For the field configuration defined by the ansatz \eqref{ansatz}, the dimensionless Hamiltonian density $\widetilde{\cal H}$ can be expressed in terms of the non-zero components of the derivative-coupling-tensor $\tau_{\mu\nu}(\xi)$ and the coefficient tensor $\widetilde{C}_{\mu\nu}(\xi)$ as:
\be
\widetilde{\cal H}=-2[\tilde C_{tt}\tau_{tt}+\widetilde C_{tr}\tau_{tr}]-\tilde{\cal L}
\ee
where the full dimensionless Lagrangian density $\widetilde{\cal L}$ is:
\begin{align}
\widetilde{\cal L}=&-\frac{1}{2}\Big[\tau_{tt}-\tau_{rr}-\frac{2}{r^2}\tau_{\theta\theta}+\widetilde C_{tt}\tau_{tt}+\widetilde C_{rr}\tau_{rr}+2\widetilde C_{tr}\tau_{tr}+\frac{2}{r^4}\widetilde C_{\theta\theta}\tau_{\theta\theta}\Big].\nonumber
\end{align}
The dimensionless Hamiltonian density $\widetilde{\cal H}$ naturally decomposes into a quadratic part ($\widetilde{\cal H}_{02}$) and a quartic part ($\widetilde{\cal H}_{4}$), such that $\widetilde{\cal H}=\widetilde{\cal H}_{02}+\widetilde{\cal H}_{4}$. The part of the Hamiltonian density originating from the quadratic terms is given by:
\begin{align}
\widetilde {\cal H}_{02}=\frac{4\omega^2f^2}{(1+f^2)^2}+\frac{l(l+1)}{r^2}\frac{4f^2}{1+f^2}+\frac{4f'^2}{(1+f^2)^2}+\tilde\mu^2V.\label{H0}
\end{align}
The portion of the Hamiltonian density arising from the quartic interactions is initially expressed as:
\begin{align}
\widetilde {\cal H}_4&=\kappa\Bigg[\frac{6\omega^2f^2}{(1+f^2)^2}F_1(r)-\frac{2f'^2}{(1+f^2)^2}F_2(r)\nonumber\\&+\frac{12\omega^2 f'f}{(1+f^2)^2}F_3(r)-\frac{l(l+1)}{r^2}\frac{2f^2}{1+f^2}F_4(r)\Bigg].\label{H1a}
\end{align}
By substituting the explicit forms of the functions $F_n(r)$ into the above expression, the quartic Hamiltonian density $\widetilde{\cal H}_4$ is obtained in the fully expanded form:
\begin{align}
\widetilde{\cal H}_4=&-\kappa\Bigg[-2\overbrace{(\beta e^2+2\gamma e^2)}^{0}\frac{3\omega^4f^2-f'^4}{(1+f^2)^4}\nonumber\\&+4(\beta e^2-4)\frac{\omega^2f^2f'^2}{(1+f^2)^4}\nonumber\\&+4(\beta e^2-1)\frac{l(l+1)}{r^2}\frac{f^2(\omega^2f^2+f'^2)}{(1+f^2)^3}\nonumber\\
&+\underbrace{(2\beta e^2+2\gamma e^2-1)}_{\beta e^2-1}\frac{l^2(l+1)^2}{r^4}\frac{f^4}{(1+f^2)^2}\Bigg].\label{H1b}
\end{align}
The total energy $E$ of the field configuration is determined by integrating the dimensionless Hamiltonian density over the volume, incorporating the $\sqrt{-g}$ factor for volume element transformation (which simplifies to $r^2 \sin\theta$ in spherical coordinates):
\begin{align*}
E=\int_{{\mathbb{R}}^3}d^3\xi\sqrt{-g}\,\Big(\widetilde {\cal H}_{02}+\widetilde {\cal H}_4\Big)
=4\pi\int_0^{\infty}dr\, r^2\Big(\widetilde {\cal H}_{02}+\widetilde {\cal H}_4\Big).
\end{align*}

Following our previous analysis of the extended Skyrme-Faddeev model \cite{Ferreira:2010jb}, we assume that $e^2<0$ and $M^2>0$. The Hamiltonian density of the quadratic terms, $\widetilde{\cal H}_{02}$ (Equation \eqref{H0}), is inherently positive definite due to its structure. Upon examining the full Hamiltonian density $\widetilde{\cal H}$ (Equation \eqref{H2}), specifically the quartic contribution $\widetilde{\cal H}_4$ (Equation \eqref{H1b}), we observe that most terms contribute positively to the energy density under the following constraints on the coupling constants:
\be
2 \beta e^2+2\gamma e^2-1\ge 0,\qquad\text{and}\qquad \beta e^2-4\ge 0.\label{leftbottom}
\ee
The only term in $\widetilde{\cal H}_4$ that is not intrinsically guaranteed to be positive definite is the one proportional to the combination of coupling constants $\beta e^2+2\gamma e^2$. To definitively ensure the overall positive definiteness of the total Hamiltonian density $\widetilde{\cal H}$, we impose the strong constraint:
\be
 \beta e^2+2\gamma e^2=0.\label{upper}
 \ee
This condition not only nullifies the potentially negative term but also simplifies the combination of coupling constants appearing in the last term of Equation \eqref{H1b} from $(2\beta e^2+2\gamma e^2-1)$ to ${\beta e^2-1}$.

\begin{figure}[h!]
\centering
\subfigure{\includegraphics[width=0.4\textwidth,height=0.4\textwidth, angle =0]{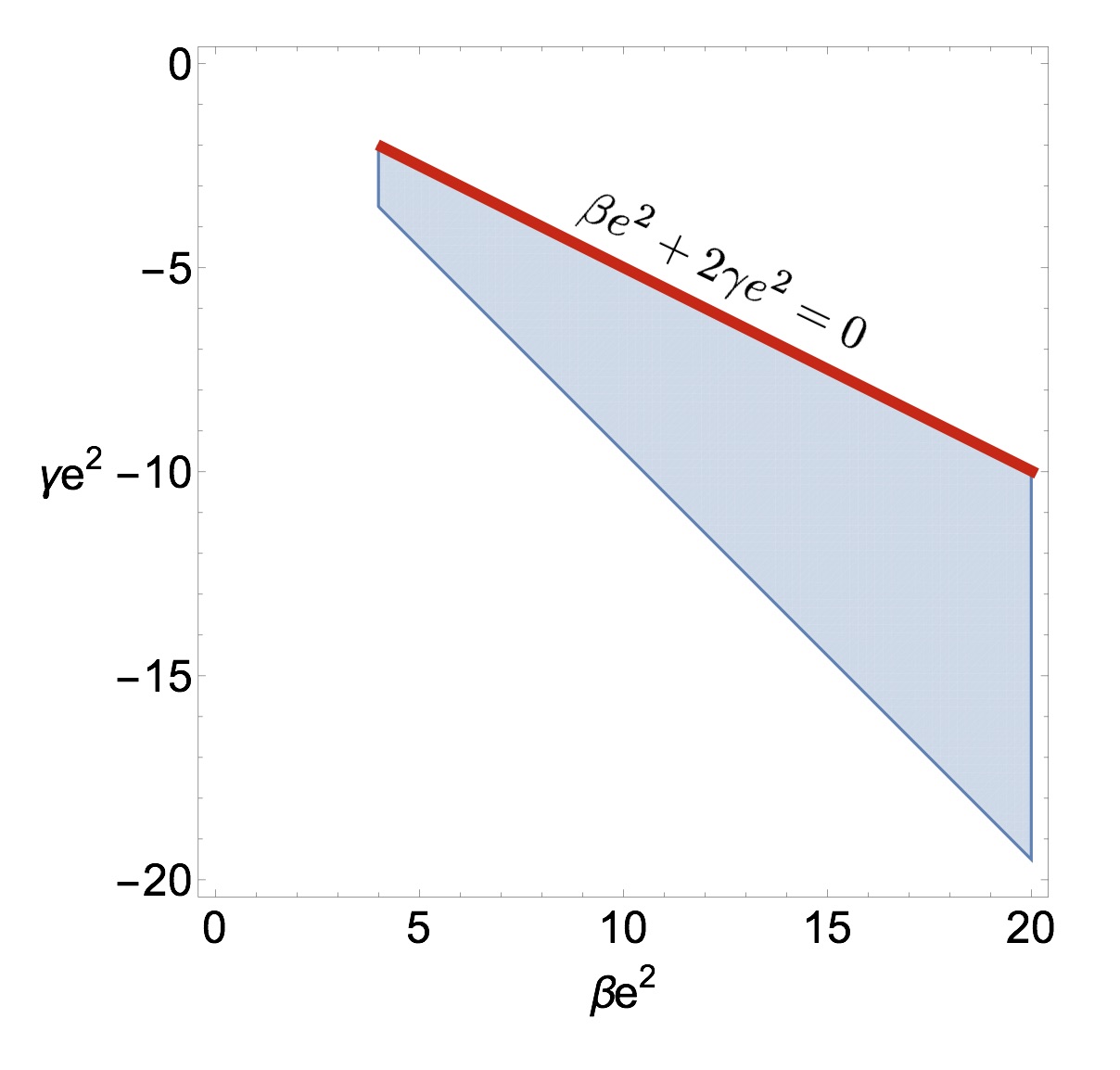}}
\caption{The coupling constant space delimited by the conditions $\beta e^2+2\gamma e^2\le 0$, $2\beta e^2+2\gamma e^2-1\ge 0$, and $\beta e^2-4\ge 0$, ensuring a non-negative contribution from the quartic terms.}\label{bound}
\end{figure}

The parameter space defined by the constraints \eqref{leftbottom} and $\beta e^2+2\gamma e^2\le 0$ is visually represented in FIG.~\ref{bound}. The upper boundary of this region, which is demarcated by the red line, corresponds precisely to the equality condition established in Equation \eqref{upper} ($\beta e^2+2\gamma e^2=0$). This boundary line represents the specific combination of coupling constants that guarantees a positive definite contribution of the quartic terms to the total energy density.

Notably, within the regime where ${e^2 < 0}$, the requirement for positive definiteness imposes constraints on the coupling constants such that ${\beta < 0}$ (which implies $\beta e^2 > 0$) and ${\gamma > 0}$ (which implies $\gamma e^2 < 0$). Below the upper boundary (the red line)--that is, for ${\beta e^2 + 2\gamma e^2 < 0}$?solutions with a non-negative energy density may still be permissible, provided that the additional condition ${3\omega^4 f^2 > f'^4}$ is locally satisfied.

By applying these constraints ($\beta e^2+2\gamma e^2=0$, $e^2<0$, $M^2>0$), the radial equation of motion (Equation \eqref{eom1}) is simplified and can be expressed in the form:
\begin{align}
&f''+\frac{2}{r}f'+\left(\omega^2\frac{1-f^2}{1+f^2}-\frac{l(l+1)}{r^2}\right)f-\frac{2ff'^2}{1+f^2}\nonumber\\&-\frac{\tilde\mu^2}{8}{\rm sgn}(f)\sqrt{1+f^2}+\kappa \,G(r) =0.\label{eq:radial}
\end{align}
Here, the function $G(r)$ represents the non-linear contributions resulting from the quartic terms and is defined as:
\begin{widetext}
\begin{align}
G(r)=&-\underbrace{(\beta e^2+2\gamma e^2)}_{0}\frac{1}{(1+f^2)^2}\Bigg[3f'^2f''+\frac{2}{r}f'^3-\frac{6ff'^4}{1+f^2}-\omega^4\frac{(1-f^2)f^3}{1+f^2}\Bigg]\nonumber\\
&-(\beta e^2-1)\frac{l(l+1)}{r^2}\frac{1}{1+f^2}\Bigg[f^2f''+\frac{(1-2f^2)ff'^2}{1+f^2}+\omega^2\frac{(2-f^2)f^3}{1+f^2}\Bigg]\nonumber\\
&+(\beta e^2-4)\frac{\omega^2}{(1+f^2)^2}\left[f^2f''+\frac{2}{r}f^2f'+\frac{(1-3f^2)ff'^2}{1+f^2}\right]\nonumber\\
&+\underbrace{{\textstyle\frac{1}{2}}(2\beta e^2+2\gamma e^2-1)}_{\frac{1}{2}(\beta e^2-1)}\frac{l^2(l+1)^2}{r^4}\frac{f^3}{1+f^2},
\end{align}
\end{widetext}
It is important to observe that the first term in the expression for $G(r)$ vanishes due to the imposed constraint $\beta e^2+2\gamma e^2=0$, and the coefficient of the final term simplifies to $\frac{1}{2}(\beta e^2-1)$.

\section{Noether currents}
\label{sec3}
A primary concern when studying Q-balls is to ensure the stability of their solutions. Given the non-topological nature of Q-balls, their stability must be generated by a mechanism other than topology. Consequently, this section focuses on the model's global symmetry, as it is responsible for the existence of the conserved Noether charges that are crucial for determining stability.

The stability of Q-balls, a class of non-topological solitons, is critically dependent upon the relationship between their energy ($E$) and their conserved Noether charge ($Q$), typically expressed as $E \sim Q^{\delta}$ \cite{Arodz:2008nm}. The exponent $\delta$ is of crucial importance for determining stability. Specifically, when the exponent $\delta$ is less than unity ($\delta < 1$), the energy of a single Q-ball of charge $Q$ is lower than the combined energy of a collection of smaller Q-balls possessing the same total charge. This illustrates that if two compact Q-balls with positive charges, $Q_1$ and $Q_2$, were to merge into a single Q-ball of charge $Q=Q_1+Q_2$, the total energy would decrease. Conversely, this implies that a single Q-ball will not spontaneously undergo fission into smaller Q-balls, as such a decay mechanism would necessitate an increase in the total energy, rendering it energetically unfavorable. This stability criterion, $\delta < 1$, is known to hold specifically for compact Q-balls within the $\mathbb{C}P^N$ model \cite{Klimas:2017eft}.

The presence of additional terms in the Lagrangian, specifically the quartic interactions, can potentially alter this relationship ($E \sim Q^{\delta}$). It is not immediately obvious that such terms would necessarily preserve the critical inequality $\delta < 1$. Consequently, a thorough examination of the precise scaling relationship between the energy and the Noether charge is essential to ascertain the stability of Q-balls in models that incorporate these more complex interactions.

The Noether charge of the field configurations is introduced by considering the Lagrangian density's invariance under the $U(1)^{N}$ global symmetry transformation:
\[
u_m\rightarrow e^{i\alpha_m}u_m,\qquad m=-l,\ldots,l.
\]
Here, the $\alpha_m$ are $2l+1$ independent, real-valued parameters. The Noether currents $J_{\mu}^{(m)}(x)$ associated with this symmetry are first rewritten in the dimensionless Cartesian coordinates $\tilde{x}^{\mu}$. The relationship is given by $\widetilde{J}_{\mu}^{(m)}(\tilde{x})=r_0^3J_{\mu}^{(m)}(x)$, where:
\be
\widetilde J_{\mu}^{(m)}(\tilde x)=-\frac{4i}{(1+u^{\dagger}\cdot u)^2}\sum_{m'=-l}^l \left[u^*_m \widetilde C_{\mu\nu}(\widetilde x)\Delta^2_{mm'}\frac{\partial u_{m'}}{\partial \tilde x_{\nu}}-{\rm c.c}\right].\label{NC}
\ee
When subsequently expressed in the curvilinear coordinates $\xi^{\mu}=(t,r,\theta,\phi)$, the currents from Equation \eqref{NC} transform as a linear combination: $\widetilde{J}_{\mu}^{(m)}(\tilde{x})=\frac{\partial\xi^{\alpha}}{\partial \tilde{x}^{\mu}}\widetilde{J}_{\alpha}^{(m)}(\xi)$, where $\widetilde{J}_{\alpha}^{(m)}(\xi)$ is defined by the expression:
\be
\widetilde J_{\alpha}^{(m)}(\xi)=-\frac{4i}{(1+u^{\dagger}\cdot u)^2}\left[ \widetilde C_{\alpha\beta}(\xi)g^{\beta\sigma}K^{(m)}_{\sigma}-{\rm c.c}\right],\label{NC1}
\ee
with the tensor $K^{(m)}_{\sigma}$ (with further specification provided in  Appendix \eqref{appendB}) given by:
\be
K^{(m)}_{\sigma}=u^*_m\sum_{m'=-l}^l \Delta^2_{mm'}\frac{\partial u_{m'}}{\partial\xi^{\sigma}}.\nonumber
\ee
The symbol $\widetilde{C}_{\alpha\beta}(\xi)$ is defined by Equation \eqref{Cxi}.

For the chosen radial ansatz, the spatial components $\widetilde{J}_{r}^{(m)}(\xi)$ and $\widetilde{J}_{\theta}^{(m)}(\xi)$ vanish. The remaining non-zero components are the time and angular currents:
\begin{align}
\widetilde J_{t}^{(m)}(\xi)&=8\omega\frac{f^2}{(1+f^2)^2}\left[1+\kappa\left(F_1(r)+3\frac{f'}{f}F_3(r)\right)\right]\frac{(l-m)!}{(l+m)!}(P^m_l)^2,\\
\widetilde J_{\phi}^{(m)}(\xi)&=8m\frac{f^2}{1+f^2}(1-\kappa F_4(r))\,\frac{(l-m)!}{(l+m)!}(P^m_l)^2.
\end{align}
Here, $P^m_l(\cos\theta)$ denotes the associated Legendre polynomials. The continuity equation, 
$$\frac{1}{\sqrt{-g}}\partial_{\alpha}(\sqrt{-g}g^{\alpha\beta}\widetilde{J}_{\beta}^{(m)})=0,$$ simplifies to $\partial_t\widetilde{J}_{t}^{(m)}+\frac{1}{r^2\sin^2\theta}\partial_{\phi}\tilde{J}_{\phi}^{(m)}=0$. This condition is explicitly satisfied because both $\widetilde{J}_{t}^{(m)}$ and $\widetilde{J}_{\phi}^{(m)}$ are exclusively functions of the spatial coordinates $r$ and $\theta$, rendering them independent of $t$ and $\phi$.

Based on the work of \cite{Klimas:2017eft}, the Noether charges $Q_t^{(m)}$ are defined as the integral of the time component of the dimensionless Noether current, $\widetilde{J}_t^{(m)}(\xi)$, over the entire spatial volume:
\be
Q_t^{(m)}:=\frac{1}{2}\int_{{\mathbb{R}}^3}d^3\xi\sqrt{-g}\widetilde J^{(m)}_t(\xi)
\ee
Upon applying the ${Q}$-ball ansatz from Equation \eqref{ansatz} and integrating over the angular coordinates, the expression for the Noether charge simplifies to:
{\small
\begin{align}
Q_t^{(m)}&=\omega\frac{16\pi}{2l+1}\int_0^{\infty}dr\, r^2\frac{f^2}{(1+f^2)^2}\left[1+\kappa\left(F_1+3\frac{f'}{f}F_3\right)\right]\label{Qt}
\end{align}}
It is important to note that all Noether charges $Q_t^{(m)}$ are independent of the field index $m$. When the term proportional to $\kappa$ is excluded (i.e., $\kappa \to 0$), this expression recovers the Noether charges previously derived for the ${C}P^N$ model in the absence of quartic interactions (as discussed following the ansatz definition).

For completeness, we also define the quantities $Q_{\phi}^{(m)}$; these, however, do not represent conserved Noether charges. They are defined by the volume integral:
\begin{align*}
Q_\phi^{(m)}:=\frac{3}{2}\int_{{\mathbb{R}}^3}d^3\xi\sqrt{-g}\,\frac{\widetilde J^{(m)}_\phi(\xi)}{r^2}.
\end{align*}
Integrating this expression over the angular coordinates yields the final radial form:
\be
Q_\phi^{(m)}=m\frac{48\pi}{2l+1}\int_0^{\infty}dr\,\frac{f^2}{1+f^2}\left[1-\kappa F_4\right].\label{Qphi}
\ee

The total energy $E$, defined by the volume integral $E=\int_{{\mathbb{R}}^3}d^3\xi\sqrt{-g}\,\widetilde {\cal H}$, can be expressed in terms of the Noether charge $Q_t^{(m)}$ and the non-conserved charge $Q_\phi^{(m)}$, as detailed in Equation \eqref{engC2}. Under the constraint ${\beta e^2+2\gamma e^2=0}$, the expression for the total energy is:
\begin{align}
E&=\sum_{m=-l}^l\omega Q_t^{(m)}+\sum_{m=-l}^lm Q_\phi^{(m)}+\int_{{\mathbb{R}}^3}d^3\xi\sqrt{-g}\left[\frac{4f'^2}{(1+f^2)^2}+\tilde\mu^2V\right]\nonumber\\
&-\kappa(\beta e^2-1)\int_{{\mathbb{R}}^3}d^3\xi\sqrt{-g}\,\Bigg[\frac{l(l+1)}{r^2}\frac{4\omega^2f^4}{(1+f^2)^3}-\frac{l^2(l+1)^2}{r^4}\frac{f^4}{(1+f^2)^2}\Bigg].
\end{align}

\section{Compactons}
\label{sec4}

In \cite{Klimas:2017eft}, the authors demonstrated that the $\mathbb{C}P^N$ model, utilizing the potential defined in Equation \eqref{potencjal}, yields compact solutions for odd values of $N$. This implies that the field solutions possess a finite spatial size, characterized by a specific outer radius, $R_{\text{out}}$. At this radius, the field solution and its first derivative smoothly approach a vacuum state (i.e., they approach zero). Beyond this boundary ($r > R_{\text{out}}$), the field is identically zero, meaning the solution is non-trivial only within the region $r<R_{\text{out}}$. These compact solutions are further classified into two distinct types--Q-balls and Q-shells--based on the behavior of their radial functions.

The Q-ball solutions are specifically observed for odd field numbers, namely $N=1$ and $N=3$. For the $N=1$ case, the radial function $f(r)$ possesses a non-zero value at the center and a negative slope there, defined by the boundary conditions $f(0)>0$ and $f'(0)<0$. Conversely, when $N=3$, the radial function is zero at the origin but starts with a positive slope, characterized by the boundary conditions $f(0)=0$ and $f'(0)>0$.

For higher odd values of $N$, specifically $N=5, 7, 9, \ldots$, the resulting solutions are qualitatively distinct from Q-balls and are consequently designated as Q-shells. A key characteristic of these solutions is the presence of a vacuum state within a spherical region, spanning from the origin up to a radius $R$ ($0 \le r \le R$).

\subsection{Expansion at the compacton boundary}
In the vicinity of the external compacton radius, $R_{\text{out}}$, the field values approach the vacuum state of zero. Consequently, the behavior of the radial function is governed by the quadratic terms in the Lagrangian. It is therefore anticipated that the leading-order behavior of the compacton's radial function will be quadratic, mirroring the characteristics observed in the previously examined ${C}P^N$ model \cite{Klimas:2017eft}. To analyze the behavior for $r < R_{\text{out}}$, we substitute the following power series expansion into the field equation, imposing the condition $f(r)=0$ for $r \ge R_{\text{out}}$:
\[
f(r)=b_2(R_{\rm out}-r)^2+b_3(R_{\rm out}-r)^3+\ldots
\]  
The radial equation yields the following expressions for the first three coefficients $b_k$:
\begin{align}
b_2&=\frac{\tilde \mu^2}{16}\label{eq:b2}\\
b_3&=\frac{\tilde \mu^2}{24R_{\rm out}}\label{eq:b3}\\
b_4&=\frac{\tilde \mu^2}{192 R_{\rm out}^2}\Big(8+l(l+1)-R_{\rm out}^2\omega^2\Big)+\kappa(\beta e^2+2\gamma e^2)\frac{\tilde\mu^6}{2048}\label{eq:b4}
\end{align}
This result demonstrates that the inclusion of quartic terms in the model does not affect the two leading coefficients, $b_2$ and $b_3$. Furthermore, if the constraint $\beta e^2+2\gamma e^2=0$ is imposed, the contribution to the $b_4$ term resulting from the quartic interactions also vanishes, making $b_4$ independent of the quartic terms. It is noteworthy that the $b_4$ term is the first coefficient to exhibit an explicit dependence on the number $l$.

\subsection{Expansion at the center}
To obtain a solution for the radial function $f(r)$, we assume it can be expressed as a power series expansion around the center, $r=0$: 
\begin{equation}
f(r)=\sum_{k=0}^{\infty}a_k r^k.\label{expansion-center}
\end{equation}
By substituting this series into the field equation \eqref{eq:radial} and expanding the equation into a power series in $r$, we can solve for the coefficients, $a_k$. This is achieved by setting each coefficient of the resulting series to zero, which allows us to determine the unique values of the $a_k$ coefficients that satisfy the field equation. For our calculations, we will focus on the first few coefficients, $a_k$, as they determine the leading behavior of the radial function near the center. The values of these coefficients are heavily dependent on the number $l$, which determines the number of scalar fields, $N=2l+1$.

\subsubsection{Q-ball case: $N=1$, $(l=0)$}
Unlike the  $CP^N$ model, which lacks quartic terms, the ESF model yields more than one solution for the coefficients $a_k$. Upon substituting the power series expansion from Equation \eqref{expansion-center} into the radial equation \eqref{radialeq} and expanding the resulting expression in powers of $r$, we obtain:
\begin{equation}
-\frac{l(l+1)}{r^2}a_0-\frac{l(l+1)-2}{r}a_1+{\cal O}(1)=0.\label{eq:expansion}
\end{equation}
The singular terms, which are proportional to $r^{-2}$ and $r^{-1}$, must be eliminated to ensure a finite solution. This is achieved by an appropriate selection of the coefficients $a_0$ and $a_1$.

The first case, $l=0$, represents a direct generalization of previously studied Q-ball solutions (corresponding to the $N=1$ case). In this instance, the constraint requires that the coefficient $a_1$ vanishes ($a_1=0$), while $a_0$ remains a free constant determined by the initial conditions at the center. Consequently, the radial function $f(r)$ near the center, $r=0$, behaves like a parabola, and the second coefficient, $a_2$, is determined by the following complex expression:
\begin{align}
a_2&=\frac{\left(1+a_0^2\right){}^{7/2} \tilde\mu ^2-8  \omega ^2 a_0
   \left(1-a_0^2\right) \left(1+a_0^2\right)^2-8 \kappa  \omega ^4 \left(\beta  e^2+2 \gamma  e^2\right)a_0^3 \left(1-a_0^2\right) }{48 \left(1+a_0^2\right) \Big((1+a_0^2)^2+\kappa  \omega ^2 \left(\beta 
   e^2-4\right)a_0^2\Big)}.\label{CP1_a2}
\end{align}
The terms containing the coefficient $\kappa$ explicitly represent the modifications to the central behavior that arise from the presence of quartic terms within the model's Lagrangian.

There is a second solution that exhibits a linear behavior near the origin, as determined by the coefficient $a_1$:
\begin{equation}
a_1=\pm\sqrt{\frac{\left(1+a_0^2\right){}^2+\kappa  \omega ^2 \left(\beta  e^2-4\right)a_0^2 }{\kappa  \left(\beta  e^2+2 \gamma  e^2\right)}}\label{CP1_a1}
\end{equation}
This solution is mathematically valid only when $\kappa\neq 0$ and  $\beta  e^2+2 \gamma  e^2\neq 0$. This latter constraint implies that the energy density, as defined in Equation \eqref{H1b}, includes a term of the form $\sim\frac{3\omega^4 f^2-f'^4}{(1+f^2)^4}$ which is not guaranteed to be positive definite. It is crucial to note that the mere existence of a series expansion does not guarantee that the resulting numerical solution to the field equations will possess the desired physical properties (such as finiteness or non-negative energy). Therefore, a full numerical solution of the radial equation is required for verification.

\subsubsection{Q-ball case: $N=3$, $(l=1)$}
For Q-balls with $l=1$ (corresponding to $N=3$), the coefficient $a_0$ must be zero, while $a_1$ is a free constant. The coefficient $a_2$ is determined by the following expression:
\begin{equation}
a_2=\frac{\widetilde\mu ^2}{32 \Big(1+\kappa  \left(2-5 \beta  e^2-6 \gamma  e^2\right)a_1^2 \Big)}\label{CP3a2}
\end{equation}
The explicit dependence of $a_2$ on $a_1$ arises exclusively from the quartic terms present in the model's Lagrangian. A detailed analysis of these specific types of solutions is presented in the forthcoming section.

\subsubsection{Q-shell case: $N=5,7,9,\ldots$, $(l=2,3,4,\ldots)$}
Assuming $\text{sgn}(f)=+1$ and substituting the power series expansion \eqref{expansion-center} with the boundary conditions $a_0=a_1=0$ (consistent with $N \ge 3$ solutions) into the radial equation \eqref{radialeq}, we obtain the following expansion around $r=0$:
\begin{align*}
&-\Big(a_2 \left(l^2+l-6\right)+\frac{\tilde\mu ^2}{8}\Big)+(l-3)(l+4)a_3 r+\Big(\omega^2a_2+(l-4)(l+5)a_4\Big)r^2+\\
&+\Big(\omega^2a_3+(l-5)(l+6)a_5\Big)r^3+{\cal O}(r^4)=0.
\end{align*}
To satisfy this equation, every coefficient must vanish. When $l=2$ (i.e., $N=5$), the constant term simplifies to $-\frac{\tilde{\mu}^2}{8}$, which cannot be canceled (since $\tilde{\mu} \ne 0$). Thus, the radial equation cannot be satisfied for $l=2$. Furthermore, for values of $l$ greater than or equal to $2$, the equation for the $r^0$ coefficient yields $a_2 = -\frac{\tilde{\mu}^2}{8(l^2+l-6)}$. Since $l^2+l-6 > 0$ for $l \ge 2$, $a_2$ must be negative. This implies that $f(r)$ would be negative in the immediate vicinity of $r=0$, which contradicts our initial assumption that $f(r) \ge 0$. Consequently, it is not possible to obtain a regular radial function as a power series near $r=0$ for $l \ge 2$, meaning the model does not possess regular Q-ball solutions for these higher field numbers.

An alternative solution structure exists by considering a vacuum core, where the field $f=0$ within the sphere $r \le R_{\text{in}}$. This leads to Q-shell solutions. The expansion near this internal radius, $R_{\text{in}}$, is essentially identical to the expansion performed near the external compacton radius, $R_{\text{out}}$, since the function is approaching zero quadratically. Therefore, we assume the series expansion:
\[
f(r)=c_2(r-R_{\rm in})^2+c_3(r-R_{\rm in})^3+\ldots
\]  
Substituting this into the field equation yields the relationship $c_k=(-1)^k b_k$, where $b_k$ are the coefficients defined previously in Equations \eqref{eq:b2}-\eqref{eq:b4}, but with $R_{\text{out}}$ formally replaced by $R_{\text{in}}$.

\section{Numerical solutions}
In this section, we present several numerical solutions for Q-balls and Q-shells within the ESF model. The parameter space for the coupling constants is extensive, and a detailed investigation of all possible cases would be exceptionally time-consuming and exceeds the primary scope of this paper. Our objective here is to showcase some novel solutions that arise specifically due to the presence of quartic terms. We will also thoroughly examine the resulting relationship between energy and Noether charge, which is the key characteristic defining the stability of these solutions.

\subsection{Shooting method}
To determine the radial profile function, we numerically integrate the radial equation \eqref{radialeq}, starting from either the center ($r=0$) or the inner radius ($r=R_{\text{in}}$). The resulting numerical solution is dependent on a single free parameter: $a_0$ for $l=0$, $a_1$ for $l=1$, or $R_{\text{in}}$ for $l\ge 2$. By selecting an appropriate value for this parameter, the radial function can simultaneously satisfy the external compacton boundary conditions, ${f(R_{\text{out}})=0}$ and ${f'(R_{\text{out}})=0}$.

It is important to note that the numerical value of the parameter $\omega$ is significant. As with the  $CP^N$ model, not all values of $\omega$ are permissible. For certain values, the radial function may become unbounded, which indicates a non-physical solution. Our analysis will focus exclusively on finite-energy solutions.

\subsection{$CP^1$ Q-balls}

When the leading behavior of the radial function $f(r)$ in the vicinity of the origin ($r=0$) is described by a parabolic function, $f(r)=a_0+a_2r^2+\ldots$, we can search for compact Q-balls similar to the established ${C}P^1$ Q-balls discussed in \cite{Klimas:2017eft}. This approach is particularly relevant as the model approaches the limit $1/e^2\to 0$, $\beta\to 0$, and $\gamma\to 0$. However, being interested in a wider range of coupling constants, we analyze several cases, including those with non-trivial parameter values where the quartic terms play an active role. Given the extensive size of the complete parameter space, a detailed study of the entire region is beyond the scope of this paper. Instead, our aim is to highlight a few interesting cases that differ significantly from established models, thereby demonstrating the specific influence of the quartic terms in the Lagrangian.

\begin{figure}[h!]
\centering
\subfigure[]{\includegraphics[width=0.32\textwidth,height=0.2\textwidth, angle =0]{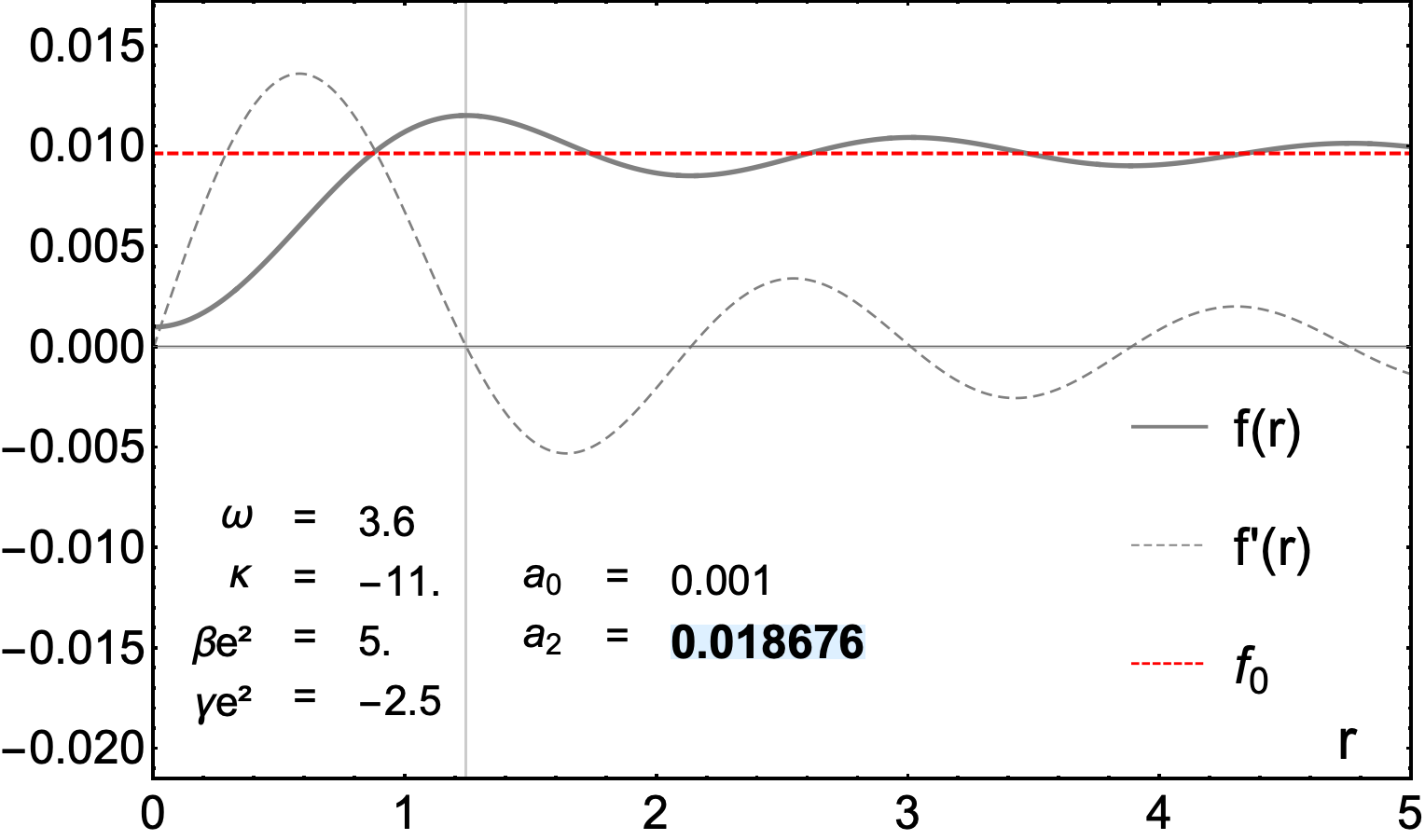}}
\hskip 0.2cm
\subfigure[]{\includegraphics[width=0.32\textwidth,height=0.2\textwidth, angle =0]{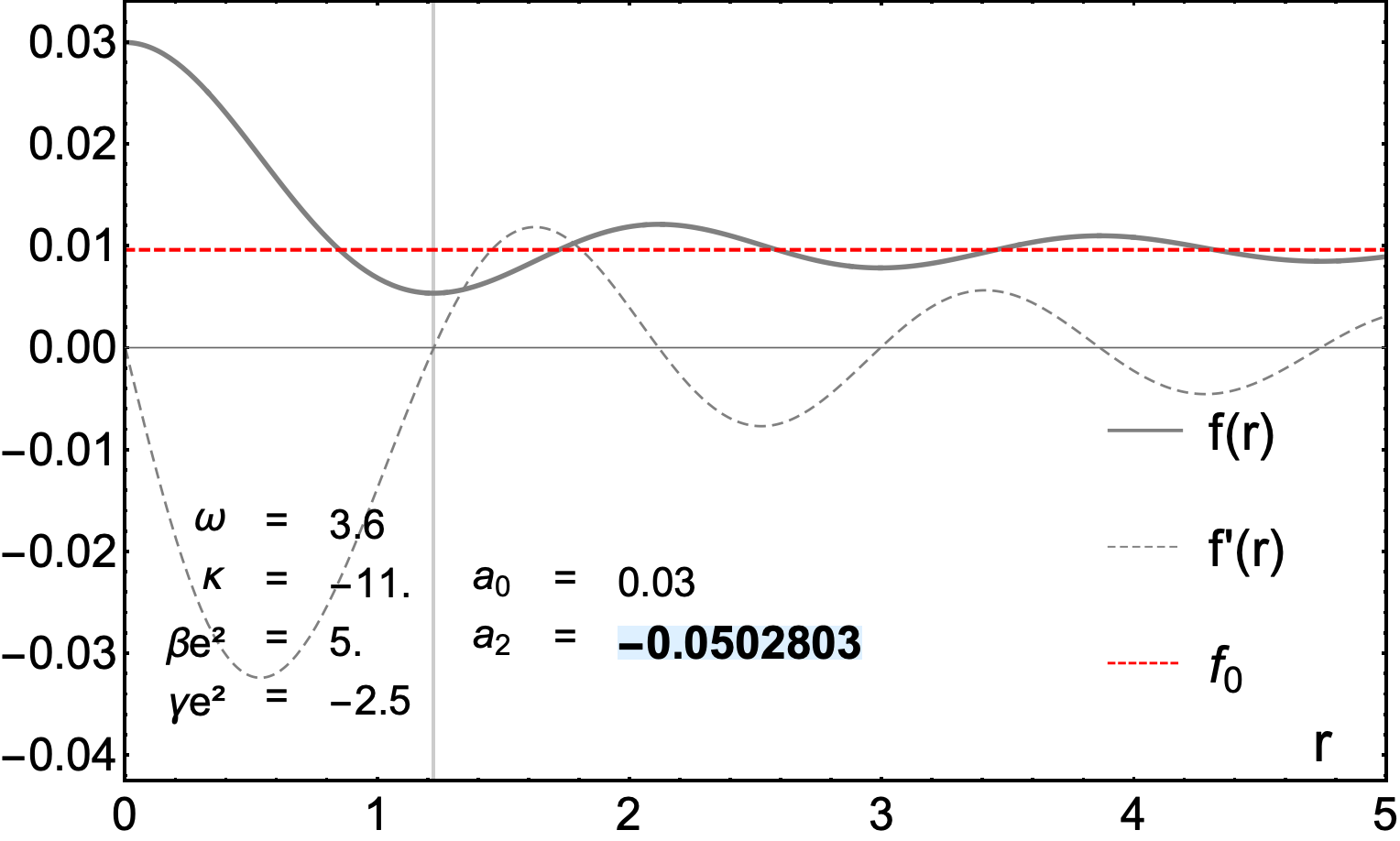}}
\hskip 0.2cm
\subfigure[]{\includegraphics[width=0.32\textwidth,height=0.2\textwidth, angle =0]{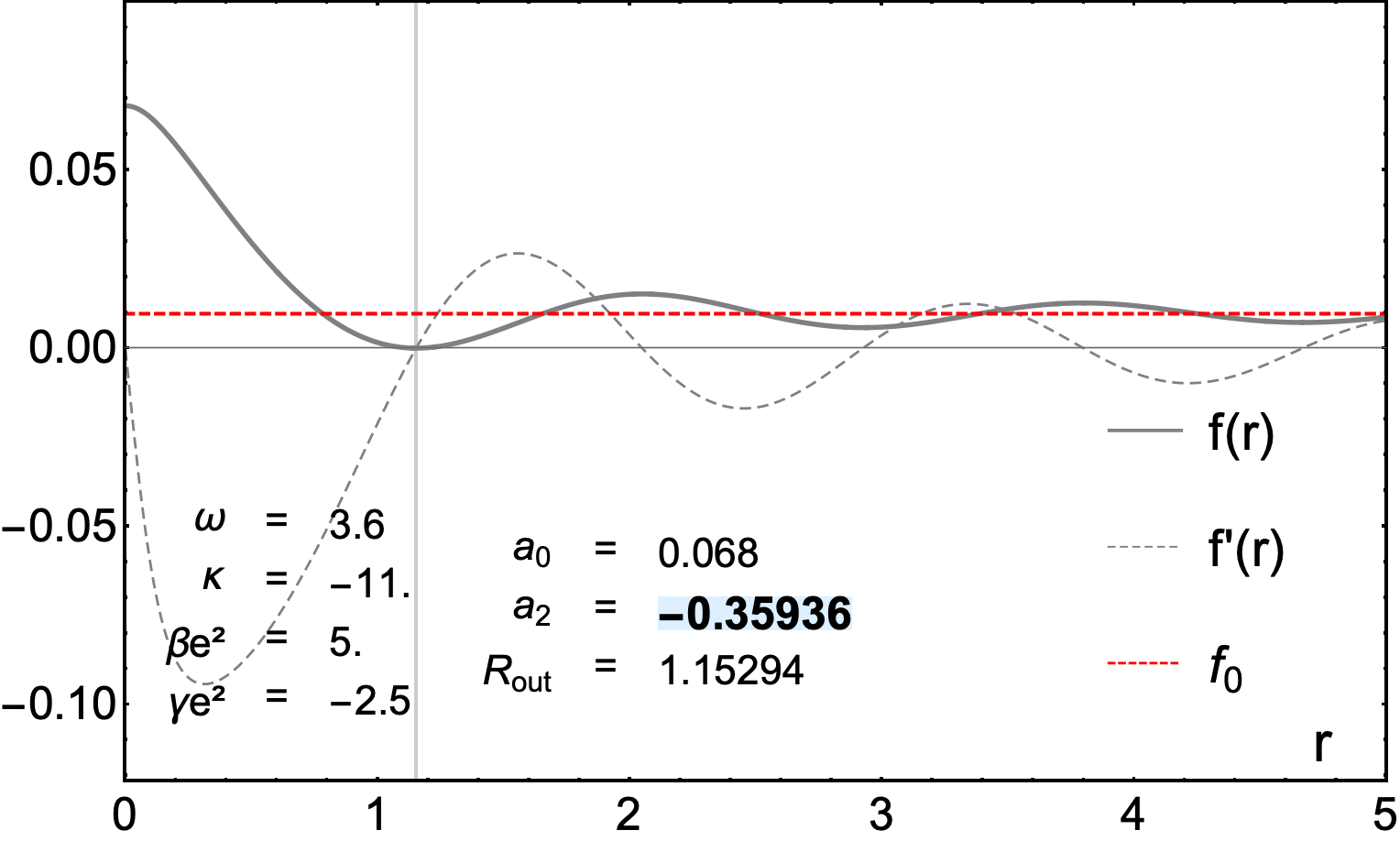}}
\caption{Radial solutions for the $CP^1$ model are characterized by the value of $a_0$. The solutions fall into three categories:  (a) those with $a_2>0$, (b) those with  $a_2<0$ and (c) the compacton case, where the radial function and its derivative simultaneously vanish at the external radius, i.e., $f(R_{\rm out})=0=f'(R_{\rm out})$.}
\label{fig:CP1radialsolutions}
\end{figure}

Before discussing compactons, we present three characteristic examples of radial functions obtained by numerically integrating the radial equation as a function of the free parameter $a_0$. For the first case, depicted in panel (a) of Figure \ref{fig:CP1radialsolutions}, the radial curve exhibits a positive second derivative at the center ($a_2>0$). This results in a curve that never reaches the vacuum value of $f=0$. Consequently, the solution is physically unacceptable as it corresponds to infinite total energy.

The stationary constant solution, $f_0$, is depicted by the red dashed line in the figure. It is determined as the solution to the algebraic equation derived from the constant-field limit of the radial equation:
\[
\frac{1-f_0^2}{(1+f_0^2)^{3/2}}f_0=\frac{\tilde\mu^2}{8\omega^2}.
\]
The exact analytic solution for $f_0$ was used to plot this line; however, due to its highly complex nature, it will not be presented here. For small values of $f_0$, the solution can be accurately approximated by the simpler linear relation $f_0 \approx \frac{\tilde{\mu}^2}{8\omega^2}$.

For the second case, depicted in panel (b) of FIG.~\ref{fig:CP1radialsolutions}, the solution is characterized by a negative value for the second derivative ($a_2<0$). The radial curve shown, however, still does not represent an acceptable physical solution.

Finally, the third panel (c) represents a curve for which the value of $a_0$ has been precisely adjusted (via the shooting method) to satisfy the compacton boundary conditions: $f(R_{\text{out}})=0$ and $f'(R_{\text{out}})=0$.  In this case, the non-trivial solution can be smoothly matched with the constant vacuum solution, $f(r)=0$. This type of solution, understood as a weak solution to the differential equation, possesses a finite energy value and is therefore physically acceptable. The matching point is denoted by $R_{\text{out}}$ and is referred to as the outer radius of the compacton.

In the example provided, a compacton solution can be obtained for a specific value of the parameter $\omega$. However, this is not a general case. For certain values of $\omega$, it is possible that as the central field value $a_0$ is increased, and the first local minimum approaches the horizontal axis, the second derivative at the center, $f''(0)$, may reach negative infinity before the field $f(r)$ reaches zero at the radius $R$ determined by the condition $f'(R)=0$. Consequently, a more detailed analysis of the central coefficient $a_2$, as given by Equation \eqref{CP1_a2}, is necessary to precisely define the parameter regions where physical solutions exist. 

\begin{figure}[h!]
\centering
{\includegraphics[width=0.8\textwidth,height=0.5\textwidth, angle =0]{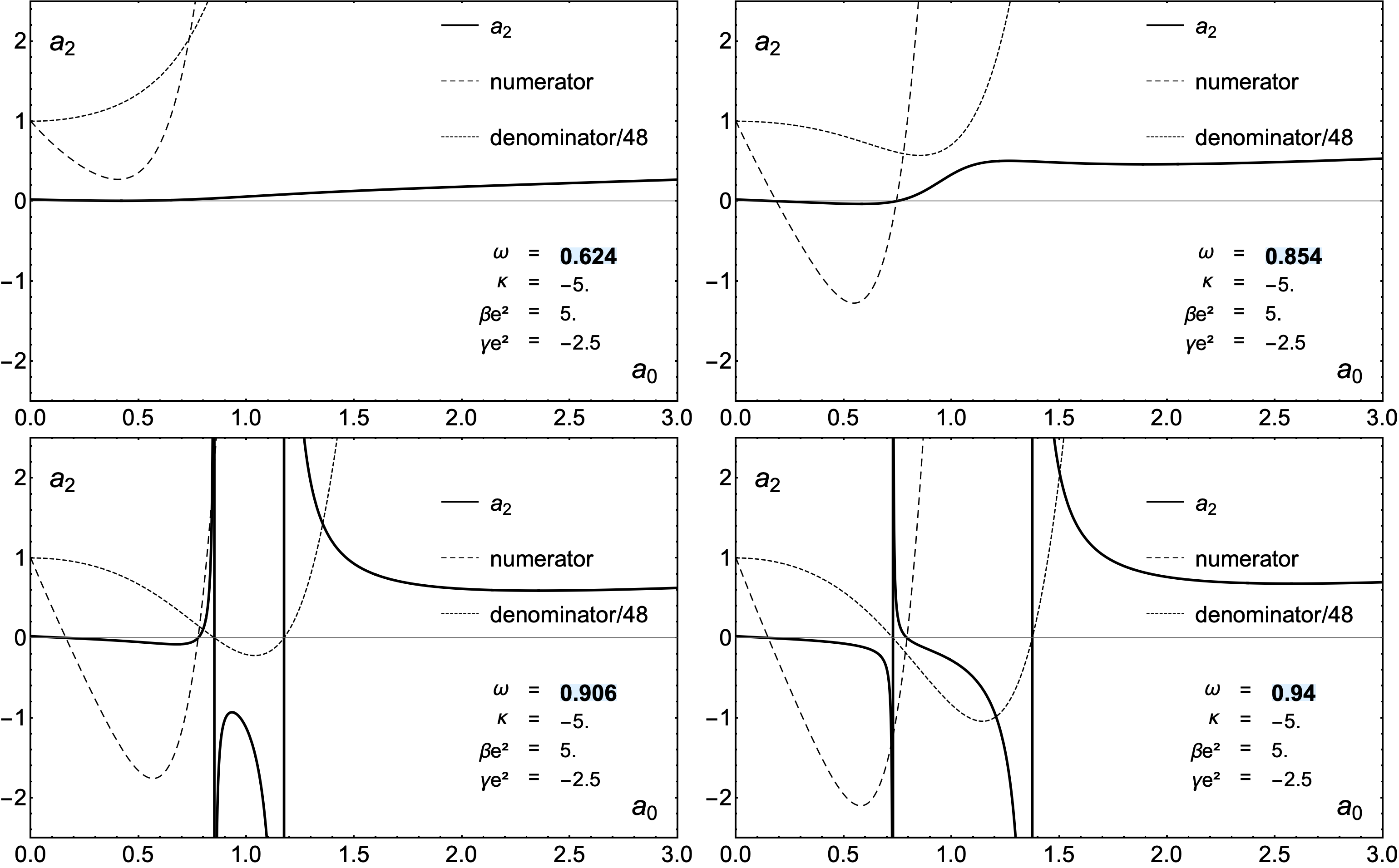}}
\caption{The dependence of the coefficient $a_2$ on the parameter $a_0$ is shown for increasing values of $\omega$ for the $CP^1$ case.}
\label{fig:CP1grid}
\end{figure}

In FIG.~\ref{fig:CP1grid}, we plot the coefficient $a_2$ as a function of the central amplitude $a_0$, varying the frequency $\omega$ while keeping all other coupling constants fixed. To improve the clarity of the analysis, we also plot the numerator and the denominator of the expression for $a_2$ (Equation \eqref{CP1_a2}) separately. For better visibility in the plot, the denominator has been scaled by a factor of $48$.

The top-left panel of FIG.~\ref{fig:CP1grid} illustrates the coefficient $a_2$ for a low frequency, specifically $\omega=0.624$, where $a_2$ is strictly positive ($a_2>0$). In this case, both the numerator and the denominator of Equation \eqref{CP1_a2} are positive for all relevant $a_0$. We therefore do not expect a compacton solution for parameter values that result in a positive $a_2$. As $\omega$ increases, we observe that the numerator becomes a function exhibiting two zeros. Due to the complexity of the equation, we cannot provide an exact analytic expression for these zeros. As $\omega$ increases, the range of $a_0$ for which $a_2$ is negative also expands. It is possible for this region to contain a compact solution, provided that, within the admissible values of $a_0$, there exists one for which the profile function $f(r)$ and its first derivative $f'(r)$ simultaneously vanish at some radius $R_{\text{out}}$. This situation is depicted in the top-right panel. For higher values of $\omega$, the situation changes further, as the denominator also develops a negative part. The zeros of the denominator lead to vertical asymptotes for $a_2$. This scenario is depicted in the bottom-left panel. The zeros of the denominator are given by the expression:
\begin{equation}
a_0^{(\pm)}=\sqrt{\frac{-(2+\alpha)\pm\sqrt{\alpha(4+\alpha)}}{2}},\qquad \alpha\equiv \kappa \omega^2(\beta e^2-4)\label{asymptota}
\end{equation}
where the existence of real solutions requires $\alpha\le -4$, providing that:
\begin{equation}
\kappa \omega^2\le -\frac{4}{\beta e^2-4}.\label{kappaomega}
\end{equation}
We note that $a_0^{(-)}<a_0^{(+)}$, and we have omitted the unphysical solutions corresponding to negative values of $a_0$. Further increasing $\omega$ causes the left zero of the denominator ($a_0^{(-)}$) to become smaller than the right zero of the numerator. As a consequence, a window where $a_2>0$ is created. This window expands as $\omega$ increases, progressively excluding more values of $a_0$ from the set of permissible initial conditions that could yield a compact solution. The bottom-right panel illustrates this situation. It is important to note that condition \eqref{kappaomega} establishes a limit on the relationship between parameters $\kappa$ and $\omega$. For a fixed coupling constant $\kappa$, for instance, the parameter $\omega$ has a minimum permissible value given by $\omega_{\min}=\frac{2}{\sqrt{-\kappa(\beta e^2-4)}}$.

The condition $\alpha\le -4$ is not a significant restriction on the parameter space; it simply corresponds to the existence of vertical asymptotes in the plot of $a_2$.  When the parameter is in the range $-4<\alpha\le 0$, the denominator of $a_2$ has no real zeros, and consequently, the formula \eqref{asymptota} does not apply. This latter scenario arises in the limit where the quartic terms decouple, leading back toward the pure ${C}P^N$ model.

\begin{figure}[h!]
\centering
{\includegraphics[width=1.0\textwidth,height=0.25\textwidth, angle =0]{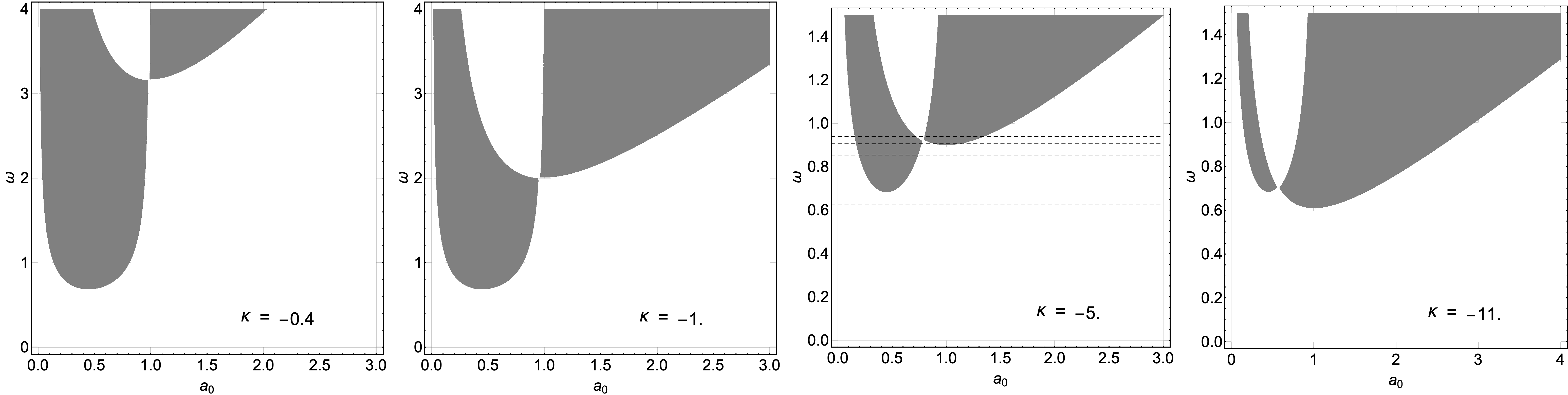}}
\caption{The sign of the coefficient $a_2$ is shown as a function of the parameters $a_0$ and $\omega$ for the $CP^1$ case with $\beta e^2=5.0$ and $\gamma e^2=-2.5$. The gray color indicates regions where ${\rm sgn}(a_2)=-1$. The horizontal lines that correspond to the values of the parameter $\omega$ used to generate Fig. \ref{fig:CP1grid}, specifically $\omega={0.624}$, $\omega={0.854}$, $\omega={0.906}$ and $\omega={0.94}$.}
\label{fig:CP1grid2}
\end{figure}

In FIG.~\ref{fig:CP1grid2}, we plot the sign of the coefficient $a_2$ as a function of the central amplitude $a_0$ and the frequency $\omega$, for several fixed values of the coupling $\kappa$. The gray color denotes the region where $\text{sgn}(a_2)=-1$. These regions are the only parts of the parameter space that may contain compacton solutions, although their existence is not guaranteed. In the third panel, we plot horizontal lines that correspond to the specific values of $\omega$ used to generate FIG.~\ref{fig:CP1grid}. This composite figure clearly demonstrates that the parameter $\omega$ always has a minimum permissible value ($\omega_{\min}$), below which no compact solutions exist.

The behavior of both the numerator and the denominator of $a_2$ (Equation \eqref{CP1_a2}) is highly dependent on the model's parameters. For instance, when we change the coupling constant to $\beta e^2=6.0$ (compared to $\beta e^2=5.0$ used in the example shown in FIG.~\ref{fig:CP1grid}), the denominator reaches zero before the numerator as $\omega$ is increased. This shift alters the range of $a_0$ that can yield compact solutions. While a detailed parameter map covering all possible solutions is desirable, obtaining such a map is far beyond the scope of this paper.

\begin{figure}[h!]
\centering
\subfigure[]{\includegraphics[width=0.32\textwidth,height=0.38\textwidth, angle =0]{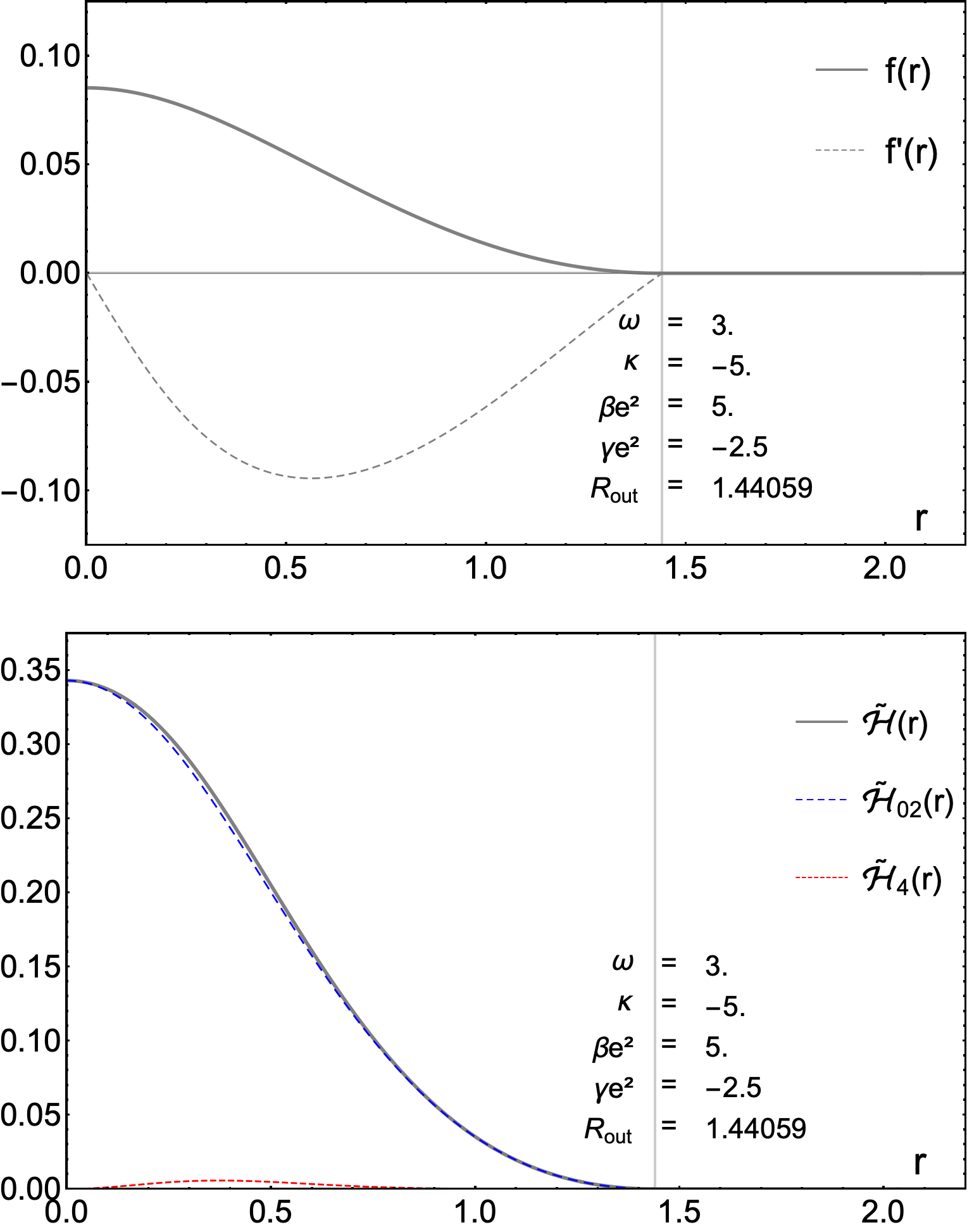}\label{fig:CP1a}}
\hskip 0.2cm
\subfigure[]{\includegraphics[width=0.32\textwidth,height=0.38\textwidth, angle =0]{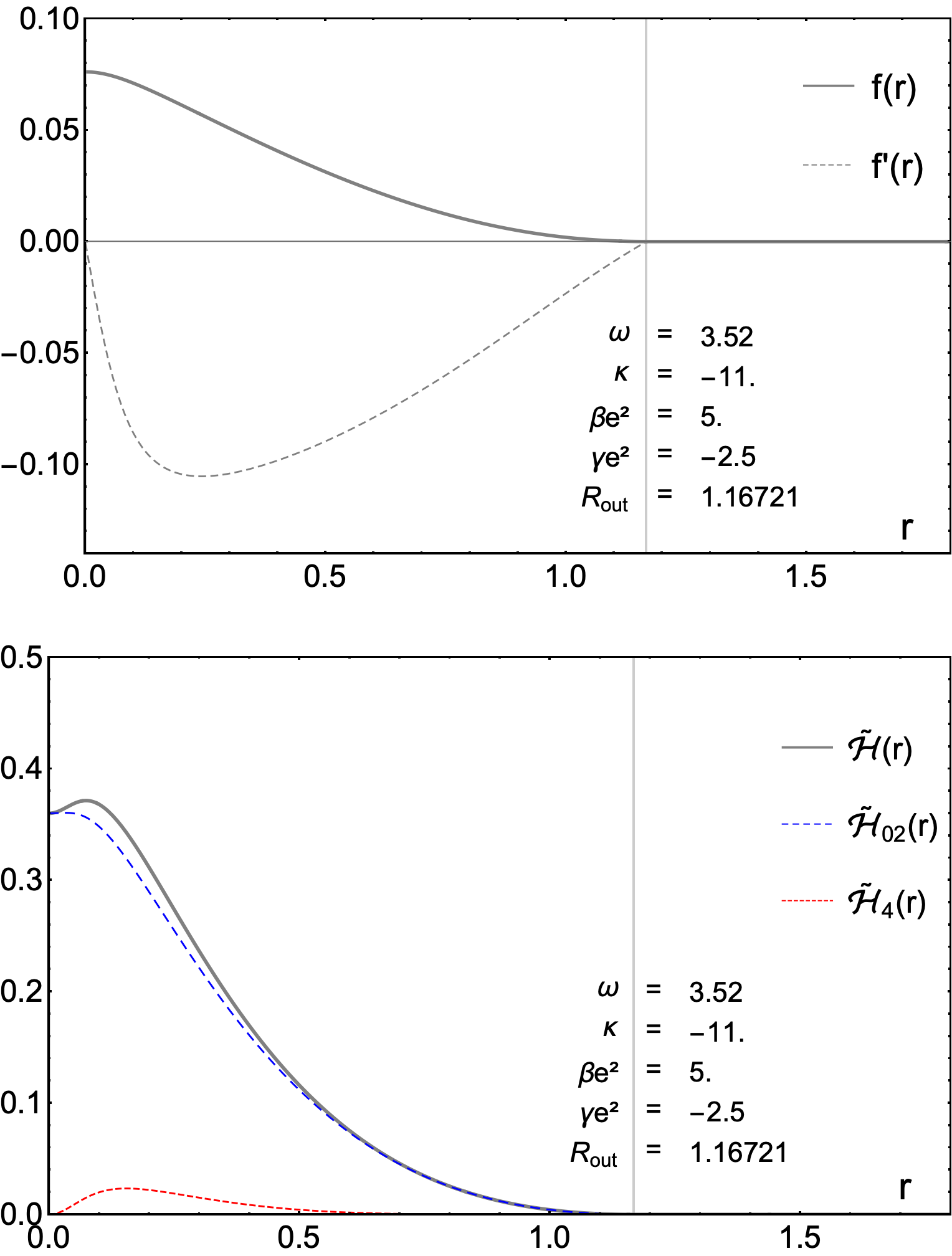}\label{fig:CP1b}}
\hskip 0.2cm
\subfigure[]{\includegraphics[width=0.32\textwidth,height=0.38\textwidth, angle =0]{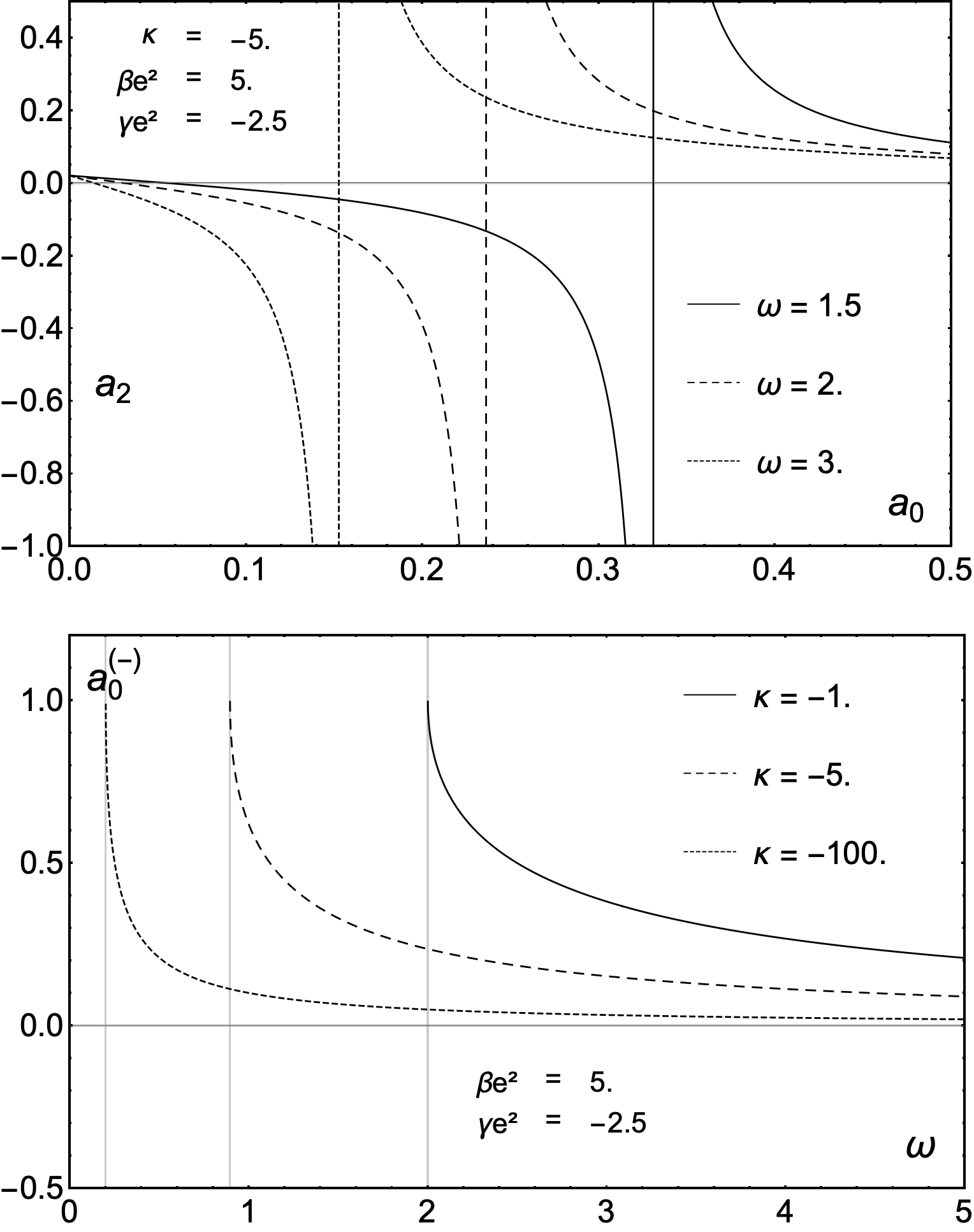}\label{fig:CP1c}}
\caption{$CP^1$ Q-balls. (a) Radial profiles and energy density for a solution with a minimal contribution from quartic terms. (b) Radial profiles and energy density for a solution with a significant contribution from quartic terms, demonstrating a shallow local minimum at the origin. (c) The top panel shows the coefficient $a_2$ as a function of $a_0$ in the region where $a_0<a^{(-)}_0$. The bottom panel shows the position of the left vertical asymptote, $a^{(-)}_0$, as a function of $\omega$.}
\label{fig:CP1}
\end{figure}

Returning to the case with $\beta e^2=5.0$ and $\kappa=-5.0$, we vary the parameter $a_0$ in an attempt to obtain compact Q-balls. The first scenario, illustrated in the top-right panel of FIG.~\ref{fig:CP1grid}, does not yield acceptable compact solutions. For this case, the values of $a_0$ that result in a negative $a_2$ form a very shallow minimum, and within this permissible range, the minimum of $f(r)$ is unable to reach zero. Similarly, for the case shown in the bottom-left panel, we also did not obtain a solution that satisfies the compacton criteria. Finally, we consider the case illustrated in the bottom-right panel of FIG.~\ref{fig:CP1grid}. For $\omega=0.94$, we are still unable to obtain the desired solution. However, as $\omega$ is increased further, the left vertical asymptote shifts toward $a_0=0$. This allows the negative coefficient $a_2$ to reach larger absolute values for small $a_0$, which in turn enables the minimum of $f(r)$ to reach zero. An example of a Q-ball obtained for ${\omega=3.0}$ is shown in FIG.~\ref{fig:CP1a}. The upper panel displays the functions $f(r)$ and $f'(r)$, while the bottom panel shows the total energy density, $\widetilde{\cal H}(r)$. Additionally, the contributions from the quadratic and potential terms, $\widetilde{\cal H}_{02}(r)$, and the quartic terms, $\widetilde{\cal H}_{4}(r)$, which constitute the total energy density, are plotted separately. It is worth noting that there are two values of $a_0$ for which $a_2$ has the same negative value. The compacton was obtained for the smaller of these two values. For the larger value, we were unable to obtain a compact solution.

In the example presented in FIG.~\ref{fig:CP1a}, the contribution of the quartic terms is minimal. To investigate a more significant influence from these terms, we increase the magnitude of the coupling constant $\kappa$. For instance, setting ${\kappa=-11}$ yields another compact solution. However, to obtain this solution, the value of the frequency $\omega$ must be increased. This adjustment is necessary because, at lower frequencies (e.g., $\omega=3.0$), the second derivative $f''(0)$ approaches negative infinity before the local minimum of $f(r)$ reaches the required zero value for a compacton. This new solution, depicted in FIG.~\ref{fig:CP1b}, demonstrates a more pronounced contribution of the quartic terms, $\widetilde{\cal H}_4(r)$, to the energy density. A notable difference, when compared to solutions in models lacking quartic interactions, is the presence of a shallow local minimum in the total energy density located precisely at the center of the Q-ball. The necessity of increasing $\omega$ can be understood by examining the top panel of FIG.~\ref{fig:CP1c}. The maximum permissible value of the shooting parameter $a_0$ is constrained by a vertical asymptote, whose position shifts to higher values as $\omega$ increases. For certain values of $\omega$, the second derivative of the radial function becomes singular before the compacton condition ($f(R_{\text{out}})=f'(R_{\text{out}})=0$) can be satisfied. The position of the left asymptote, $a_0^{(-)}$, as defined in Equation \eqref{asymptota}, is further illustrated as a function of $\omega$ for several values of the parameter $\kappa$ in the bottom panel of FIG.~\ref{fig:CP1c}.

We also examined numerical solutions based on the expansion $f(r)=a_0+a_1 r+\ldots$, where $a_1$ is given by Equation \eqref{CP1_a1}. Despite the existence of this linear expansion around the center, we were unable to obtain compact solutions (Q-balls) in this scenario. Many of the numerical curves generated in this case grew indefinitely with radius, rendering them unphysical.

\subsection{$CP^3$ Q-balls}
In this section, we present some compact Q-ball solutions for the $\mathbb{C}P^3$ case (corresponding to $l=1$ or $N=3$). The radial solution for these Q-balls exhibits a dominant linear behavior at the center, specifically $f(r)=a_1 r+a_2 r^2+\ldots$, where the coefficient $a_2$ is determined by Equation \eqref{CP3a2}.
\begin{figure}[h!]
\centering
\subfigure[]{\includegraphics[width=0.32\textwidth,height=0.35\textwidth, angle =0]{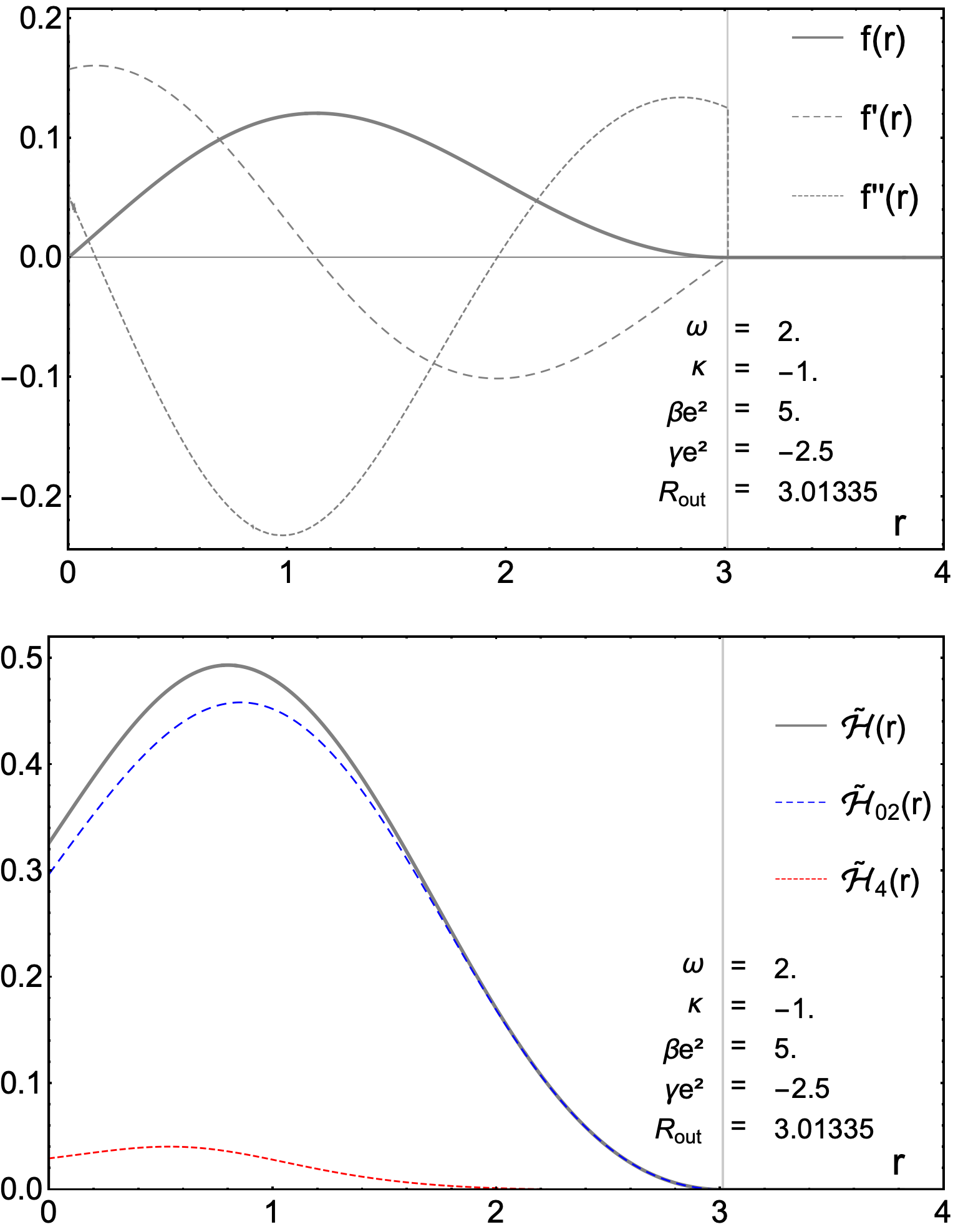}\label{fig:CP3a}}
\hskip 0.2cm
\subfigure[]{\includegraphics[width=0.32\textwidth,height=0.35\textwidth, angle =0]{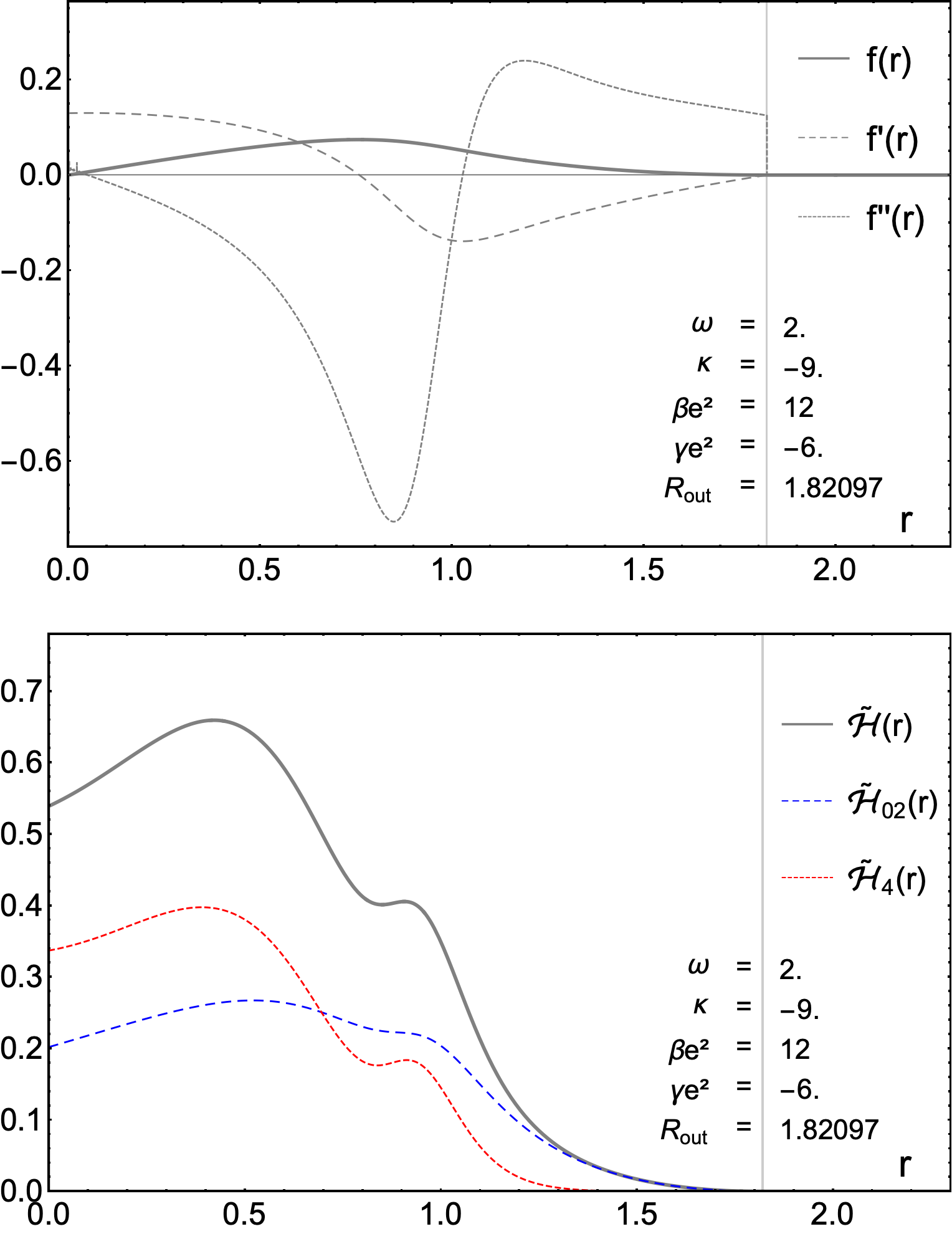}\label{fig:CP3b}}
\hskip 0.2cm
\subfigure[]{\includegraphics[width=0.32\textwidth,height=0.35\textwidth, angle =0]{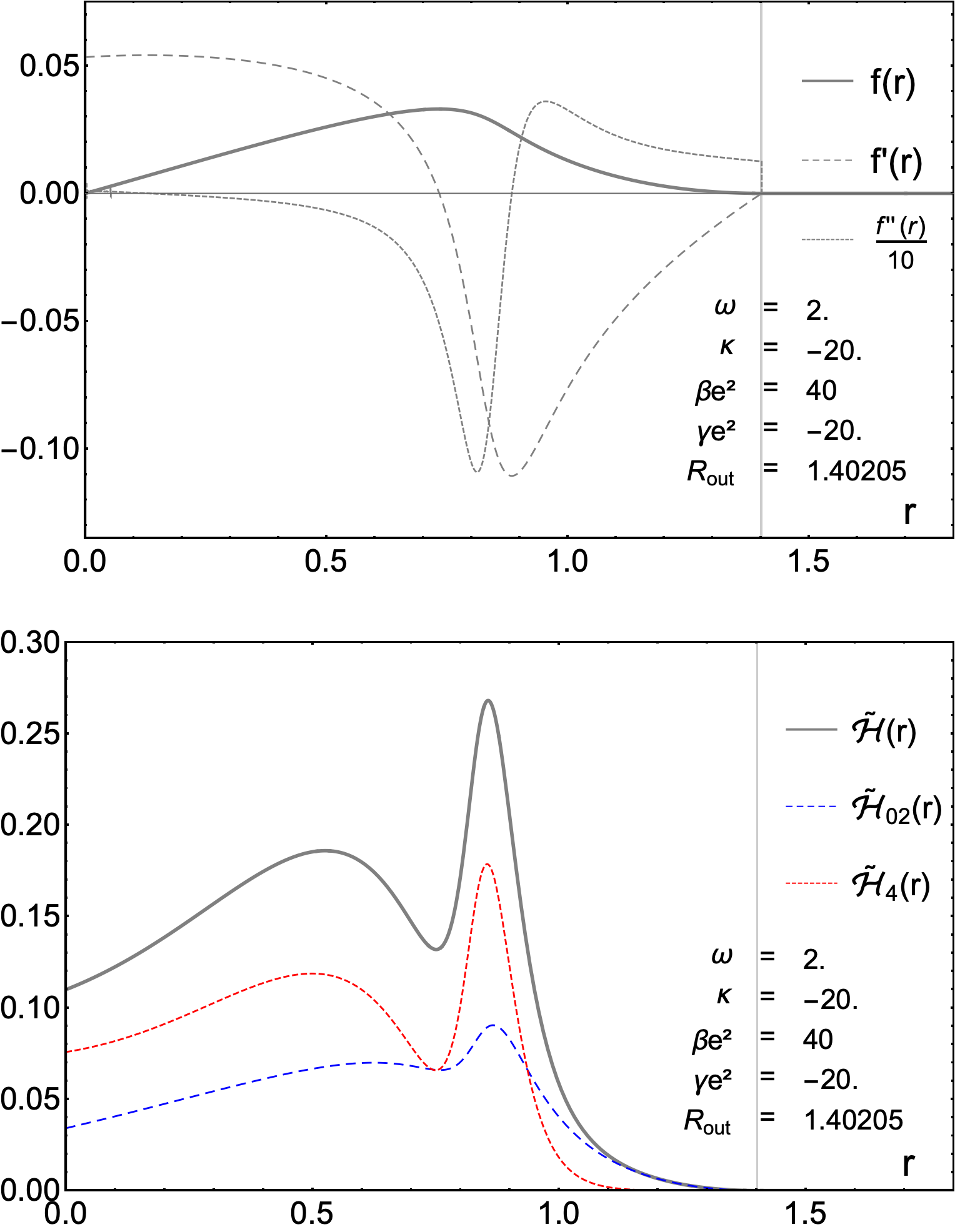}\label{fig:CP3c}}
\caption{$CP^3$ Q-balls satisfying the constraint $\beta e^2+2\gamma e^2=0$. The discontinuity of the second derivative, $f''(R)$ precisely matches the discontinuity caused by the signum function in the field equation.}
\label{fig:CP3}
\end{figure}

The Q-ball solution presented in FIG.~\ref{fig:CP3a} is qualitatively similar to those found in the ${C}P^N$ model with a sharp potential, where the initial contribution from the quartic term to the energy density is small. However, increasing the coupling constant ($\kappa$) for the quartic terms significantly modifies the energy density plot, as shown in FIG.~\ref{fig:CP3b}. Further increases to the coupling constants lead to the development of an extra peak in the energy density, clearly visible in FIG.~\ref{fig:CP3c} . This characteristic behavior of the energy density, including the formation of additional peaks, is a direct consequence of the quartic terms present in the model's Lagrangian.

\begin{figure}[h!]
\centering
\subfigure[]{\includegraphics[width=0.31\textwidth,height=0.35\textwidth, angle =0]{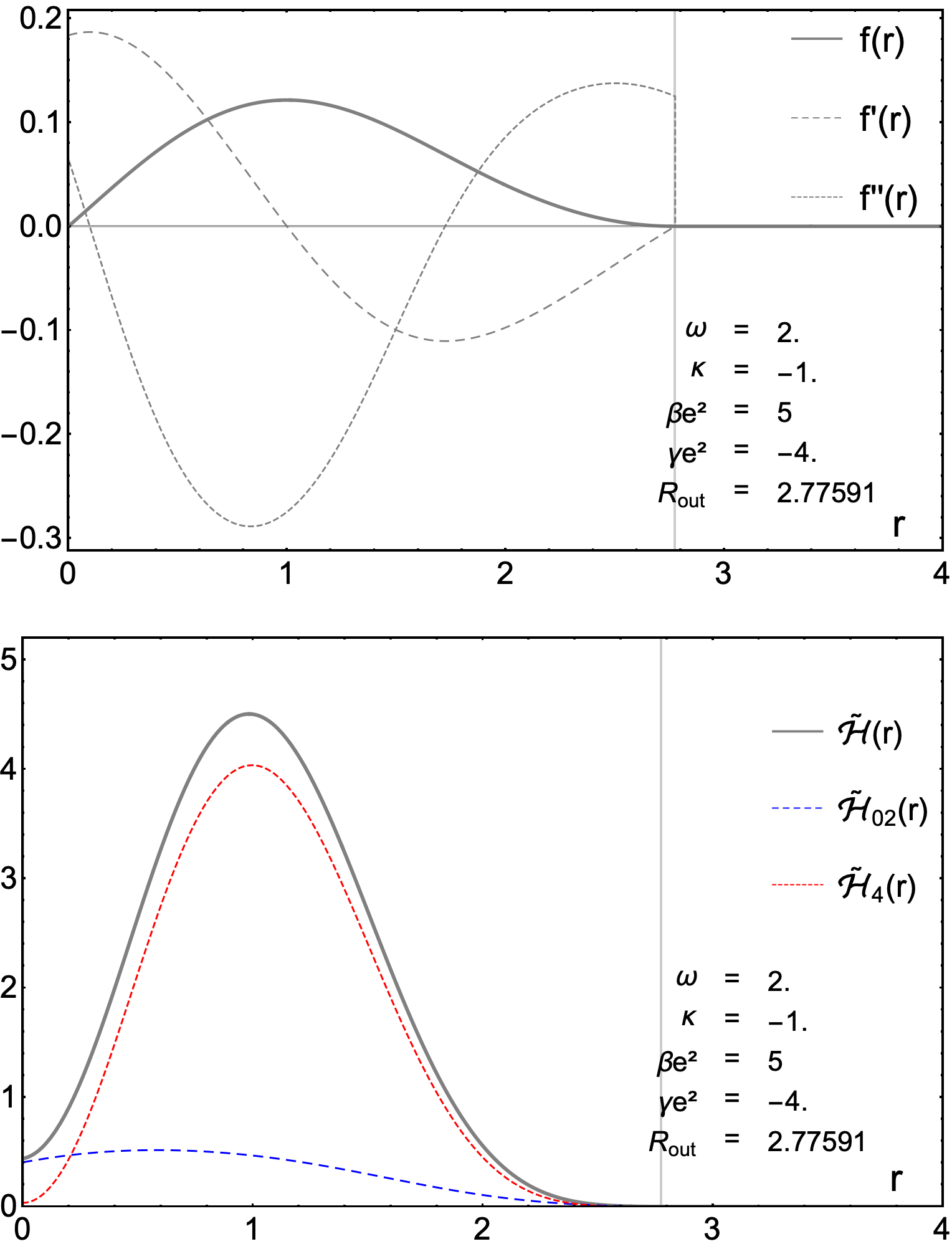}\label{fig:CP3noconstraint-a}}
\hskip 0.2cm
\subfigure[]{\includegraphics[width=0.31\textwidth,height=0.35\textwidth, angle =0]{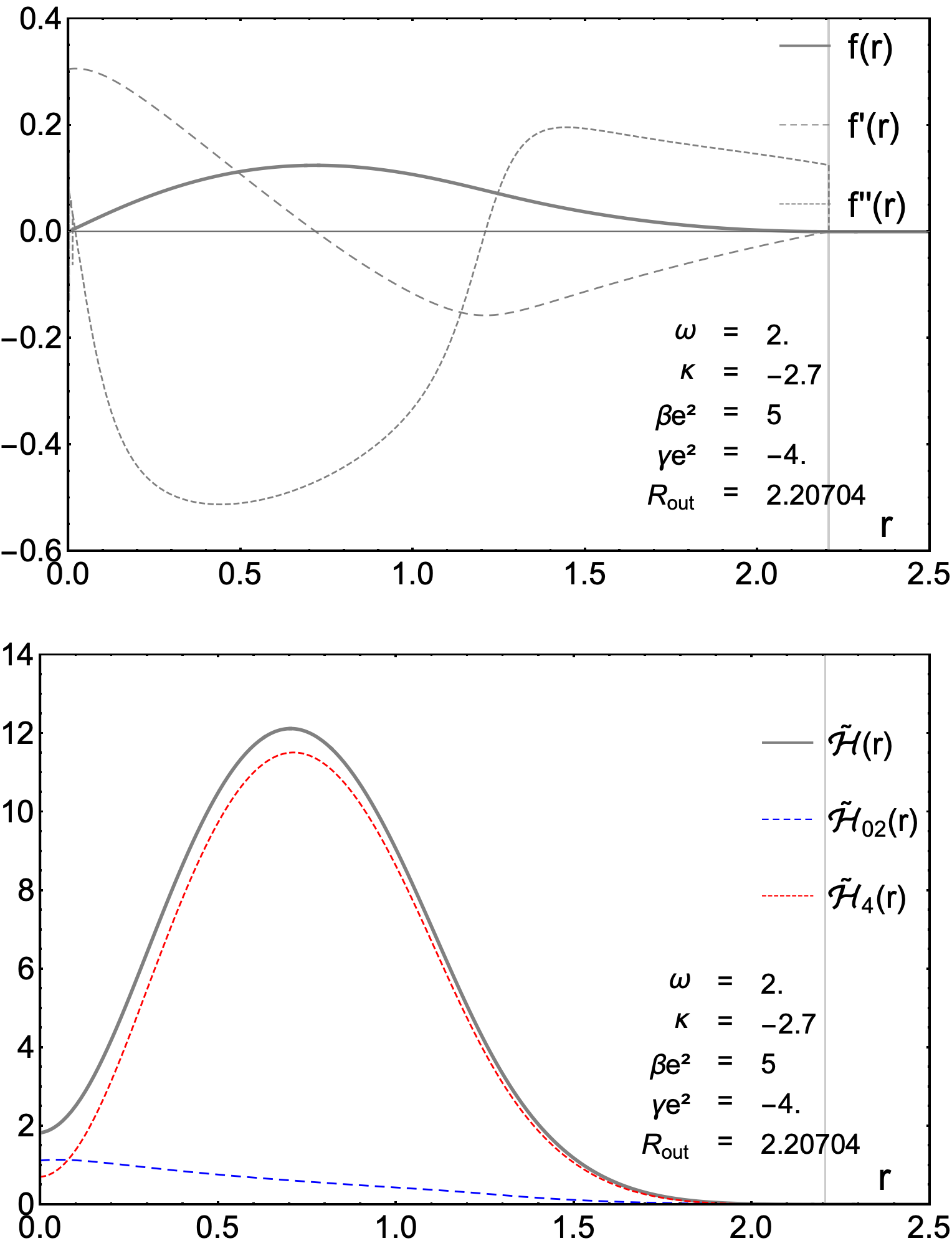}\label{fig:CP3noconstraint-b}}
\hskip 0.2cm
\subfigure[]{\includegraphics[width=0.31\textwidth,height=0.35\textwidth, angle =0]{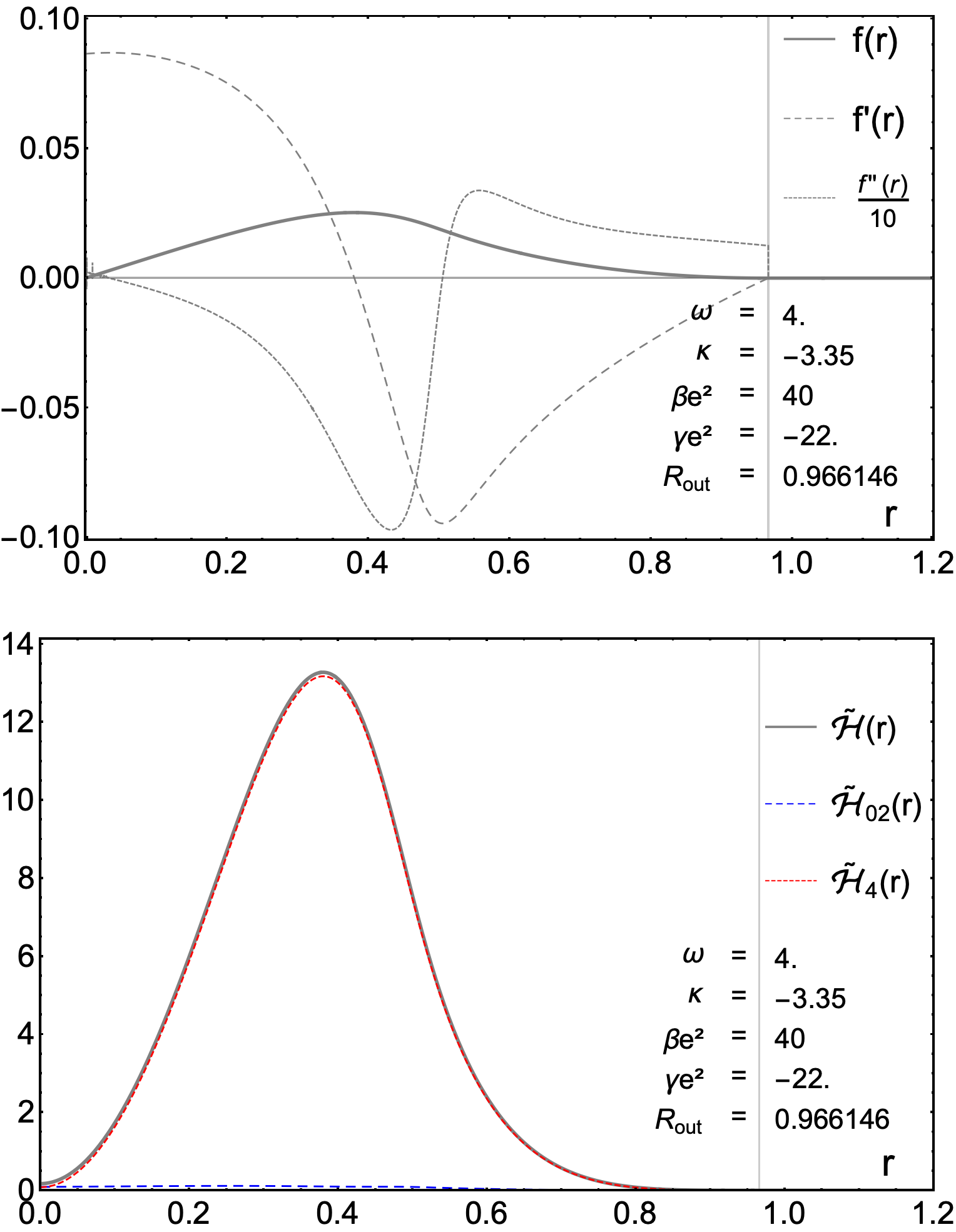},\label{fig:CP3noconstraint-c}}
\caption{$CP^3$ Q-balls for $\beta e^2+2\gamma e^2\neq 0$}
\label{fig:CP3noconstraint}
\end{figure}
The requirement for the positive definiteness of the energy density led us to impose the constraint $\beta e^2+2\gamma e^2=0$, which explicitly eliminates the term proportional to $2\kappa (\beta e^2+2\gamma e^2)\frac{3\omega^4 f^2-f'^4}{(1+f^2)^4}$ from the Hamiltonian density. However, this eliminated term does not always result in a negative contribution to the energy, as its sign depends on the relative magnitudes of $3\omega^4 f^2$ and $f'^4$. Therefore, some solutions that do not satisfy this constraint ($\beta e^2+2\gamma e^2=0$) may still be considered physically acceptable provided the total energy density remains positive. We present a few such solutions in FIG.~\ref{fig:CP3noconstraint}. In the first example, shown in FIG.~\ref{fig:CP3noconstraint-a}, the coupling constants of the quartic terms are relatively small; however, the contribution to the energy density originating from these terms is already dominant. This signifies that the term proportional to $\beta e^2+2\gamma e^2$ provides the main contribution. Increasing the absolute values of the parameters $\kappa$, $\beta e^2$, and $\gamma e^2$ leads to a further increase in the contribution from the quartic terms, as shown in FIG.~ \ref{fig:CP3noconstraint-b} and \ref{fig:CP3noconstraint-c} . This example successfully demonstrates that it is reasonable to explore compact solutions within the shadowed region below the line $\beta e^2+2\gamma e^2=0$, as shown in FIG.~\ref{bound}, broadening the permissible parameter space.

\subsection{Q-shells}
The Q-shell solutions are characterized by an internal vacuum region and possess an expansion around the inner radius, $R_{\text{in}}$, given by:
\[
f(r)=b_2(r-R_{\rm in})^2-b_3(r-R_{\rm in})^3+b_4(r-R_{\rm in})^4+\ldots
\]
where the coefficients $b_2$, $b_3$, and $b_4$ are determined by Equations \eqref{eq:b2}-\eqref{eq:b4}, with the external radius $R_{\text{out}}$ formally replaced by the internal radius $R_{\text{in}}$.
\begin{figure}[h!]
\centering
\subfigure[]{\includegraphics[width=0.32\textwidth,height=0.35\textwidth, angle =0]{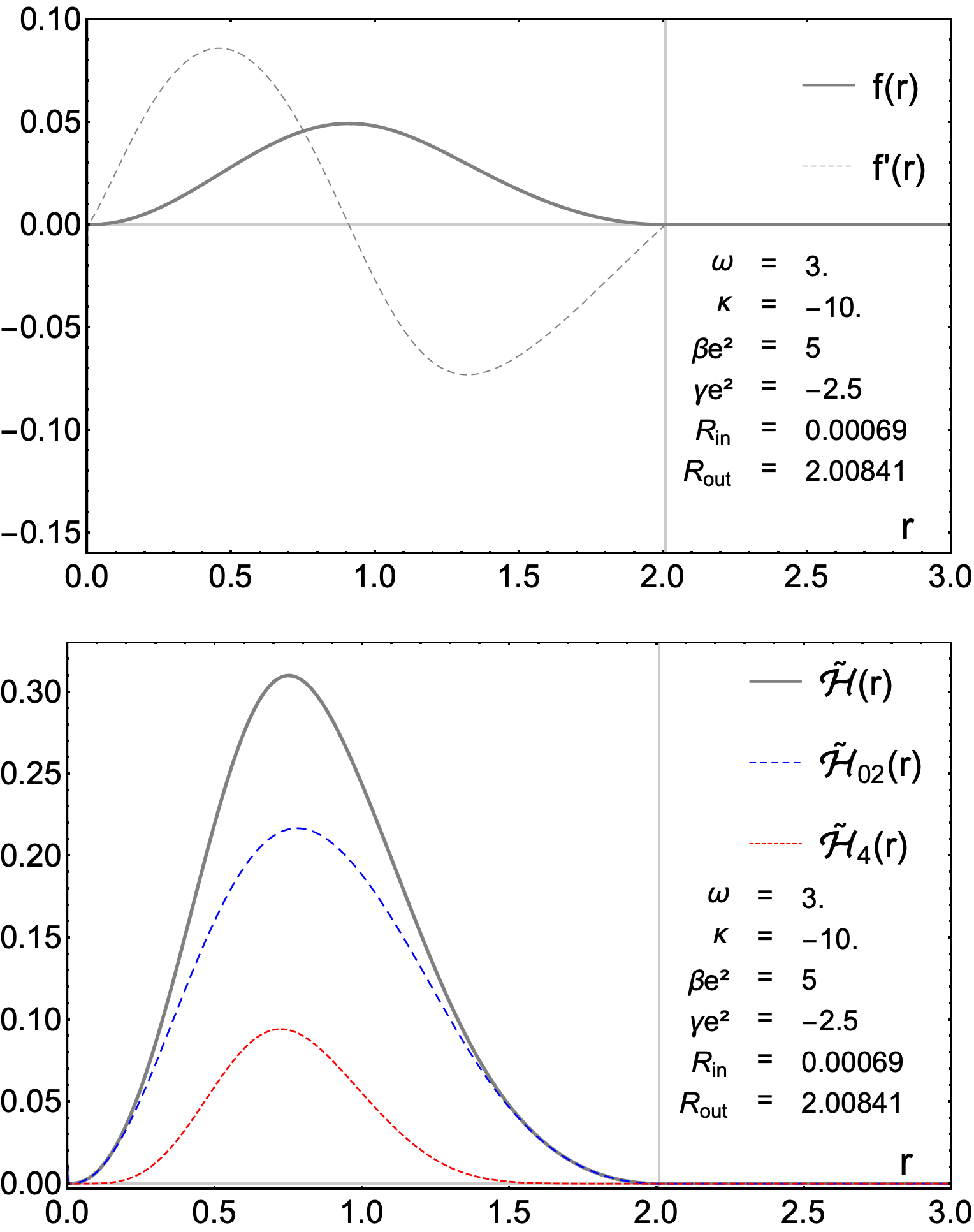}}
\hskip 0.2cm
\subfigure[]{\includegraphics[width=0.32\textwidth,height=0.35\textwidth, angle =0]{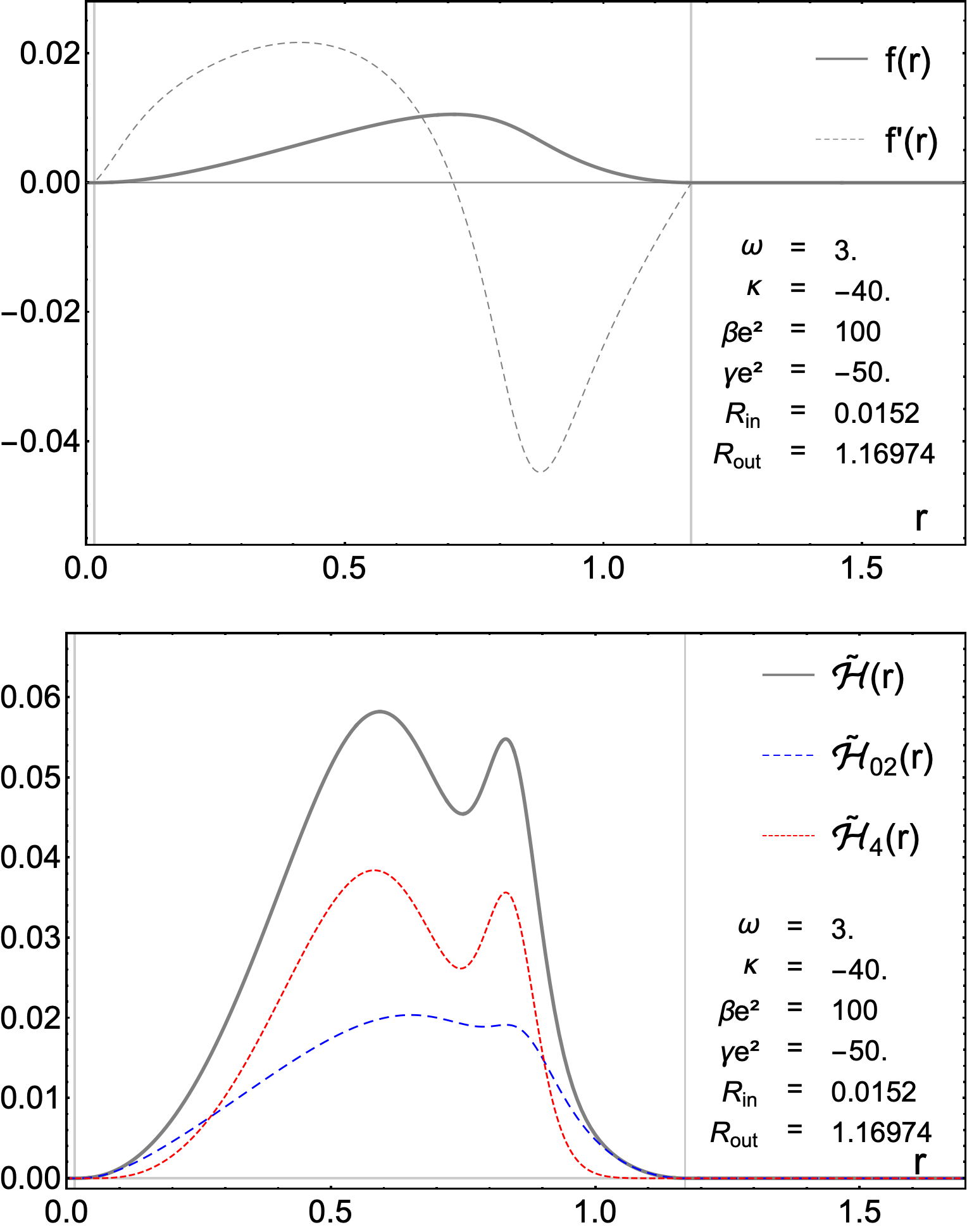}\label{fig:CP5-b}}
\hskip 0.2cm
\subfigure[]{\includegraphics[width=0.32\textwidth,height=0.35\textwidth, angle =0]{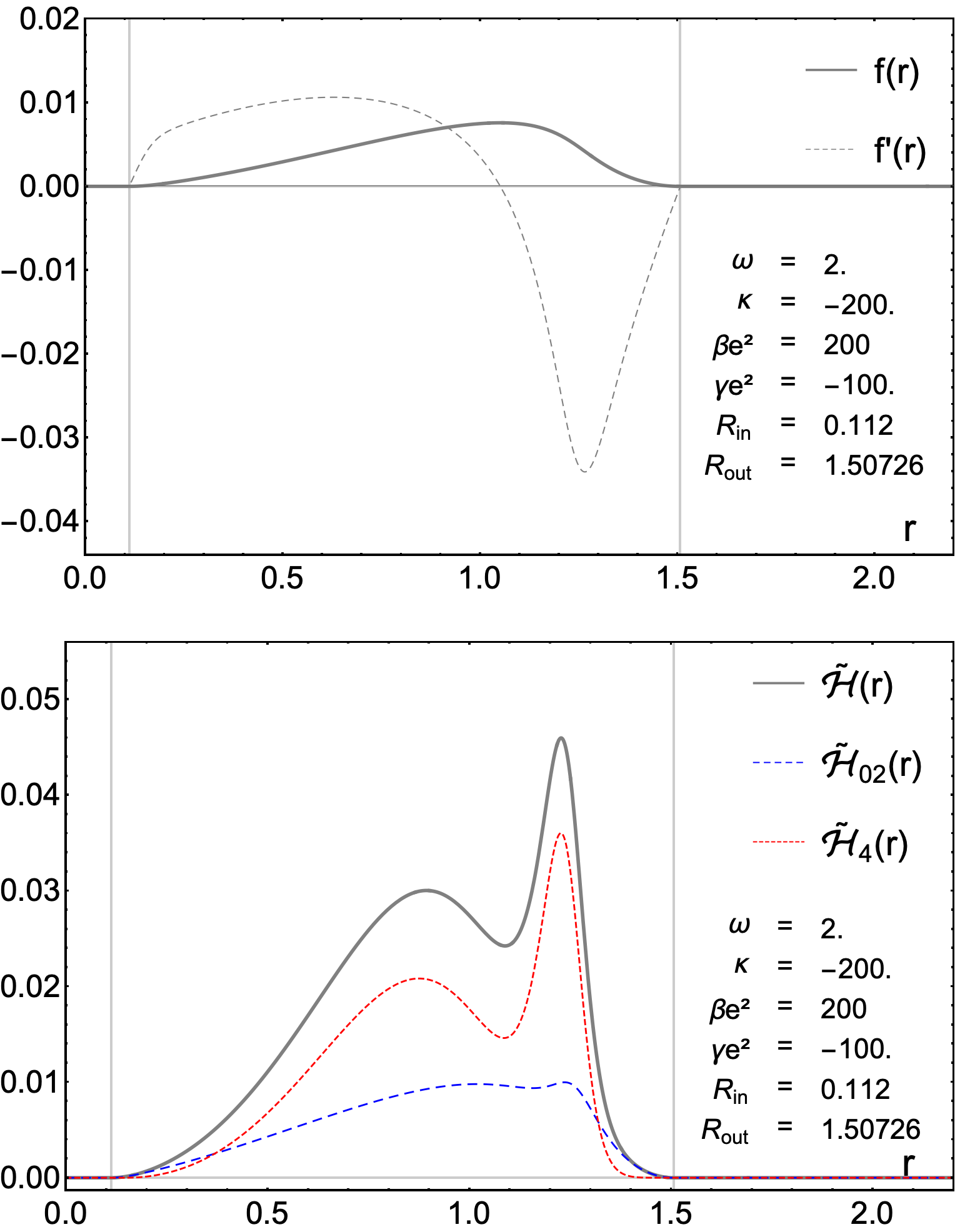}\label{fig:CP5-c}}
\caption{$CP^5$ Q-shells}
\label{fig:CP5}
\end{figure}

As a first example, we present three ${C}P^5$ Q-shell solutions in FIG.~\ref{fig:CP5} . These Q-shells are characterized by a very small inner radius, $R_{\text{in}}$. By increasing parameters derived from the coupling constants of the quartic terms, we obtain a solution that develops an additional peak in the energy density, as shown in FIG.~\ref{fig:CP5-b} and \ref{fig:CP5-c}. These solutions are particularly interesting because they are qualitatively different from those previously discussed within the ${C}P^N$ model with a sharp potential. When these shell-like solutions are coupled to gravity, their spherical configuration and significant concentration of energy in an outer shell make them viable candidates for boson stars.

\begin{figure}[h!]
\centering
\subfigure[]{\includegraphics[width=0.32\textwidth,height=0.35\textwidth, angle =0]{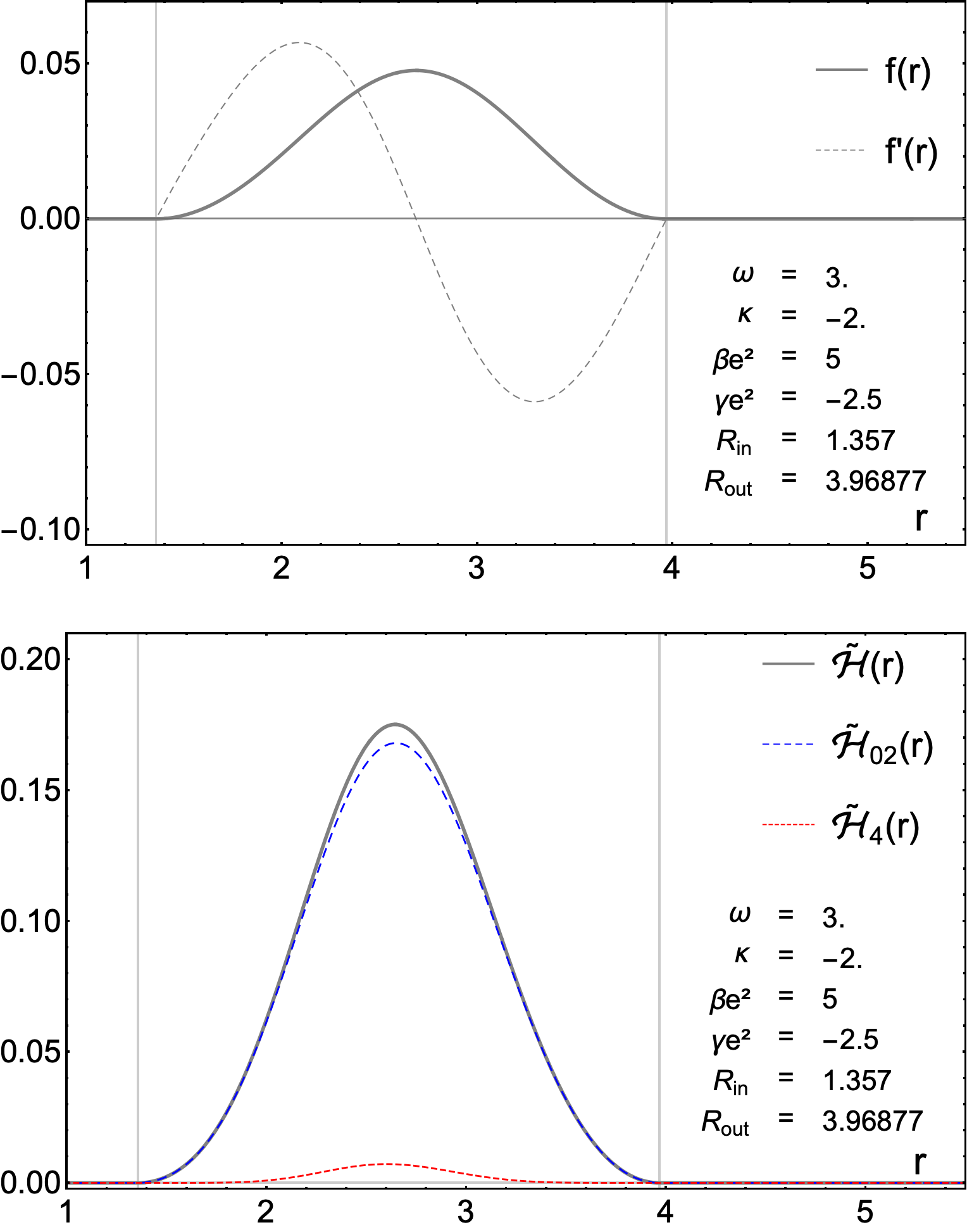}\label{fig:CP11-a}}
\hskip 0.2cm
\subfigure[]{\includegraphics[width=0.32\textwidth,height=0.35\textwidth, angle =0]{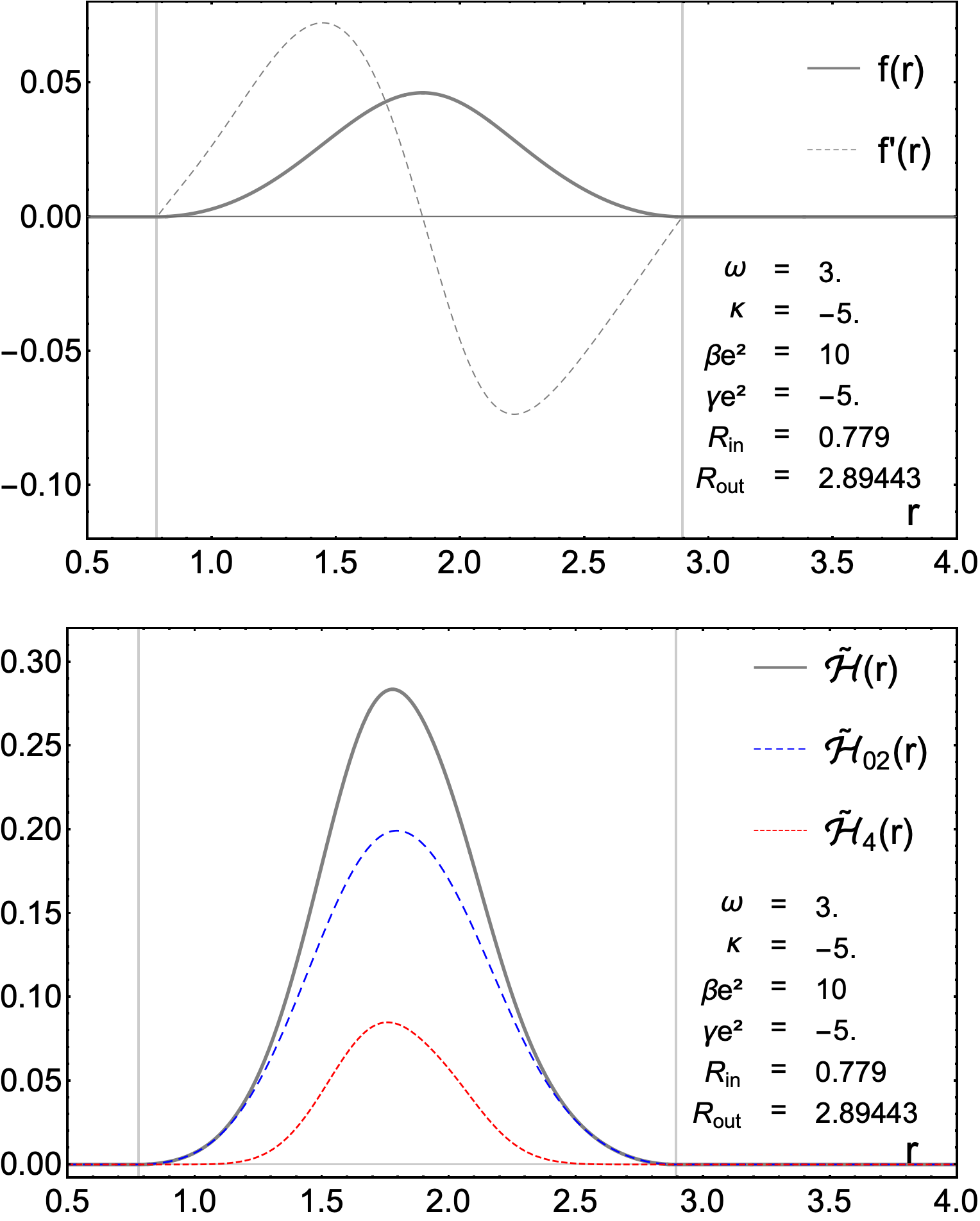}\label{fig:CP11-b}}
\hskip 0.2cm
\subfigure[]{\includegraphics[width=0.32\textwidth,height=0.35\textwidth, angle =0]{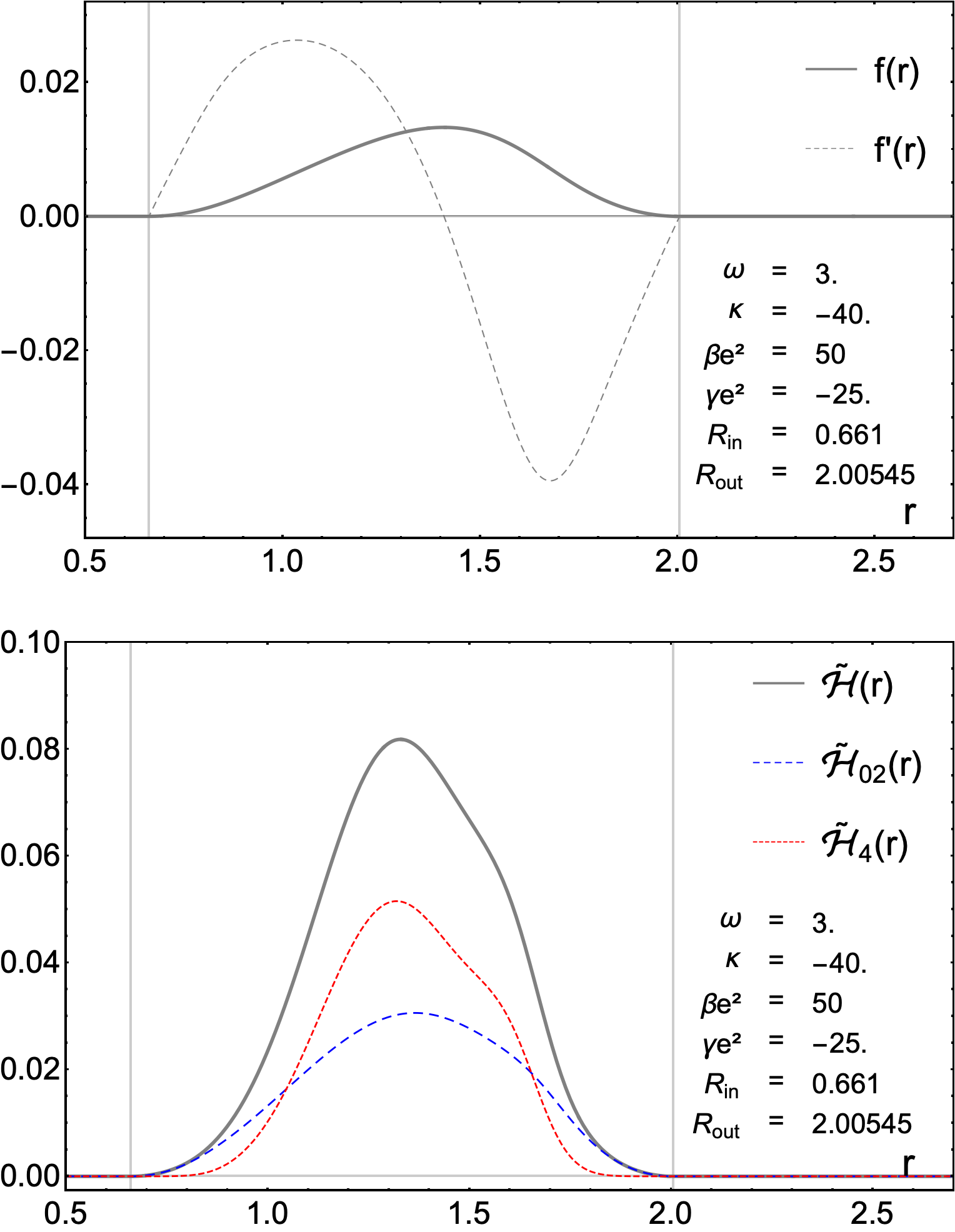}\label{fig:CP11-c}}
\caption{$CP^{11}$ Q-shells}
\label{fig:CP11}
\end{figure}

In FIG.~\ref{fig:CP11}, we plot other Q-shell solutions, specifically those corresponding to higher odd values of $N$. These solutions exhibit an inner radius, $R_{\text{in}}$, that is significantly larger than the $R_{\text{in}}$ found for the ${C}P^5$ Q-shells. We have studied numerous solutions of this kind and observed that it is more difficult to obtain an additional energy density peak for Q-shells associated with a high value of $N$. While the Q-shell in FIG.~\ref{fig:CP11-a} has a dominant contribution to its energy density from the quadratic terms, the solutions shown in FIG.~\ref{fig:CP11-b} and \ref{fig:CP11-c} contain a significant contribution originating from the quartic terms.

To this point, we have focused on specific examples of compact Q-balls, highlighting the behavior of their profile functions and energy density distributions. In the next section, we provide a more systematic numerical investigation of both Q-balls and Q-shells across a broader parameter space and varying coupling constants. Furthermore, we analyze the functional relationship between the total energy ($E$) and the Noether charge ($Q_t^{(m)}$) to characterize the stability and properties of these compact solutions.


\subsection{Energy, Noether charge and radius of the compactons for $l=0,\,1,\,2,\,5$}

The solutions of the ordinary differential equation \eqref{eq:radial} were obtained using a shooting method combined with a fourth-order explicit Runge--Kutta method with step size $\Delta r = 10^{-4}$. The radial coordinate $r$ was restricted to the interval $[R_{\text{in}}, R_{\text{out}}]$, with compacton width $R \equiv R_{\text{out}} - R_{\text{in}} \leq 20$. For fixed coupling constants and $\omega$, the solutions are obtained by varying a single shooting parameter $p(0)$, defined as $f(0)$ for $l=0$, $f'(0)$ for $l=1$, and $R_{\text{in}}$ for $l \geq 2$, until the boundary conditions $\mid f(R_{\text{out}})\mid < 10^{-12}$  and $\mid f'(R_{\text{out}})\mid < 10^{-5}$ are satisfied. The parameter $\omega$ varies from $0$ to $100$ with a step size of $\Delta\,\omega=0.02$. Solutions exist only for $\omega$ above a lower bound $\omega_c$, which depends on the model parameters. Note that $\omega_c$, determined within our discrete $\omega$ grid and limited by the numerical accuracy of the method, should be regarded as an upper estimate of the true critical value. Our approach first varies the initial parameter $p(0)$ from $0$ (undershoot) up to the first value corresponding to an overshoot configuration by increasing it successively in small steps $\Delta \,p$, and then applies binary search to find the value of $p(0)$ that satisfies the boundary conditions.\footnote{In cases with multiple solutions, some of which may be unphysical, this approach selects the smallest value of the parameter that satisfies the boundary conditions. Typically, the maximum value of the shooting parameter starts (for $\omega=0$) at $p_{\rm max} = 1$, although it depends on the values of $\kappa$, $\beta e^2 = -2\,\gamma\,e^2$, and $l$, and $\Delta\,p = p_{\rm max}/600$ or even smaller. After finding the lower value of $\omega$ that leads to a solution, for each larger value of $\omega$ the value of $p_{\rm max}$ is updated to twice the value of $p(0)$ obtained for $\omega - 0.02$, which also updates the corresponding step size $\Delta\,p$.}

The fractional power $n$ that best linearizes the relationship between $E^{1/n}$ and $Q_t^{(m)}$ is determined by minimizing the residual $e_{\rm P} = 1 - |r_{\rm P}|$ via simulated annealing. Here, $r_{\rm P}$ is the Pearson correlation coefficient defined as:$$r_{\rm P} = \frac{\sum_{i=1}^{N_P} (Q_t^{(m)}(\omega_i) - \overline{Q_t^{(m)}})(E^{1/n}(\omega_i) - \overline{E^{1/n}})}{\sigma_{Q_t^{(m)}} \sigma_{E^{1/n}}},$$where the mean $\overline{X}$ and the variation $\sigma_X$ are given by:$$\overline{X} \equiv \frac{1}{N_P} \sum_{i=1}^{N_P} X(\omega_i) \qquad\text{and}\qquad \sigma_X \equiv \sqrt{\sum_{i=1}^{N_P} (X(\omega_i) - \overline{X})^2},$$ for $X \in \{Q_t^{(m)}, E^{1/n}\}$.

 The index $i$ labels the discrete values of $\omega$ for which solutions were obtained. These are restricted to the range $Q_t^{(m)} \leq 120$, resulting in a total of $N_P$ solutions for each set of parameters $\{l, \kappa, \beta e^2 = -2\gamma e^2\}$. By definition, the coefficient satisfies $|r_{\rm P}| \leq 1$. The identity $|r_{\rm P}| = 1$ (corresponding to $e_{\rm P} = 0$) holds if and only if there exists an exact linear relationship $E^{1/n} = a + b Q_t^{(m)}$ with real parameters $a$ and $b$ ($b \neq 0$). In our analysis, $N_P$ is typically on the order of $4800$, and the maximum value of $e_{\rm P}$ across all studied cases is less than $3 \times 10^{-5}$.

The procedure described above is repeated for three distinct subsets of the data: $\omega \geq 20$, $\omega \geq 50$, and the interval $\omega_c \leq \omega \leq \omega_c + 2$, yielding the exponents $n_{\omega \geq 20}$, $n_{\omega \geq 50}$, and $n_S$, respectively. The values of $n_{\omega \geq 20}$ and $n_{\omega \geq 50}$ provide estimates for $n$ as $\omega$ approaches the large-$\omega$ regime, whereas $n_S$ characterizes $n$ in the lowest accessible frequency region. Although these results suggest that $n$ may vary across different regimes, a sufficiently small variation allows the power law $E \sim Q^n$ to be maintained as a robust approximation of the energy--Noether charge relation.\footnote{In certain cases, the $\omega$-interval associated with $n_S$ may overlap with those defined for the large-$\omega$ regimes?for instance, when $\omega_c > 20$, as observed in configurations with large values of $\beta e^2$ and $|\kappa|$. Nevertheless, the interval corresponding to $n_S$ remains more restrictive, provided that $\omega_c < 98$. This latter condition is satisfied across all configurations examined in this study.}

Assume that $E = \alpha (Q_t^{(m)})^{n}$ holds locally near each grid point, with the parameters $n$ and $\alpha$ varying slowly. To capture the local variation of $n$ across different $\omega$ regimes more explicitly, we introduce an effective exponent: $n^{{\rm eff}} = \frac{d \ln E}{d \ln Q_t^{(m)}}$.
For the numerical evaluation, we apply a natural cubic spline to the data points $(x_i, y_i)$, where $x_i = \ln Q_t^{(m)}(\omega_i)$ and $y_i = \ln E(\omega_i)$. On each sub-interval, the spline is represented by the polynomial $y = \sum_{j=0}^{3} b_i^j (x - x_i)^j$. Once the real-valued coefficients $b_i^j$ are determined, the effective exponent at each point is simply given by the first-order coefficient: $n^{{\rm eff}}(x_i) = \frac{dy}{dx}\big|_{x = x_i}=b_i^1$.

\begin{figure}[h!]
\centering
\includegraphics[scale=0.355]{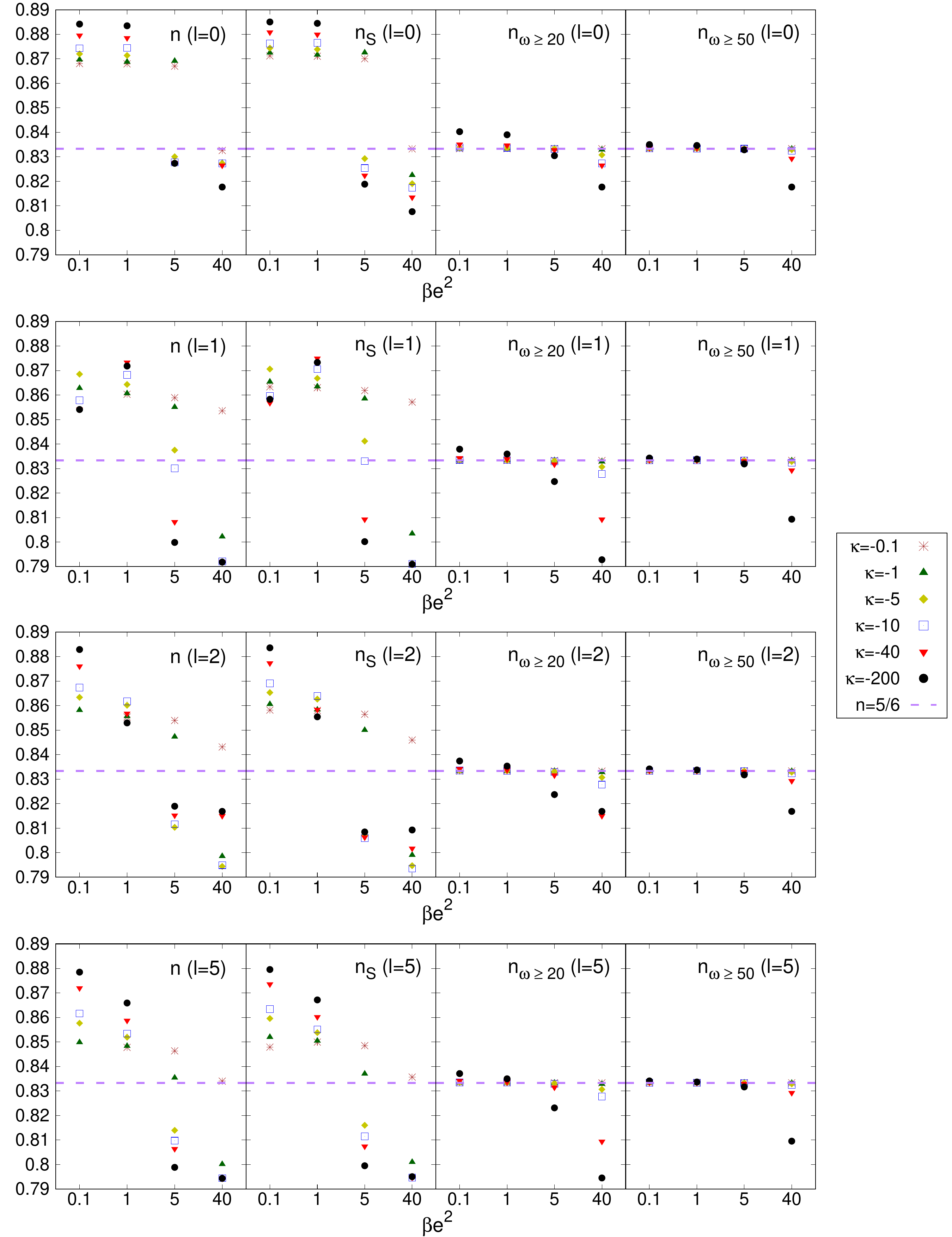}
\caption{The values of $n$, $n_S$, $n_{\omega \geq 20}$ and $n_{\omega \geq 50}$ that minimizes $e_P$ for Q-balls $(l=0,\,1)$ and Q-shells $(l=2,\,5)$, for some values of $\kappa$ and $\beta e^2 = -2\,\gamma e^2$.}\label{fig:n}
\end{figure}

\begin{figure}[h!]
\centering
\includegraphics[scale=0.355]{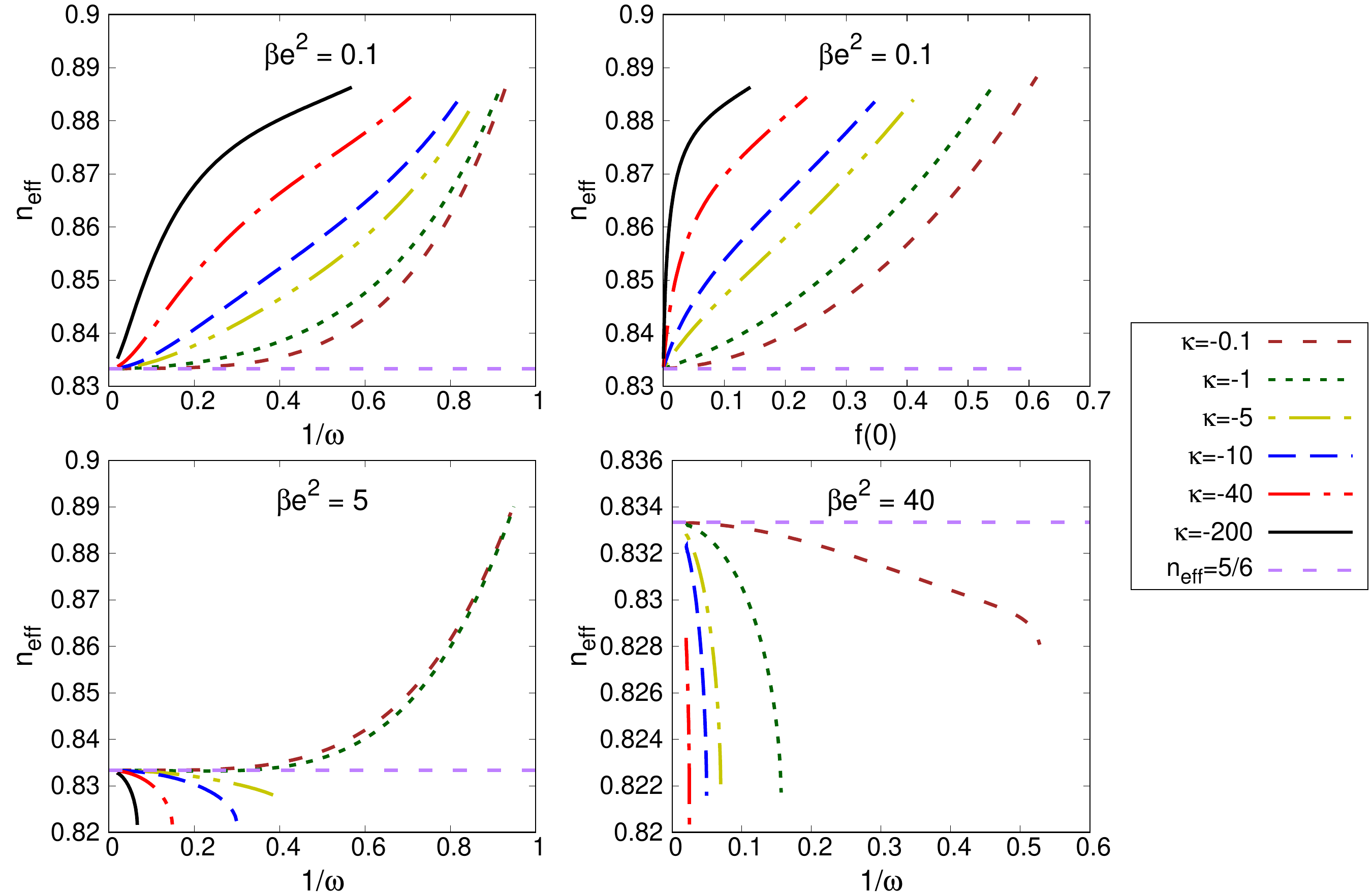}
\caption{The effective $CP^1$ exponent $n_{{\rm eff}}$ as a function of $1/\omega$ for $\beta\,e^2 = -2\,\gamma e^2 = 0.1,\,5,\,40$, and as a function of $f(0)$ for $\beta\, e^2 = -2\,\gamma e^2 = 0.1$, both  for some values of $\kappa$, restricted to $\omega \leq 50$ and $Q_t^{(m)}\leq 120$.}\label{fig:nvar1}
\end{figure}

\begin{figure}[h!]
\centering
\includegraphics[scale=0.355]{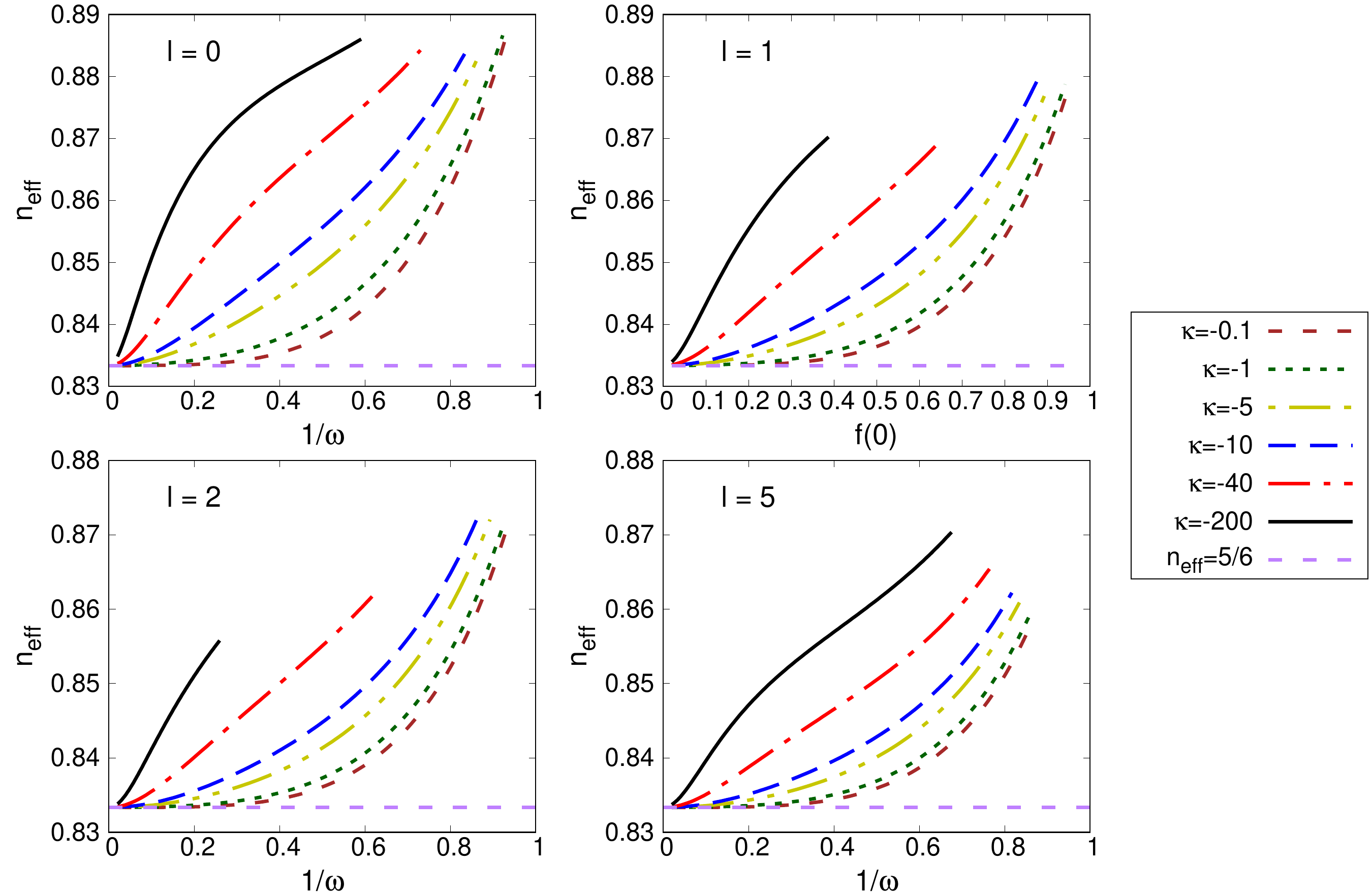}
\caption{The effective exponent $n_{eff}$ as a function of $1/\omega$ for $\beta\,e^2 = -2\,\gamma e^2 = 1$ and $l=0,\,1,\,2,\,5$, for some values of $\kappa$, restricted to $\omega \leq 50$ and $Q_t^{(m)}\leq 120$.}\label{fig:nvar2}
\end{figure}

\begin{figure}[h!]
\centering
\includegraphics[scale=0.355]{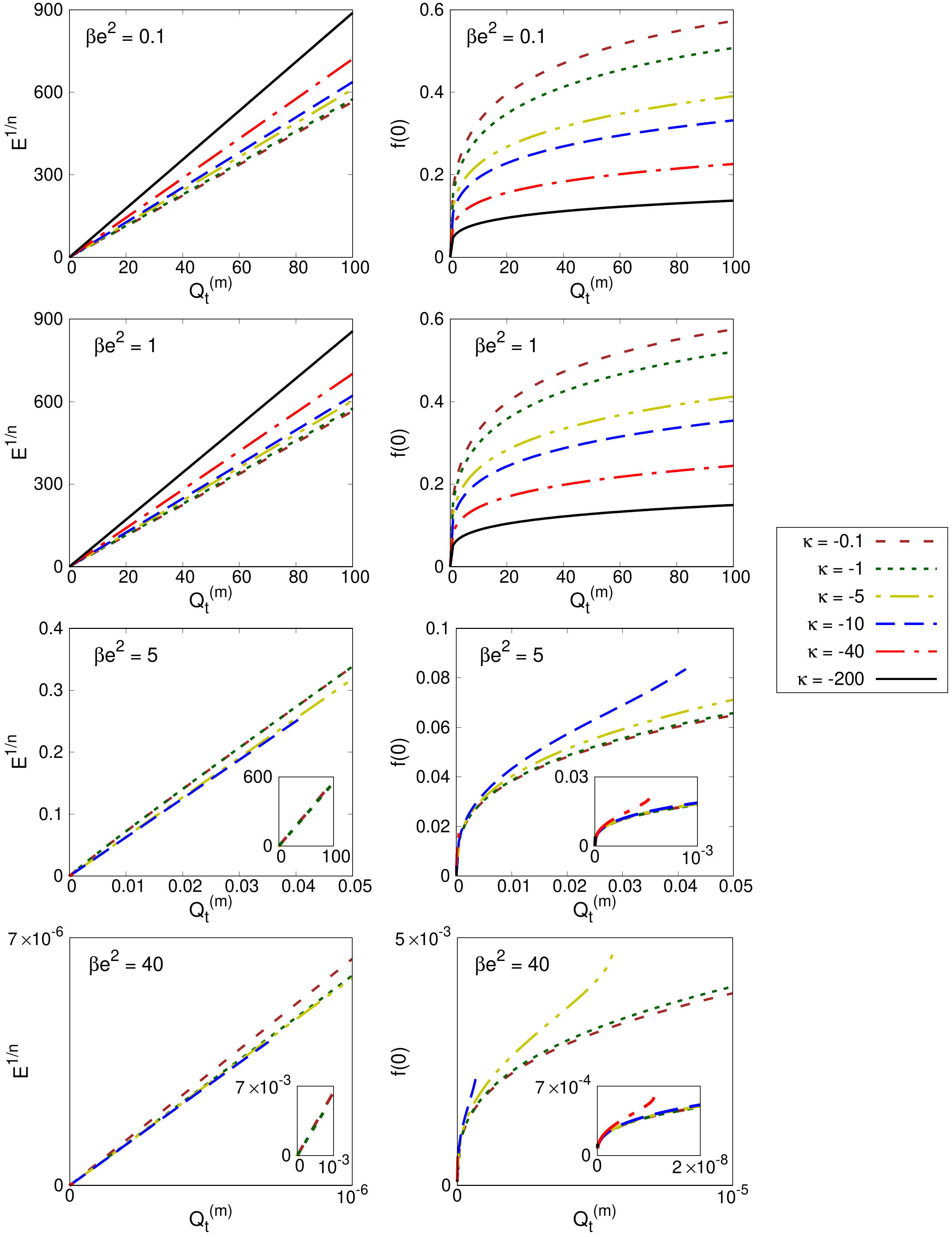}
\caption{$CP^{1}$ Q-ball fractional energy power $E^{1/n}$ and $f(0)$ as functions of $Q_t^{(m)}$, for some values of $\kappa$ and $\beta e^2 = -2\,\gamma e^2$. The corresponding values of $n$ are shown in FIG.~\ref{fig:n}.
}\label{fig:energy_cp1}
\end{figure}

\begin{figure}[h!]
\centering
\includegraphics[scale=0.355]{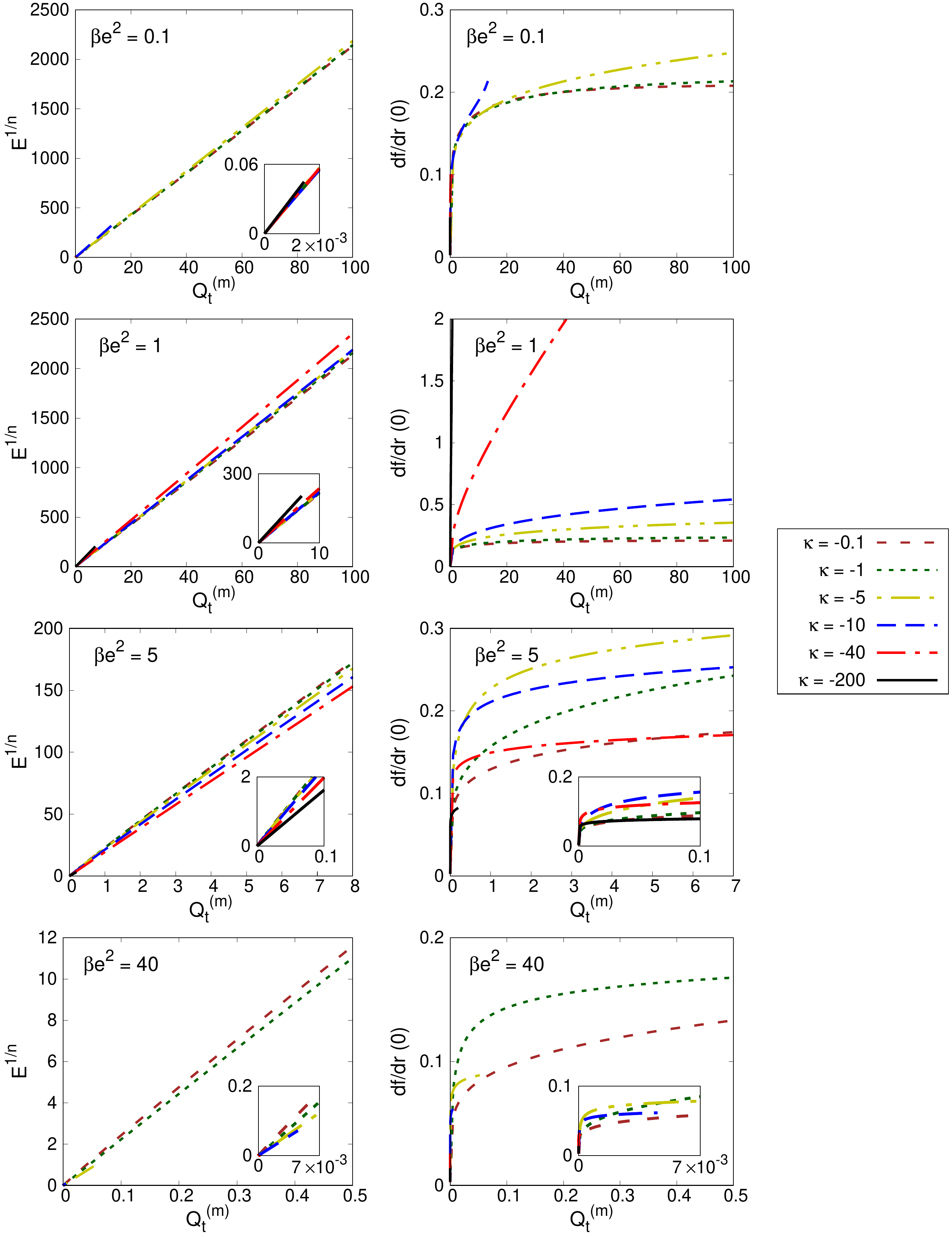}
\caption{$CP^{3}$  Q-ball fractional energy power $E^{1/n}$ and $f'(0)=df/dr(0)$ as functions of $Q_t^{(m)}$, for some values of $\kappa$ and $\beta e^2 = -2\,\gamma e^2$. The corresponding values of $n$ are shown in FIG.~\ref{fig:n}.}
\label{fig:energy_cp3}
\end{figure}

\begin{figure}[h!]
\centering
\includegraphics[scale=0.355]{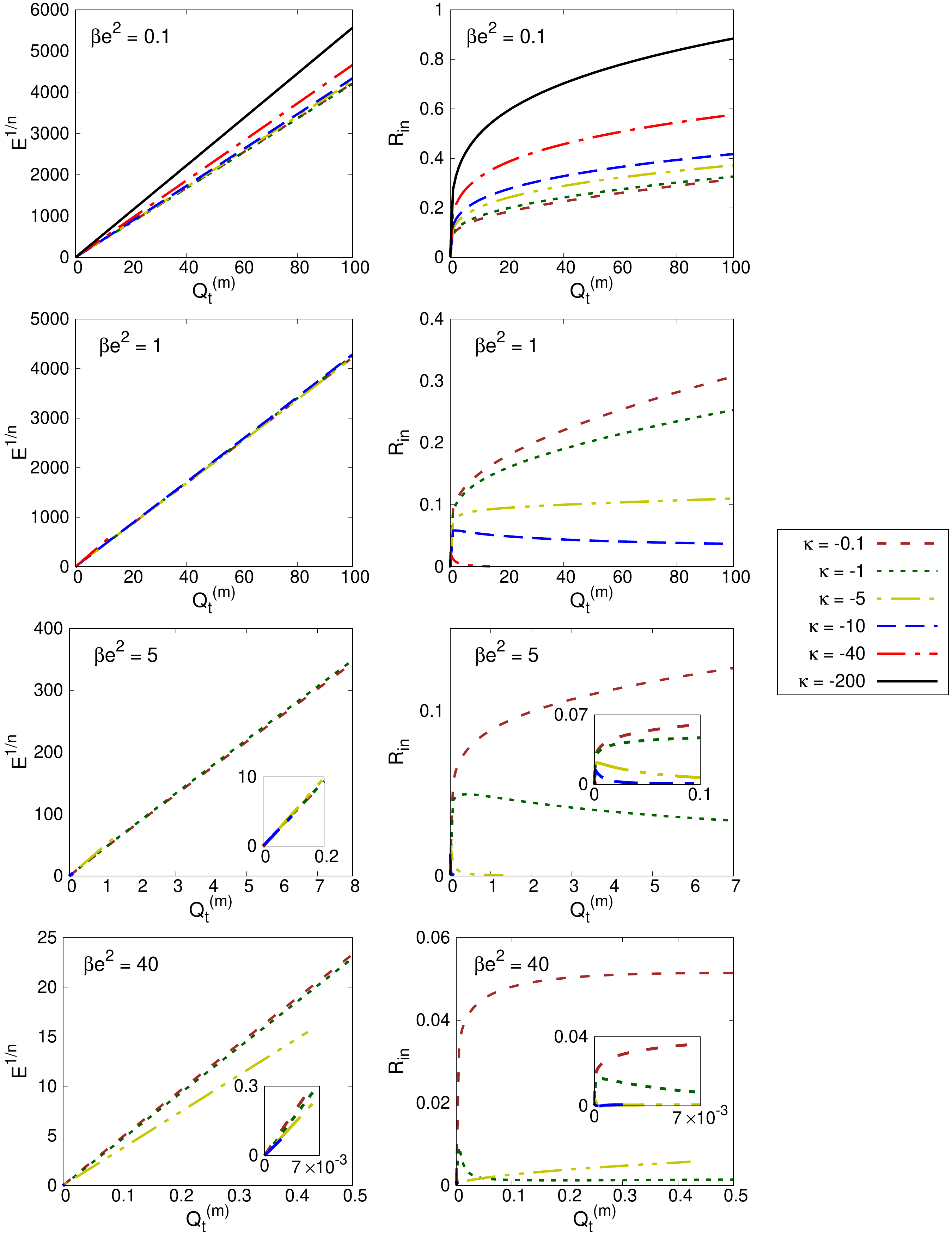}
\caption{$CP^{5}$ Q-shell fractional energy power $E^{1/n}$ and the inner radius $R_{\text{in}}$ as functions of $Q_t^{(m)}$, for selected values of $\kappa$ and $\beta e^2 = -2\,\gamma e^2$. The corresponding values of $n$ are shown in FIG.~\ref{fig:n}.}
\label{fig:energy_cp5}
\end{figure}

\begin{figure}[h!]
\centering
\includegraphics[scale=0.355]{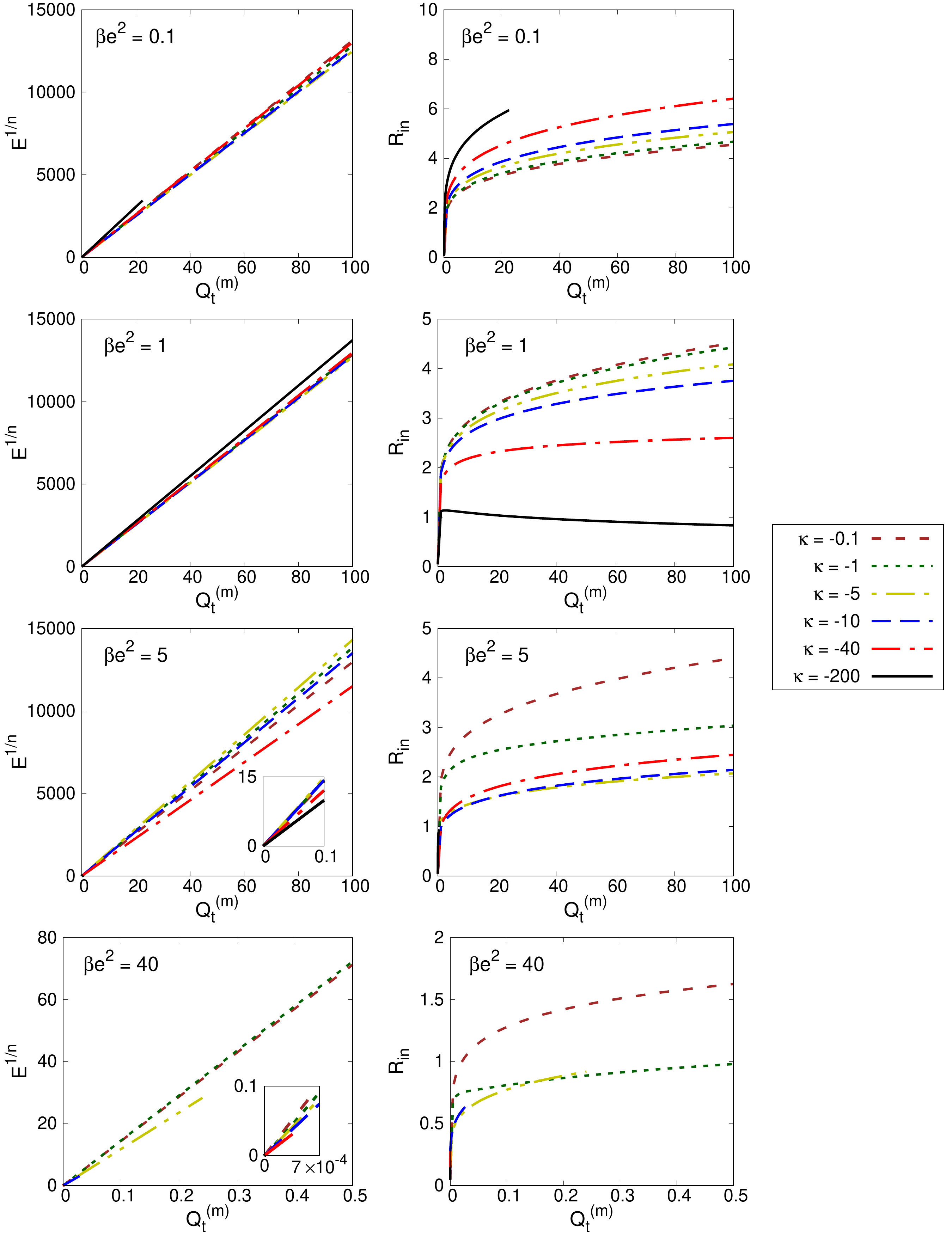}
\caption{$CP^{11}$ Q-shell fractional energy power $E^{1/n}$ and the inner radius $R_{\text{in}}$ as functions of $Q_t^{(m)}$, for some values of $\kappa$ and $\beta e^2 = -2\,\gamma e^2$. The corresponding values of $n$ are shown in FIG.~\ref{fig:n}.}
\label{fig:energy_cp11}
\end{figure}

FIG.~\ref{fig:n} shows that the values of $n$, $n_S$, $n_{\omega \geq 20}$, and $n_{\omega \geq 50}$ are consistently less than unity, oscillating within the range $0.79\!-\!0.89$ for all combinations of the parameters $\beta e^2 = -2\gamma e^2 \in \{0.1, 1, 5, 40\}$, $\kappa \in \{-0.1, -1, -5, -10, -40, -200\}$, and $l \in \{0, 1, 2, 5\}$. Generally, the values of $n$ closely track those of $n_S$. Furthermore, since $n$ varies only weakly and $E^{1/n}$ is approximately linear in $Q$ (as illustrated in FIGS.~\ref{fig:energy_cp1}, \ref{fig:energy_cp3}, \ref{fig:energy_cp5}, and \ref{fig:energy_cp11}), the scaling relation $E \sim Q^{n}$ provides a robust approximation. Although the relation is not exact, the Pearson residual $e_P$ remains small. Consequently, because $E \sim (Q_t^{(m)})^{n}$ with $n \leq 1$, all Q-balls and Q-shells are stable. Indeed, the maximum deviation of $n_S$ from the large-$\omega$ limit of $5/6$--defined as $\Delta n \equiv |n_S - 5/6|$--is $0.05169$ (approximately $6\%$). This peak deviation occurs for the configuration $l = 0$, $\beta e^2 = 0.1$, and $\kappa = -200$, where $n_S = 0.88502$.

For all cases $l \in \{0, 1, 2, 5\}$, the profile function $f(r)$ becomes small throughout its domain. This is indirectly indicated by the simultaneous decrease in compacton width ($R = R_{\rm out} - R_{\rm in}$), energy, and Noether charge within the large-$\omega$ regime. As shown in FIG.~\ref{fig:n}, all curves in this regime converge toward $n \approx 5/6$, thereby reproducing the energy--Noether charge relation $E \sim (Q_t^{(m)})^{5/6}$ characteristic of the pure $CP^N$ model with a $V$-shaped potential (i.e., in the absence of quartic terms), as derived in \cite{Klimas:2017eft}. This behavior arises because, as $f(r)$ remains small everywhere in the large-$\omega$ limit, contributions from higher-order terms are effectively suppressed. For example, in the case $l=0$, $\beta e^2 = -2\gamma e^2 = 1$, and $\kappa = -0.1$, the fitted value for the full dataset is $n = 0.86804$ ($e_P \sim 10^{-5}$). Restricting the analysis to solutions with $\omega \geq 10$ yields $n_{\omega \geq 10} = 0.83336$, which deviates from $5/6$ by only $0.003\%$, with an extremely low residual $e_P^{\omega \geq 10} \sim 10^{-12}$. In this configuration, $f(0) = 6.9879 \times 10^{-3}$ at $\omega = 10$ and continues to decrease as $\omega$ increases, confirming the suppression of the nonlinear profile.

\begin{figure}[h!]
\centering
\includegraphics[scale=0.355]{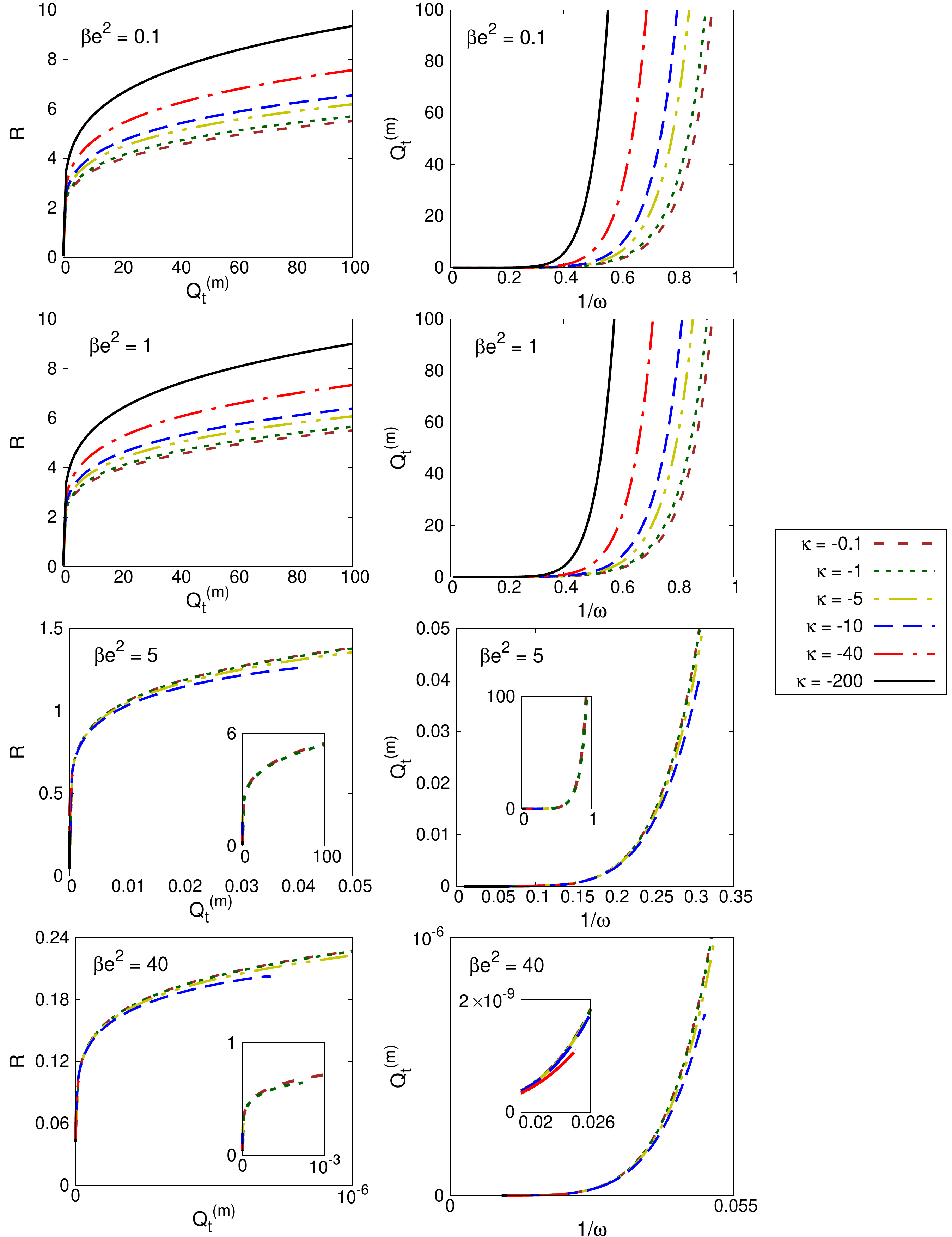}
\caption{$CP^{1}$ Q-ball radius $R$ versus the Noether charge $Q_t^{(m)}$, and $Q_t^{(m)}$ versus $1/\omega$, for some values of $\kappa$ and $\beta e^2 = -2\,\gamma e^2$.
}
\label{fig:radius_cp1}
\end{figure}

FIG.~\ref{fig:nvar1} displays $n_{\rm eff}$ as a function of $f(0)$ for $\beta e^2=0.1$ and $l=0$, as well as $n_{\rm eff}$ versus $1/\omega$ for $\beta e^2 \in \{0.1, 5, 40\}$. The corresponding results for $\beta e^2=1$ are presented in FIG.~\ref{fig:nvar2} for $l \in \{0, 1, 2, 5\}$. Across all cases studied, $n_{\rm eff}$ converges to approximately $5/6$ in the large-$\omega$ regime ($1/\omega \to 0$). In this limit, the profile function at the origin, $f(0)$, becomes small--a behavior explicitly demonstrated by the correlation between $n_{\rm eff}$ and $f(0)$ for the $\beta e^2=0.1$ case.\footnote{To construct the $n_{\rm eff}$ plots, we refined the numerical sampling by reducing the step size from $\Delta\omega = 0.02$ to $\Delta\omega = 0.002$ within the interval $\omega_c \leq \omega \leq \omega_c + 2$. This increased resolution targets the region where $E$ and $Q$ attain their maximum values. For the large-frequency regime ($\omega \geq 10$), a spatial step size of $\Delta r = 10^{-5}$ was employed. Furthermore, we imposed the constraint $Q \leq 120$ to maintain consistency with the criteria used for the global $n$ values (see Fig.~\ref{fig:n}). The analysis was also restricted to $\omega \leq 50$ to exclude solutions with vanishingly small energy and charge, as such points would otherwise introduce significant numerical uncertainty into the determination of $n_{\rm eff}$.} For the other studied values of $\beta e^2$ with $l=0$, the smallness of $f(0)$ in this regime can be inferred by cross-referencing FIGS.~\ref{fig:radius_cp1} and \ref{fig:energy_cp1}. Specifically, FIG.~\ref{fig:radius_cp1} indicates that large values of $\omega$ correspond to the lowest obtained values of the Noether charge, which in turn correspond to the lower values of $f(0)$ shown in FIG.~\ref{fig:energy_cp1}. A similar behavior is observed in FIG.~\ref{fig:nvar2} for the cases where $\beta e^2=1$ and $\kappa \in \{-0.1, -1, -5, -10, -40, -200\}$ across $l \in \{0, 1, 2, 5\}$. Consequently, the analysis of $n_{\rm eff}$ is consistent with and reproduces the results obtained from the study of $n$, $n_S$, $n_{\omega \geq 20}$, and $n_{\omega \geq 50}$.

FIGS.~\ref{fig:nvar1} and \ref{fig:nvar2} also demonstrate that $\omega_c$--corresponding to the maximum value of $1/\omega$ at which the curves terminate--tends to increase with larger values of $|\kappa|$ or $\beta e^2$. Consequently, for large $|\kappa|$ and $\beta e^2$, the domain of existence in $1/\omega$ can shrink substantially, as exemplified by the $\beta e^2 = 40$ plot in FIG.~\ref{fig:nvar1}. This contraction of the domain helps explain the reduced variation of $n_{\rm eff}$ observed in this regime. For instance, in the case $l = 0$, $\beta e^2 = 40$, and $\kappa = -200$, the threshold frequency reaches $\omega_c > 50$. This leads to the identity $n_{\omega \geq 20} = n_{\omega \geq 50}$, as illustrated in FIG.~\ref{fig:n}.

\begin{figure}[h!]
\centering
\includegraphics[scale=0.355]{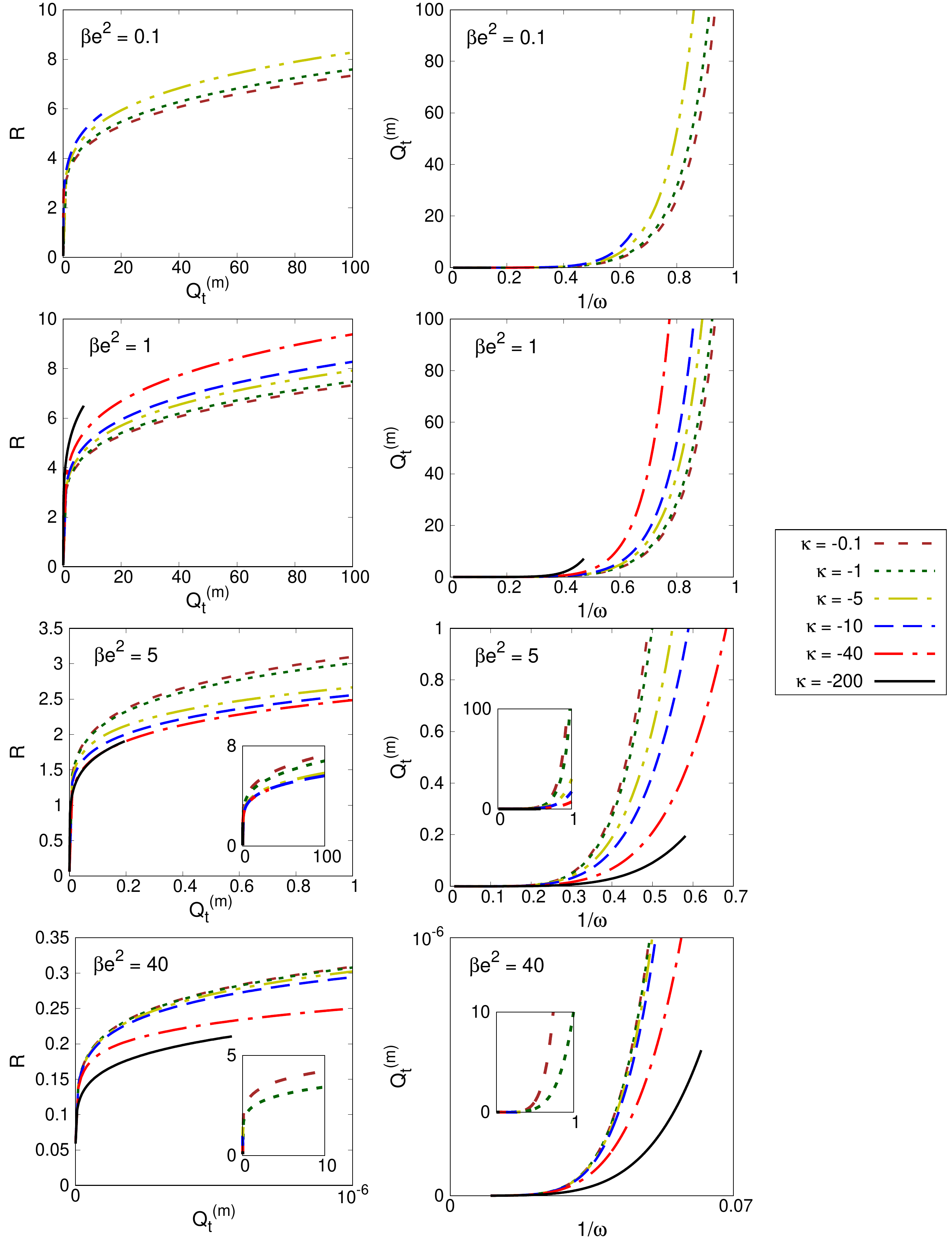}
\caption{$CP^{3}$ Q-ball radius $R$ versus the Noether charge $Q_t^{(m)}$, and $Q_t^{(m)}$ versus $1/\omega$, for some values of $\kappa$ and $\beta e^2 = -2\,\gamma e^2$}
\label{fig:radius_cp3}
\end{figure}

\begin{figure}[h!]
\centering
\includegraphics[scale=0.355]{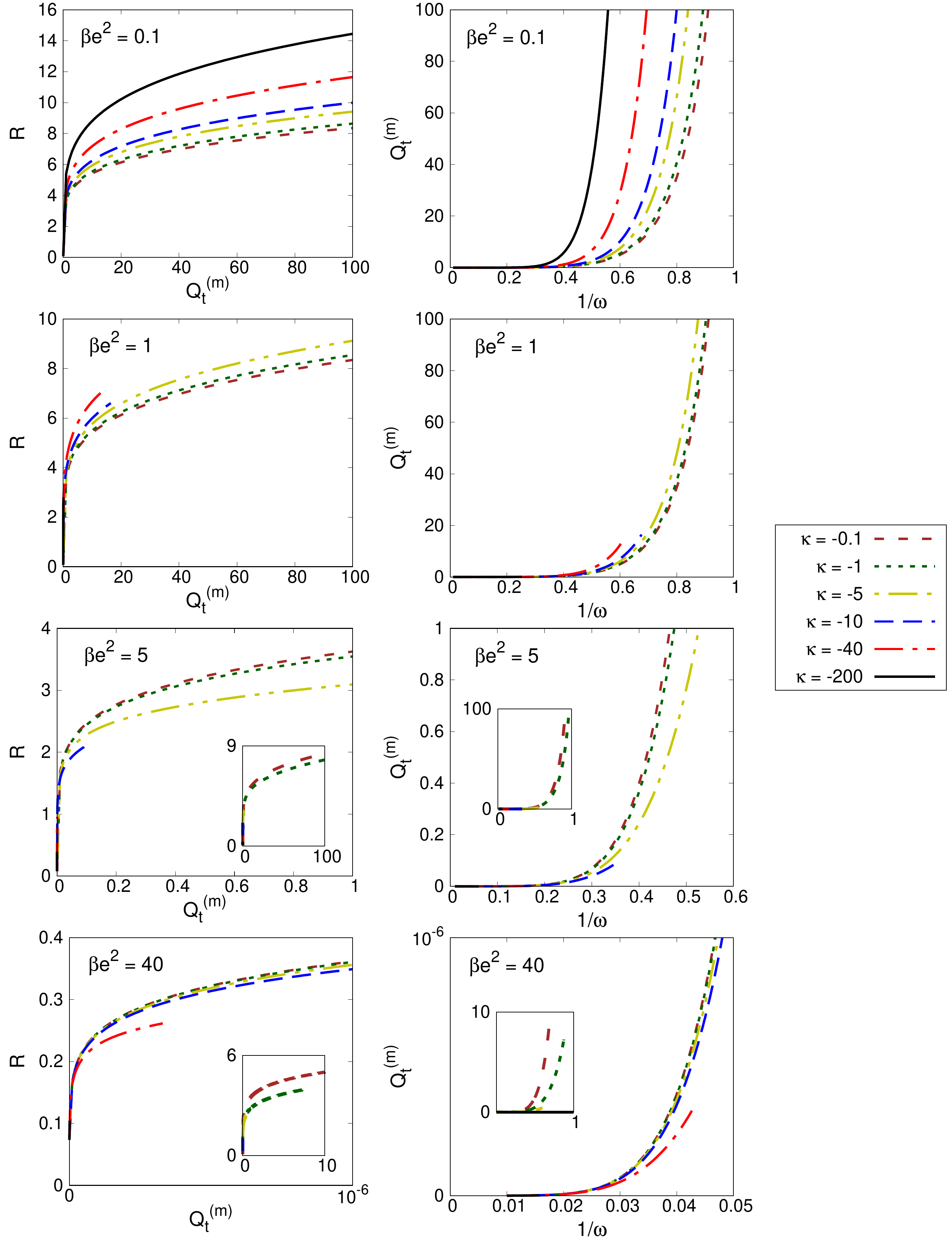}
\caption{$CP^{5}$ Q-shell radius $R\equiv R_{{\rm out}}-R_{{\rm in}}$ versus the Noether charge $Q_t^{(m)}$, and $Q_t^{(m)}$ versus $1/\omega$, for some values of $\kappa$ and $\beta e^2 = -2\,\gamma e^2$.}
\label{fig:radius_cp5}
\end{figure}

\begin{figure}[h!]
\centering
\includegraphics[scale=0.355]{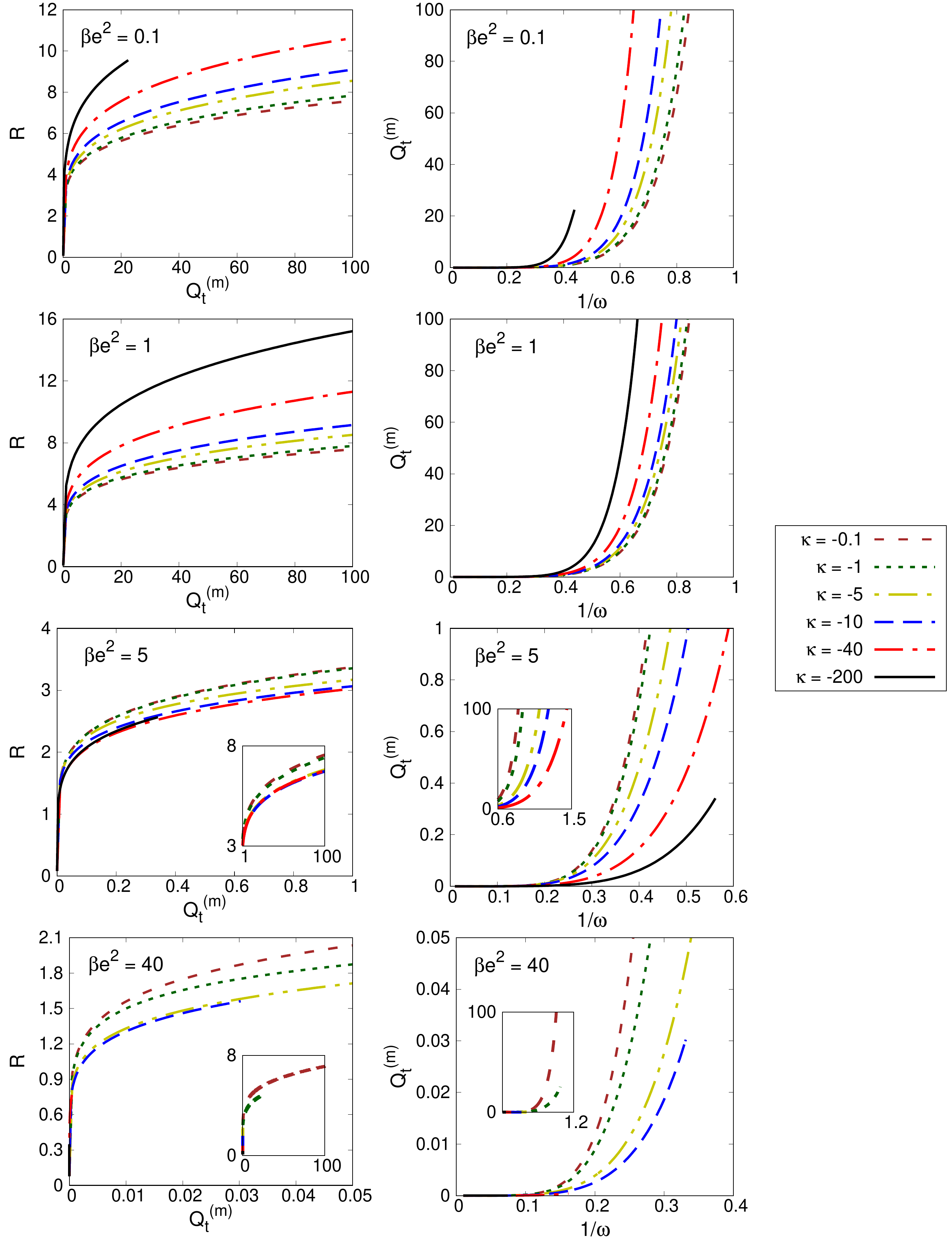}
\caption{$CP^{11}$  Q-shell radius $R\equiv R_{{\rm out}}-R_{{\rm in}}$ versus the Noether charge $Q_t^{(m)}$, and $Q_t^{(m)}$ versus $1/\omega$, for some values of $\kappa$ and $\beta e^2 = -2\,\gamma e^2$.}
\label{fig:radius_cp11}
\end{figure}

Across all parameter sets examined, solutions exist for $\omega \geq \omega_c$. However, for a fixed relation $\beta e^2 = -2\gamma e^2$, variations in $\kappa$ can lead to significant differences in the threshold frequency $\omega_c$. Under these conditions, the maximum values of the energy and Noether charge often vary by several orders of magnitude. This wide disparity in scales can render certain curves less prominent in the figures, as seen, for example, in the configurations with $\kappa = -200$ (see FIGS.~\ref{fig:energy_cp1}--\ref{fig:radius_cp11}).


Regarding the $CP^3$ solutions illustrated in FIGS. \ref{fig:energy_cp3} and \ref{fig:radius_cp3}, it should be noted that for high values of $|\kappa|$ and small values of $\beta e^2$, the solution curves are confined to the regime of small Noether charges. To ensure these features are discernible, a magnified view is provided in the upper-left panel of FIG. \ref{fig:energy_cp3}. Furthermore, it should be noted that the $CP^3$ $Q$-ball solution exhibits a higher sensitivity to variations in the parameter $a_1$ than the solutions presented in \cite{Klimas:2017eft}. In the latter case, the quadratic coefficient is constant, specifically $a_2 = \frac{\widetilde \mu^2}{32}$. However, the inclusion of quartic terms--under the condition $\beta e^2 = -2\gamma e^2$ modifies this coefficient to:
\[
a_2 = \frac{\widetilde \mu^2}{32} \frac{1}{1 - 2\kappa (\beta e^2 - 1) a_1}.
\]
Consequently, $a_2$ becomes explicitly dependent on $a_1$ for all cases where $\beta e^2 \neq 1$, introducing a coupling between these expansion coefficients that is absent in purely quadratic models.

The energy, Noether charge, and the maximum value of the profile function $f(r)$ within the compacton domain ($R_{\rm in} \leq r \leq R_{\rm out}$) can vary by several orders of magnitude as $\omega$ increases from the threshold $\omega_c$ to the maximum considered value of $\omega = 100$. Specifically, these quantities tend to decrease monotonically with increasing $\omega$. For example, in the case $l=0$, $\beta e^2=1$, and $\kappa = -0.1$, the value at the origin is $f(0) = 0.73799$ at $\omega = 0.94$; this value decreases to $7.0040 \times 10^{-5}$ by $\omega = 100$.

In the large-$\omega$ regime, differences in physical quantities such as the Noether charge and compacton width (see FIGS.~\ref{fig:energy_cp1}--\ref{fig:radius_cp11}) become less pronounced for a fixed value of $l$, provided that $\beta e^2$ and $\kappa$ are not varied drastically. For example, with $l=0$ and $\beta e^2=1$ at $\omega=30$, one finds $f(0)=7.7803 \times 10^{-4}$ for $\kappa=-0.1$ and $f(0)=7.3264 \times 10^{-4}$ for $\kappa=-200$; this represents a reduction of only approximately $5.8\%$ relative to the $\kappa=-0.1$ case. In contrast, at $\omega=1.40$, the values are $f(0)=0.338301$ for $\kappa=-0.1$ and $f(0)=0.229725$ for $\kappa=-200$, a significantly larger reduction of about $32.1\%$. This behavior is expected, as the diminishing field amplitudes at large $\omega$ effectively suppress the contribution of the quartic terms, leading to a convergence of the solutions.

The energy and Noether charge decrease monotonically with $\omega$ for every parameter set $\{l, \kappa, \beta e^2 = -2\gamma e^2\}$ examined. In the low-$\omega$ regime--where $\omega$ approaches the threshold $\omega_c$ and $1/\omega$ consequently reaches its maximum--the Noether charge increases rapidly, as illustrated in FIGS.~\ref{fig:energy_cp1}, \ref{fig:energy_cp3}, \ref{fig:energy_cp5}, and \ref{fig:energy_cp11}. Conversely, the critical value $\omega_c$ tends to increase with the magnitude of $|\kappa|$ or $\beta e^2$ across all cases ($l \in \{0, 1, 2, 5\}$). Indeed, the $Q_t^{(m)}$ curves shift toward smaller values of $1/\omega$ as $|\kappa|$ increases, indicating that the system enters the critical regime at higher frequencies. This behavior is explicitly demonstrated in the plots for $\beta e^2 =-2\gamma e^2= \{5, 40\}$, where the $Q_t^{(m)}$ versus $1/\omega$ curves are seen to terminate. These plots clearly demonstrate that the curves terminate at progressively smaller values of $1/\omega$ as $|\kappa|$ increases.

FIGS.~\ref{fig:radius_cp1}, \ref{fig:radius_cp3}, \ref{fig:radius_cp5}, and \ref{fig:radius_cp11} demonstrate that the compacton width $R$ increases more rapidly at low Noether charge (the low-$\omega$ regime), whereas this growth rate diminishes at larger values of the charge. In the large-$\omega$ regime, the $R$ versus $Q_t^{(m)}$ curves tend to converge, making the differences between them less pronounced. In this limit, the compacton width decreases monotonically, becoming significantly smaller as $\omega$ continues to increase.

\section{Summary and remarks}
This paper successfully generalizes the compact Q-ball type solutions for scalar field models possessing a ${C}P^N$ target space. Specifically, we have investigated the extended Skyrme-Faddeev (ESF) model, which augments the traditional Skyrme-Faddeev framework with the inclusion of a potential and quartic interaction terms. The main objective of this study was to analyze the influence of these quartic terms on the properties of compact Q-balls.

Similar to our previous results for models without quartic terms, we found two qualitatively distinct types of solutions: Q-balls and Q-shells. These solutions were obtained via a symmetry reduction of the model, which parameterizes the $N=2l+1$ complex scalar fields $u_m$ using the product of a time-dependent term $e^{i\omega t}$, a radial function $f(r)$, and spherical harmonics $Y_{lm}(\theta,\phi)$.   Q-balls are compact solutions obtained for ${C}P^N$ models containing fields with $N=1$ and $N=3$. For higher odd values of $N$, the compactons are qualitatively different, possessing a hollow spherical region at the center where the scalar field assumes the vacuum value. For this reason, these solutions are designated as Q-shells. The final radial profile was determined by numerically solving the resulting radial equation of motion. 

The increased complexity of the ESF model, particularly due to the presence of terms quadratic in powers of derivatives compared to the simpler ${C}P^N$ model, introduces three additional coupling constants ($e^{-2}$, $\beta$, and $\gamma$). This higher complexity necessitated an extra analysis regarding the existence of mathematical solutions and their physical admissibility (specifically, the positive definiteness of the energy density). For this reason, our primary focus was on solutions where the Hamiltonian density is strictly positive definite. However, we also identified cases where solutions exhibiting non-negative total energy density may still exist, even when the stricter positive-definite constraint on the Hamiltonian's quartic terms is not met.

The analysis of the solutions revealed that the existence of compactons satisfying the boundary conditions $f(R_{\text{out}})=0$ and $f'(R_{\text{out}})=0$ depends strongly on the intricate interplay between the model's coupling constants, the frequency parameter $\omega$, and the shooting parameters ($a_0$,  $a_1$ or $R_{\text {in}}$). We found that acceptable compact solutions often exist within certain windows in the parameter space, which are separated by "forbidden regions" where solutions are either non-compact or singular. Within the range of acceptable solutions, we noted interesting modifications to the compacton shapes and the energy density profiles that originate directly from the quartic terms of the model. 

The most striking difference observed is the modification of the bell-shaped profiles of the energy density, which now develop an extra peak localized in the outer region of the compacton (in the vicinity of $R_{\text{out}}$). Such solutions are characterized by a high concentration of energy in this region, which may be interpreted as a "dense crust."
The presence of a dense crust--characterized by a localized extra peak in the energy density near the compacton's boundary ($R_{\text{out}}$)--is a significant finding that extends to both Q-shells and Q-balls (such as the $\mathbb{C}P^3$ case). When these extended solutions are coupled to gravity, they become candidates for boson stars. This crust feature is particularly interesting because it suggests the possibility of self-gravitating objects with distinct mass-energy layering. Unlike simplified boson star models that often exhibit smoothly decreasing density profiles, the ESF model, via its quartic terms, provides a mechanism for generating stars with a highly concentrated, dense outer layer or shell structure. This complex internal makeup could lead to unique astrophysical signatures, such as altered mass-radius relations or distinct gravitational wave profiles during merger events, thus offering richer theoretical models for exotic compact objects.

We have also investigated the stability of Q-balls and Q-shells by analyzing the energy--Noether charge relation, which is well approximated by $E \sim Q^{\delta}$. For our numerical solutions, the parameter $\delta$, which is typically found in the range $\delta \simeq 0.79\text{--}0.89$, is always significantly smaller than one, implying the stability of such compactons.

\section*{Acknowledgmets}
This study was financed in part by the Coordenação de Aperfeiçoamento de Pessoal de Nível Superior -- Brasil (CAPES) -- Finance Code 001. N.S. was supported in part by JSPS KAKENHI Grant No. JP23K02794. L.R.L. is supported by the grant 2025/20538-2, São Paulo Research Foundation (FAPESP).
\appendix
\section{}
\label{appendA}

Many simplifications follow directly from the addition theorem for spherical harmonics:
\be
P_l(\cos\gamma)=\frac{4\pi}{2l+1}\sum_{m=-l}^lY^*_{lm}(\theta',\phi')Y_{lm}(\theta,\phi),\nonumber
\ee
where $\cos\gamma=\cos\theta\cos\theta'+\sin\theta\sin\theta'\cos(\phi-\phi')$. In particular, when $\theta=\theta'$ and $\phi=\phi'$ (which implies $\gamma=0$), this equality yields$$\sum_{m=-l}^l\frac{(l-m)!}{(l+m)}[P_l(\cos\theta)]^2=1.$$ Taking the derivative with respect to $\theta$ on both sides of the last equality gives
\be
\sum_{m=-l}^l\frac{(l-m)!}{(l+m)}P_l(\cos\theta)\partial_{\theta}P_l(\cos\theta)=0. \label{a3}
\ee
Applying the ansatz from equation \eqref{ansatz}results in the following expressions:
\begin{align}
u^{\dagger}\cdot \partial_tu&=i\omega f^2& u^{\dagger}\cdot \partial_ru&=f'f & u^{\dagger}\cdot \partial_{\theta} u&=0& u^{\dagger}\cdot \partial_{\phi} u&=0\nonumber\\
\partial_{\theta}u^{\dagger}\cdot \partial_{\theta}u&=\frac{l(l+1)}{2}f^2& \partial_ru^{\dagger}\cdot \partial_ru&=f'^2& \partial_{t}u^{\dagger}\cdot \partial_{r}u&=-i\omega f'f&\partial_{\theta}u^{\dagger}\cdot \partial_{\phi}u &=0\nonumber\\
\partial_{\phi}u^{\dagger}\cdot \partial_{\phi}u&=\frac{l(l+1)}{2}\sin^2\theta f^2&\partial_{t}u^{\dagger}\cdot \partial_{t}u &=\omega^2 f^2&\partial_ru^{\dagger}\cdot \partial_{\alpha}u&=0&\partial_tu^{\dagger}\cdot \partial_{\alpha}u&=0\nonumber
\end{align}

\section{}
\label{appendB}
The expressions for $K_{\sigma}$ are given by:
\begin{align} 
K^{(m)}_{\sigma}&:=u^*_m\sum_{m'=-l}^l \Delta^2_{mm'}\partial_{\sigma}u_{m'}\label{b1}\\
&=u^*_m\sum_{m'=-l}^l \left[(1+u^{\dagger}\cdot u)\delta_{mm'}-u_mu^*_{m'}\right]\partial_{\sigma}u_{m'}\nonumber\\
&=(1+u^{\dagger}\cdot u)\,u_m^*\partial_{\sigma}u_{m}-|u_m|^2\sum_{m'=-l}^lu_{m'}^*\partial_{\sigma}u_{m'}.\label{b2}
\end{align}
Expression  \eqref{b1}  is more convenient when $\partial_tu_{m'}=i\omega u_{m'}$ and $\partial_{r}u_{m'}=\frac{f'}{f} u_{m'}$, as $u_m'$ is an eigenvector of the matrix $\Delta^2_{mm'}u_{m'}$, with an eigenvalue of $\lambda=1$. For the case where $\sigma=\theta$, the term $\sum_{m'=-l}^lu_{m'}^*\partial_{\theta}u_{m'}$ vanishes according to equation  \eqref{a3}. Similarly, for $\sigma=\phi$, one obtains  $\sum_{m'=-l}^lu_{m'}^*\partial_{\phi}u_{m'}=i\sum_{m'=-l}^l |u_{m'}|^2=0$. In these instances, it is more convenient to apply expression \eqref{b2}. This leads to the following results:
\begin{align}
K^{(m)}_t&=i\omega f^2\frac{(l-m)!}{(l+m)!}(P^m_l)^2,\nonumber\\
K^{(m)}_r&=f'f\frac{(l-m)!}{(l+m)!}(P^m_l)^2,\nonumber\\
K^{(m)}_{\theta}&= f^2(1+f^2)\frac{(l-m)!}{(l+m)!}P^m_l\partial_{\theta}P^m_l,\nonumber\\
K^{(m)}_{\phi}&=im f^2(1+f^2)\frac{(l-m)!}{(l+m)!}(P^m_l)^2.\nonumber
\end{align}

\section{}
\label{appendC}

The sum $\sum_{m=-l}^l\omega Q_t^{(m)}$ is given by a radial integral, and this quantity can be expressed as a volume integral of some density. This yields:
\begin{align}
\sum_{m=-l}^l\omega Q_t^{(m)}=&\int_{{\mathbb{R}}^3}d^3\xi\sqrt{-g}\Bigg[\overbrace{\frac{4\omega^2f^2}{(1+f^2)^2}}^A+\kappa\frac{4\omega^2f^2}{(1+f^2)^2}F_1+\overbrace{\kappa\frac{12\omega^2f'f}{(1+f^2)^2}F_3}^C\Bigg]
\end{align}
A similar approach provides the expression for $\sum_{m=-l}^lm Q_\phi^{(m)}$ in the form:
\begin{align}
\sum_{m=-l}^lm Q_\phi^{(m)}=&\int_{{\mathbb{R}}^3}d^3\xi\sqrt{-g}\Bigg[\overbrace{\frac{l(l+1)}{r^2}\frac{4f^2}{1+f^2}}^B-\kappa\frac{l(l+1)}{r^2}\frac{4f^2}{1+f^2}F_4\Bigg].
\end{align}
The total energy is given by:
\begin{align}
E=&\int_{{\mathbb{R}}^3}d^3\xi\sqrt{-g}\Bigg[\overbrace{\frac{4\omega^2f^2}{(1+f^2)^2}}^A+\overbrace{\frac{l(l+1)}{r^2}\frac{4f^2}{1+f^2}}^B\Bigg]+\int_{{\mathbb{R}}^3}d^3\xi\sqrt{-g}\left[\frac{4f'^2}{(1+f^2)^2}+\tilde\mu^2V\right]\nonumber\\
&+\int_{{\mathbb{R}}^3}d^3\xi\sqrt{-g}\Bigg[\kappa\frac{6\omega^2f^2}{(1+f^2)^2}F_1(r)-\kappa\frac{2f'^2}{(1+f^2)^2}F_2(r)\nonumber\\ &+\overbrace{\kappa\frac{12\omega^2 f'f}{(1+f^2)^2}F_3(r)}^C-\kappa\frac{l(l+1)}{r^2}\frac{2f^2}{1+f^2}F_4(r)\Bigg]\label{engC1}
\end{align}
This can also be cast in the form:
\begin{align}
E=&\sum_{m=-l}^l\left(\omega Q_t^{(m)}+m Q_\phi^{(m)}\right)+\int_{{\mathbb{R}}^3}d^3\xi\sqrt{-g}\left[\frac{4f'^2}{(1+f^2)^2}+\tilde\mu^2V\right]\nonumber\\
&+\int_{{\mathbb{R}}^3}d^3\xi\sqrt{-g}\Bigg[\kappa\frac{2\omega^2f^2}{(1+f^2)^2}F_1-\kappa\frac{2f'^2}{(1+f^2)^2}F_2+\kappa \frac{l(l+1)}{r^2}\frac{2f^2}{1+f^2}F_4\Bigg].\label{engC2}
\end{align}

\bibliography{refs.bib}

\begin{thebibliography}{39}%
\makeatletter
\providecommand \@ifxundefined [1]{%
 \@ifx{#1\undefined}
}%
\providecommand \@ifnum [1]{%
 \ifnum #1\expandafter \@firstoftwo
 \else \expandafter \@secondoftwo
 \fi
}%
\providecommand \@ifx [1]{%
 \ifx #1\expandafter \@firstoftwo
 \else \expandafter \@secondoftwo
 \fi
}%
\providecommand \natexlab [1]{#1}%
\providecommand \enquote  [1]{``#1''}%
\providecommand \bibnamefont  [1]{#1}%
\providecommand \bibfnamefont [1]{#1}%
\providecommand \citenamefont [1]{#1}%
\providecommand \href@noop [0]{\@secondoftwo}%
\providecommand \href [0]{\begingroup \@sanitize@url \@href}%
\providecommand \@href[1]{\@@startlink{#1}\@@href}%
\providecommand \@@href[1]{\endgroup#1\@@endlink}%
\providecommand \@sanitize@url [0]{\catcode `\\12\catcode `\$12\catcode
  `\&12\catcode `\#12\catcode `\^12\catcode `\_12\catcode `\%12\relax}%
\providecommand \@@startlink[1]{}%
\providecommand \@@endlink[0]{}%
\providecommand \url  [0]{\begingroup\@sanitize@url \@url }%
\providecommand \@url [1]{\endgroup\@href {#1}{\urlprefix }}%
\providecommand \urlprefix  [0]{URL }%
\providecommand \Eprint [0]{\href }%
\providecommand \doibase [0]{http://dx.doi.org/}%
\providecommand \selectlanguage [0]{\@gobble}%
\providecommand \bibinfo  [0]{\@secondoftwo}%
\providecommand \bibfield  [0]{\@secondoftwo}%
\providecommand \translation [1]{[#1]}%
\providecommand \BibitemOpen [0]{}%
\providecommand \bibitemStop [0]{}%
\providecommand \bibitemNoStop [0]{.\EOS\space}%
\providecommand \EOS [0]{\spacefactor3000\relax}%
\providecommand \BibitemShut  [1]{\csname bibitem#1\endcsname}%
\let\auto@bib@innerbib\@empty
\bibitem [{\citenamefont {Higgs}(1964)}]{PhysRevLett.13.508}%
  \BibitemOpen
  \bibfield  {author} {\bibinfo {author} {\bibfnamefont {Peter~W.}\
  \bibnamefont {Higgs}},\ }\bibfield  {title} {\enquote {\bibinfo {title}
  {Broken symmetries and the masses of gauge bosons},}\ }\href {\doibase
  10.1103/PhysRevLett.13.508} {\bibfield  {journal} {\bibinfo  {journal} {Phys.
  Rev. Lett.}\ }\textbf {\bibinfo {volume} {13}},\ \bibinfo {pages} {508--509}
  (\bibinfo {year} {1964})}\BibitemShut {NoStop}%
\bibitem [{\citenamefont {Ginzburg}\ and\ \citenamefont
  {Landau}(1950)}]{Ginzburg:1950sr}%
  \BibitemOpen
  \bibfield  {author} {\bibinfo {author} {\bibfnamefont {V.~L.}\ \bibnamefont
  {Ginzburg}}\ and\ \bibinfo {author} {\bibfnamefont {L.~D.}\ \bibnamefont
  {Landau}},\ }\bibfield  {title} {\enquote {\bibinfo {title} {{On the Theory
  of superconductivity}},}\ }\href {\doibase
  10.1016/b978-0-08-010586-4.50078-x} {\bibfield  {journal} {\bibinfo
  {journal} {Zh. Eksp. Teor. Fiz.}\ }\textbf {\bibinfo {volume} {20}},\
  \bibinfo {pages} {1064--1082} (\bibinfo {year} {1950})}\BibitemShut {NoStop}%
\bibitem [{\citenamefont {Zakrzewski}(1988)}]{zakrzewski1988low}%
  \BibitemOpen
  \bibfield  {author} {\bibinfo {author} {\bibfnamefont {WJ}~\bibnamefont
  {Zakrzewski}},\ }\href@noop {} {\emph {\bibinfo {title} {Low dimensional
  sigma models}}},\ \bibinfo {type} {Tech. Rep.}\ (\bibinfo  {institution} {Los
  Alamos National Laboratory (LANL), Los Alamos, NM (United States)},\ \bibinfo
  {year} {1988})\BibitemShut {NoStop}%
\bibitem [{\citenamefont {Faddeev}(1975)}]{Faddeev:1975tz}%
  \BibitemOpen
  \bibfield  {author} {\bibinfo {author} {\bibfnamefont {L.~D.}\ \bibnamefont
  {Faddeev}},\ }\bibfield  {title} {\enquote {\bibinfo {title} {{Quantization
  of Solitons}},}\ }in\ \href@noop {} {\emph {\bibinfo {booktitle} {{18th
  International Conference on High-Energy Physics}}}}\ (\bibinfo {year}
  {1975})\BibitemShut {NoStop}%
\bibitem [{\citenamefont {Faddeev}(1995)}]{Faddeev:1995au}%
  \BibitemOpen
  \bibfield  {author} {\bibinfo {author} {\bibfnamefont {L.~D.}\ \bibnamefont
  {Faddeev}},\ }\href@noop {} {\emph {\bibinfo {title} {{40 years in
  mathematical physics}}}}\ (\bibinfo {year} {1995})\BibitemShut {NoStop}%
\bibitem [{\citenamefont {Battye}\ and\ \citenamefont
  {Sutcliffe}(1998)}]{Battye:1998pe}%
  \BibitemOpen
  \bibfield  {author} {\bibinfo {author} {\bibfnamefont {Richard~A.}\
  \bibnamefont {Battye}}\ and\ \bibinfo {author} {\bibfnamefont {Paul~M.}\
  \bibnamefont {Sutcliffe}},\ }\bibfield  {title} {\enquote {\bibinfo {title}
  {{Knots as stable soliton solutions in a three-dimensional classical field
  theory.}}}\ }\href {\doibase 10.1103/PhysRevLett.81.4798} {\bibfield
  {journal} {\bibinfo  {journal} {Phys. Rev. Lett.}\ }\textbf {\bibinfo
  {volume} {81}},\ \bibinfo {pages} {4798--4801} (\bibinfo {year} {1998})},\
  \Eprint {http://arxiv.org/abs/hep-th/9808129} {arXiv:hep-th/9808129}
  \BibitemShut {NoStop}%
\bibitem [{\citenamefont {Sutcliffe}(2007)}]{Sutcliffe:2007ui}%
  \BibitemOpen
  \bibfield  {author} {\bibinfo {author} {\bibfnamefont {Paul}\ \bibnamefont
  {Sutcliffe}},\ }\bibfield  {title} {\enquote {\bibinfo {title} {{Knots in the
  Skyrme-Faddeev model}},}\ }\href {\doibase 10.1098/rspa.2007.0038} {\bibfield
   {journal} {\bibinfo  {journal} {Proc. Roy. Soc. Lond. A}\ }\textbf {\bibinfo
  {volume} {463}},\ \bibinfo {pages} {3001--3020} (\bibinfo {year} {2007})},\
  \Eprint {http://arxiv.org/abs/0705.1468} {arXiv:0705.1468 [hep-th]}
  \BibitemShut {NoStop}%
\bibitem [{\citenamefont {Faddeev}\ and\ \citenamefont
  {Niemi}(1999{\natexlab{a}})}]{Faddeev:1998yz}%
  \BibitemOpen
  \bibfield  {author} {\bibinfo {author} {\bibfnamefont {L.~D.}\ \bibnamefont
  {Faddeev}}\ and\ \bibinfo {author} {\bibfnamefont {Antti~J.}\ \bibnamefont
  {Niemi}},\ }\bibfield  {title} {\enquote {\bibinfo {title} {{Partial duality
  in SU(N) Yang-Mills theory}},}\ }\href {\doibase
  10.1016/S0370-2693(99)00100-8} {\bibfield  {journal} {\bibinfo  {journal}
  {Phys. Lett. B}\ }\textbf {\bibinfo {volume} {449}},\ \bibinfo {pages}
  {214--218} (\bibinfo {year} {1999}{\natexlab{a}})},\ \Eprint
  {http://arxiv.org/abs/hep-th/9812090} {arXiv:hep-th/9812090} \BibitemShut
  {NoStop}%
\bibitem [{\citenamefont {Gies}(2001)}]{Gies:2001hk}%
  \BibitemOpen
  \bibfield  {author} {\bibinfo {author} {\bibfnamefont {Holger}\ \bibnamefont
  {Gies}},\ }\bibfield  {title} {\enquote {\bibinfo {title} {{Wilsonian
  effective action for SU(2) Yang-Mills theory with Cho-Faddeev-Niemi-Shabanov
  decomposition}},}\ }\href {\doibase 10.1103/PhysRevD.63.125023} {\bibfield
  {journal} {\bibinfo  {journal} {Phys. Rev. D}\ }\textbf {\bibinfo {volume}
  {63}},\ \bibinfo {pages} {125023} (\bibinfo {year} {2001})},\ \Eprint
  {http://arxiv.org/abs/hep-th/0102026} {arXiv:hep-th/0102026} \BibitemShut
  {NoStop}%
\bibitem [{\citenamefont {Ferreira}(2009)}]{Ferreira:2008nn}%
  \BibitemOpen
  \bibfield  {author} {\bibinfo {author} {\bibfnamefont {L.~A.}\ \bibnamefont
  {Ferreira}},\ }\bibfield  {title} {\enquote {\bibinfo {title} {{Exact vortex
  solutions in an extended Skyrme-Faddeev model}},}\ }\href {\doibase
  10.1088/1126-6708/2009/05/001} {\bibfield  {journal} {\bibinfo  {journal}
  {JHEP}\ }\textbf {\bibinfo {volume} {05}},\ \bibinfo {pages} {001} (\bibinfo
  {year} {2009})},\ \Eprint {http://arxiv.org/abs/0809.4303} {arXiv:0809.4303
  [hep-th]} \BibitemShut {NoStop}%
\bibitem [{\citenamefont {Faddeev}\ and\ \citenamefont
  {Niemi}(1999{\natexlab{b}})}]{Faddeev:1999cj}%
  \BibitemOpen
  \bibfield  {author} {\bibinfo {author} {\bibfnamefont {L.~D.}\ \bibnamefont
  {Faddeev}}\ and\ \bibinfo {author} {\bibfnamefont {Antti~J.}\ \bibnamefont
  {Niemi}},\ }\bibfield  {title} {\enquote {\bibinfo {title} {{Decomposing the
  Yang-Mills field}},}\ }\href {\doibase 10.1016/S0370-2693(99)01035-7}
  {\bibfield  {journal} {\bibinfo  {journal} {Phys. Lett. B}\ }\textbf
  {\bibinfo {volume} {464}},\ \bibinfo {pages} {90--93} (\bibinfo {year}
  {1999}{\natexlab{b}})},\ \Eprint {http://arxiv.org/abs/hep-th/9907180}
  {arXiv:hep-th/9907180} \BibitemShut {NoStop}%
\bibitem [{\citenamefont {Kondo}\ \emph {et~al.}(2008)\citenamefont {Kondo},
  \citenamefont {Shinohara},\ and\ \citenamefont {Murakami}}]{Kondo:2008xa}%
  \BibitemOpen
  \bibfield  {author} {\bibinfo {author} {\bibfnamefont {Kei-Ichi}\
  \bibnamefont {Kondo}}, \bibinfo {author} {\bibfnamefont {Toru}\ \bibnamefont
  {Shinohara}}, \ and\ \bibinfo {author} {\bibfnamefont {Takeharu}\
  \bibnamefont {Murakami}},\ }\bibfield  {title} {\enquote {\bibinfo {title}
  {{Reformulating SU(N) Yang-Mills theory based on change of variables}},}\
  }\href {\doibase 10.1143/PTP.120.1} {\bibfield  {journal} {\bibinfo
  {journal} {Prog. Theor. Phys.}\ }\textbf {\bibinfo {volume} {120}},\ \bibinfo
  {pages} {1--50} (\bibinfo {year} {2008})},\ \Eprint
  {http://arxiv.org/abs/0803.0176} {arXiv:0803.0176 [hep-th]} \BibitemShut
  {NoStop}%
\bibitem [{\citenamefont {Ferreira}\ and\ \citenamefont
  {Klimas}(2010)}]{Ferreira:2010jb}%
  \BibitemOpen
  \bibfield  {author} {\bibinfo {author} {\bibfnamefont {L.~A.}\ \bibnamefont
  {Ferreira}}\ and\ \bibinfo {author} {\bibfnamefont {P.}~\bibnamefont
  {Klimas}},\ }\bibfield  {title} {\enquote {\bibinfo {title} {{Exact vortex
  solutions in a $CP^N$ Skyrme-Faddeev type model}},}\ }\href {\doibase
  10.1007/JHEP10(2010)008} {\bibfield  {journal} {\bibinfo  {journal} {JHEP}\
  }\textbf {\bibinfo {volume} {10}},\ \bibinfo {pages} {008} (\bibinfo {year}
  {2010})},\ \Eprint {http://arxiv.org/abs/1007.1667} {arXiv:1007.1667
  [hep-th]} \BibitemShut {NoStop}%
\bibitem [{\citenamefont {Ferreira}\ \emph {et~al.}(2011)\citenamefont
  {Ferreira}, \citenamefont {Klimas},\ and\ \citenamefont
  {Zakrzewski}}]{Ferreira:2011ja}%
  \BibitemOpen
  \bibfield  {author} {\bibinfo {author} {\bibfnamefont {L.~A.}\ \bibnamefont
  {Ferreira}}, \bibinfo {author} {\bibfnamefont {P.}~\bibnamefont {Klimas}}, \
  and\ \bibinfo {author} {\bibfnamefont {W.~J.}\ \bibnamefont {Zakrzewski}},\
  }\bibfield  {title} {\enquote {\bibinfo {title} {{Some properties of (3+1)
  dimensional vortex solutions in the extended $CP^N$ Skyrme-Faddeev model}},}\
  }\href {\doibase 10.1007/JHEP12(2011)098} {\bibfield  {journal} {\bibinfo
  {journal} {JHEP}\ }\textbf {\bibinfo {volume} {12}},\ \bibinfo {pages} {098}
  (\bibinfo {year} {2011})},\ \Eprint {http://arxiv.org/abs/1111.2338}
  {arXiv:1111.2338 [hep-th]} \BibitemShut {NoStop}%
\bibitem [{\citenamefont {Ferreira}\ \emph {et~al.}(2012)\citenamefont
  {Ferreira}, \citenamefont {Jaykka}, \citenamefont {Sawado},\ and\
  \citenamefont {Toda}}]{Ferreira:2011mz}%
  \BibitemOpen
  \bibfield  {author} {\bibinfo {author} {\bibfnamefont {L.~A.}\ \bibnamefont
  {Ferreira}}, \bibinfo {author} {\bibfnamefont {J.}~\bibnamefont {Jaykka}},
  \bibinfo {author} {\bibfnamefont {Nobuyuki}\ \bibnamefont {Sawado}}, \ and\
  \bibinfo {author} {\bibfnamefont {Kouichi}\ \bibnamefont {Toda}},\ }\bibfield
   {title} {\enquote {\bibinfo {title} {{Vortices in the Extended
  Skyrme-Faddeev Model}},}\ }\href {\doibase 10.1103/PhysRevD.85.105006}
  {\bibfield  {journal} {\bibinfo  {journal} {Phys. Rev. D}\ }\textbf {\bibinfo
  {volume} {85}},\ \bibinfo {pages} {105006} (\bibinfo {year} {2012})},\
  \Eprint {http://arxiv.org/abs/1112.1085} {arXiv:1112.1085 [hep-th]}
  \BibitemShut {NoStop}%
\bibitem [{\citenamefont {Sawado}\ and\ \citenamefont
  {Tamaki}(2013)}]{Sawado:2013yza}%
  \BibitemOpen
  \bibfield  {author} {\bibinfo {author} {\bibfnamefont {Nobuyuki}\
  \bibnamefont {Sawado}}\ and\ \bibinfo {author} {\bibfnamefont {Yuta}\
  \bibnamefont {Tamaki}},\ }\bibfield  {title} {\enquote {\bibinfo {title}
  {{Exact, molecular-shaped vortices with fractional and integer charges in the
  extended Skyrme-Faddeev model}},}\ }\href@noop {} {\  (\bibinfo {year}
  {2013})},\ \Eprint {http://arxiv.org/abs/1309.6004} {arXiv:1309.6004
  [hep-th]} \BibitemShut {NoStop}%
\bibitem [{\citenamefont {Amari}\ \emph {et~al.}(2015)\citenamefont {Amari},
  \citenamefont {Klimas}, \citenamefont {Sawado},\ and\ \citenamefont
  {Tamaki}}]{Amari:2015sva}%
  \BibitemOpen
  \bibfield  {author} {\bibinfo {author} {\bibfnamefont {Yuki}\ \bibnamefont
  {Amari}}, \bibinfo {author} {\bibfnamefont {Pawel}\ \bibnamefont {Klimas}},
  \bibinfo {author} {\bibfnamefont {Nobuyuki}\ \bibnamefont {Sawado}}, \ and\
  \bibinfo {author} {\bibfnamefont {Yuta}\ \bibnamefont {Tamaki}},\ }\bibfield
  {title} {\enquote {\bibinfo {title} {{Potentials and the vortex solutions in
  the $CP^N$ Skyrme-Faddeev model}},}\ }\href {\doibase
  10.1103/PhysRevD.92.045007} {\bibfield  {journal} {\bibinfo  {journal} {Phys.
  Rev. D}\ }\textbf {\bibinfo {volume} {92}},\ \bibinfo {pages} {045007}
  (\bibinfo {year} {2015})},\ \Eprint {http://arxiv.org/abs/1504.02848}
  {arXiv:1504.02848 [hep-th]} \BibitemShut {NoStop}%
\bibitem [{\citenamefont {Lee}\ and\ \citenamefont
  {Pang}(1992{\natexlab{a}})}]{LEE1992251}%
  \BibitemOpen
  \bibfield  {author} {\bibinfo {author} {\bibfnamefont {T.D}\ \bibnamefont
  {Lee}}\ and\ \bibinfo {author} {\bibfnamefont {Y}~\bibnamefont {Pang}},\
  }\bibfield  {title} {\enquote {\bibinfo {title} {Nontopological solitons},}\
  }\href {\doibase https://doi.org/10.1016/0370-1573(92)90064-7} {\bibfield
  {journal} {\bibinfo  {journal} {Physics Reports}\ }\textbf {\bibinfo {volume}
  {221}},\ \bibinfo {pages} {251--350} (\bibinfo {year}
  {1992}{\natexlab{a}})}\BibitemShut {NoStop}%
\bibitem [{\citenamefont {Derrick}(1964)}]{Derrick:1964ww}%
  \BibitemOpen
  \bibfield  {author} {\bibinfo {author} {\bibfnamefont {G.~H.}\ \bibnamefont
  {Derrick}},\ }\bibfield  {title} {\enquote {\bibinfo {title} {{Comments on
  nonlinear wave equations as models for elementary particles}},}\ }\href
  {\doibase 10.1063/1.1704233} {\bibfield  {journal} {\bibinfo  {journal} {J.
  Math. Phys.}\ }\textbf {\bibinfo {volume} {5}},\ \bibinfo {pages}
  {1252--1254} (\bibinfo {year} {1964})}\BibitemShut {NoStop}%
\bibitem [{\citenamefont {Lee}\ and\ \citenamefont
  {Pang}(1992{\natexlab{b}})}]{Lee:1991ax}%
  \BibitemOpen
  \bibfield  {author} {\bibinfo {author} {\bibfnamefont {T.~D.}\ \bibnamefont
  {Lee}}\ and\ \bibinfo {author} {\bibfnamefont {Y.}~\bibnamefont {Pang}},\
  }\bibfield  {title} {\enquote {\bibinfo {title} {{Nontopological
  solitons}},}\ }\href {\doibase 10.1016/0370-1573(92)90064-7} {\bibfield
  {journal} {\bibinfo  {journal} {Phys. Rept.}\ }\textbf {\bibinfo {volume}
  {221}},\ \bibinfo {pages} {251--350} (\bibinfo {year}
  {1992}{\natexlab{b}})}\BibitemShut {NoStop}%
\bibitem [{\citenamefont {Rosen}(1968)}]{Rosen:1968mfz}%
  \BibitemOpen
  \bibfield  {author} {\bibinfo {author} {\bibfnamefont {Gerald}\ \bibnamefont
  {Rosen}},\ }\bibfield  {title} {\enquote {\bibinfo {title} {{Particlelike
  Solutions to Nonlinear Complex Scalar Field Theories with Positive-Definite
  Energy Densities}},}\ }\href {\doibase 10.1063/1.1664693} {\bibfield
  {journal} {\bibinfo  {journal} {J. Math. Phys.}\ }\textbf {\bibinfo {volume}
  {9}},\ \bibinfo {pages} {996} (\bibinfo {year} {1968})}\BibitemShut {NoStop}%
\bibitem [{\citenamefont {Werle}(1977)}]{Werle:1977cc}%
  \BibitemOpen
  \bibfield  {author} {\bibinfo {author} {\bibfnamefont {J.}~\bibnamefont
  {Werle}},\ }\bibfield  {title} {\enquote {\bibinfo {title} {{Dirac Spinor
  Solitons Or Bags}},}\ }\href {\doibase 10.1016/0370-2693(77)90238-6}
  {\bibfield  {journal} {\bibinfo  {journal} {Phys. Lett. B}\ }\textbf
  {\bibinfo {volume} {71}},\ \bibinfo {pages} {357--359} (\bibinfo {year}
  {1977})}\BibitemShut {NoStop}%
\bibitem [{\citenamefont {Arodz}\ \emph {et~al.}(2005)\citenamefont {Arodz},
  \citenamefont {Klimas},\ and\ \citenamefont {Tyranowski}}]{Arodz:2005gz}%
  \BibitemOpen
  \bibfield  {author} {\bibinfo {author} {\bibfnamefont {H.}~\bibnamefont
  {Arodz}}, \bibinfo {author} {\bibfnamefont {P.}~\bibnamefont {Klimas}}, \
  and\ \bibinfo {author} {\bibfnamefont {T.}~\bibnamefont {Tyranowski}},\
  }\bibfield  {title} {\enquote {\bibinfo {title} {{Field-theoretic models with
  V-shaped potentials}},}\ }\href@noop {} {\bibfield  {journal} {\bibinfo
  {journal} {Acta Phys. Polon. B}\ }\textbf {\bibinfo {volume} {36}},\ \bibinfo
  {pages} {3861--3876} (\bibinfo {year} {2005})},\ \Eprint
  {http://arxiv.org/abs/hep-th/0510204} {arXiv:hep-th/0510204} \BibitemShut
  {NoStop}%
\bibitem [{\citenamefont {Arodz}\ \emph {et~al.}(2006)\citenamefont {Arodz},
  \citenamefont {Klimas},\ and\ \citenamefont {Tyranowski}}]{Arodz:2005bc}%
  \BibitemOpen
  \bibfield  {author} {\bibinfo {author} {\bibfnamefont {H.}~\bibnamefont
  {Arodz}}, \bibinfo {author} {\bibfnamefont {P.}~\bibnamefont {Klimas}}, \
  and\ \bibinfo {author} {\bibfnamefont {T.}~\bibnamefont {Tyranowski}},\
  }\bibfield  {title} {\enquote {\bibinfo {title} {{Scaling, self-similar
  solutions and shock waves for V-shaped field potentials}},}\ }\href {\doibase
  10.1103/PhysRevE.73.046609} {\bibfield  {journal} {\bibinfo  {journal} {Phys.
  Rev. E}\ }\textbf {\bibinfo {volume} {73}},\ \bibinfo {pages} {046609}
  (\bibinfo {year} {2006})},\ \Eprint {http://arxiv.org/abs/hep-th/0511022}
  {arXiv:hep-th/0511022} \BibitemShut {NoStop}%
\bibitem [{\citenamefont {Arodz}\ and\ \citenamefont
  {Lis}(2008)}]{Arodz:2008jk}%
  \BibitemOpen
  \bibfield  {author} {\bibinfo {author} {\bibfnamefont {H.}~\bibnamefont
  {Arodz}}\ and\ \bibinfo {author} {\bibfnamefont {J.}~\bibnamefont {Lis}},\
  }\bibfield  {title} {\enquote {\bibinfo {title} {{Compact Q-balls in the
  complex signum-Gordon model}},}\ }\href {\doibase 10.1103/PhysRevD.77.107702}
  {\bibfield  {journal} {\bibinfo  {journal} {Phys. Rev. D}\ }\textbf {\bibinfo
  {volume} {77}},\ \bibinfo {pages} {107702} (\bibinfo {year} {2008})},\
  \Eprint {http://arxiv.org/abs/0803.1566} {arXiv:0803.1566 [hep-th]}
  \BibitemShut {NoStop}%
\bibitem [{\citenamefont {Arodz}\ \emph {et~al.}(2009)\citenamefont {Arodz},
  \citenamefont {Karkowski},\ and\ \citenamefont
  {Swierczynski}}]{Arodz:2009ye}%
  \BibitemOpen
  \bibfield  {author} {\bibinfo {author} {\bibfnamefont {H.}~\bibnamefont
  {Arodz}}, \bibinfo {author} {\bibfnamefont {J.}~\bibnamefont {Karkowski}}, \
  and\ \bibinfo {author} {\bibfnamefont {Z.}~\bibnamefont {Swierczynski}},\
  }\bibfield  {title} {\enquote {\bibinfo {title} {{Spinning Q-balls in the
  complex signum-Gordon model}},}\ }\href {\doibase 10.1103/PhysRevD.80.067702}
  {\bibfield  {journal} {\bibinfo  {journal} {Phys. Rev. D}\ }\textbf {\bibinfo
  {volume} {80}},\ \bibinfo {pages} {067702} (\bibinfo {year} {2009})},\
  \Eprint {http://arxiv.org/abs/0907.2801} {arXiv:0907.2801 [hep-th]}
  \BibitemShut {NoStop}%
\bibitem [{\citenamefont {Arodz}\ and\ \citenamefont
  {Lis}(2009)}]{Arodz:2008nm}%
  \BibitemOpen
  \bibfield  {author} {\bibinfo {author} {\bibfnamefont {H.}~\bibnamefont
  {Arodz}}\ and\ \bibinfo {author} {\bibfnamefont {J.}~\bibnamefont {Lis}},\
  }\bibfield  {title} {\enquote {\bibinfo {title} {{Compact Q-balls and
  Q-shells in a scalar electrodynamics}},}\ }\href {\doibase
  10.1103/PhysRevD.79.045002} {\bibfield  {journal} {\bibinfo  {journal} {Phys.
  Rev. D}\ }\textbf {\bibinfo {volume} {79}},\ \bibinfo {pages} {045002}
  (\bibinfo {year} {2009})},\ \Eprint {http://arxiv.org/abs/0812.3284}
  {arXiv:0812.3284 [hep-th]} \BibitemShut {NoStop}%
\bibitem [{\citenamefont {Richtmyer}(1978)}]{richtmyer1978}%
  \BibitemOpen
  \bibfield  {author} {\bibinfo {author} {\bibfnamefont {R.~D.}\ \bibnamefont
  {Richtmyer}},\ }\href@noop {} {\emph {\bibinfo {title} {Principles of
  Advanced Mathematical Physics}}}\ (\bibinfo  {publisher} {Springer Verlag},\
  \bibinfo {address} {Berlin},\ \bibinfo {year} {1978})\ Chap.\ \bibinfo
  {chapter} {17.3}\BibitemShut {NoStop}%
\bibitem [{\citenamefont {Evans}(1998)}]{evans1998}%
  \BibitemOpen
  \bibfield  {author} {\bibinfo {author} {\bibfnamefont {L.~C.}\ \bibnamefont
  {Evans}},\ }\href@noop {} {\emph {\bibinfo {title} {Partial Differential
  Equations}}}\ (\bibinfo  {publisher} {American Mathematical Society},\
  \bibinfo {address} {Providence},\ \bibinfo {year} {1998})\BibitemShut
  {NoStop}%
\bibitem [{\citenamefont {Dinda}\ and\ \citenamefont
  {Remoissenet}(1999)}]{PhysRevE.60.6218}%
  \BibitemOpen
  \bibfield  {author} {\bibinfo {author} {\bibfnamefont {P.~Tchofo}\
  \bibnamefont {Dinda}}\ and\ \bibinfo {author} {\bibfnamefont
  {M.}~\bibnamefont {Remoissenet}},\ }\bibfield  {title} {\enquote {\bibinfo
  {title} {Breather compactons in nonlinear klein-gordon systems},}\ }\href
  {\doibase 10.1103/PhysRevE.60.6218} {\bibfield  {journal} {\bibinfo
  {journal} {Phys. Rev. E}\ }\textbf {\bibinfo {volume} {60}},\ \bibinfo
  {pages} {6218--6221} (\bibinfo {year} {1999})}\BibitemShut {NoStop}%
\bibitem [{\citenamefont {Adam}\ \emph {et~al.}(2010)\citenamefont {Adam},
  \citenamefont {Sanchez-Guillen},\ and\ \citenamefont
  {Wereszczynski}}]{Adam:2010fg}%
  \BibitemOpen
  \bibfield  {author} {\bibinfo {author} {\bibfnamefont {C.}~\bibnamefont
  {Adam}}, \bibinfo {author} {\bibfnamefont {J.}~\bibnamefont
  {Sanchez-Guillen}}, \ and\ \bibinfo {author} {\bibfnamefont {A.}~\bibnamefont
  {Wereszczynski}},\ }\bibfield  {title} {\enquote {\bibinfo {title} {{A
  Skyrme-type proposal for baryonic matter}},}\ }\href {\doibase
  10.1016/j.physletb.2010.06.025} {\bibfield  {journal} {\bibinfo  {journal}
  {Phys. Lett. B}\ }\textbf {\bibinfo {volume} {691}},\ \bibinfo {pages}
  {105--110} (\bibinfo {year} {2010})},\ \Eprint
  {http://arxiv.org/abs/1001.4544} {arXiv:1001.4544 [hep-th]} \BibitemShut
  {NoStop}%
\bibitem [{\citenamefont {Hartmann}\ \emph {et~al.}(2013)\citenamefont
  {Hartmann}, \citenamefont {Kleihaus}, \citenamefont {Kunz},\ and\
  \citenamefont {Schaffer}}]{Hartmann:2013kna}%
  \BibitemOpen
  \bibfield  {author} {\bibinfo {author} {\bibfnamefont {Betti}\ \bibnamefont
  {Hartmann}}, \bibinfo {author} {\bibfnamefont {Burkhard}\ \bibnamefont
  {Kleihaus}}, \bibinfo {author} {\bibfnamefont {Jutta}\ \bibnamefont {Kunz}},
  \ and\ \bibinfo {author} {\bibfnamefont {Isabell}\ \bibnamefont {Schaffer}},\
  }\bibfield  {title} {\enquote {\bibinfo {title} {{Compact (A)dS Boson Stars
  and Shells}},}\ }\href {\doibase 10.1103/PhysRevD.88.124033} {\bibfield
  {journal} {\bibinfo  {journal} {Phys. Rev. D}\ }\textbf {\bibinfo {volume}
  {88}},\ \bibinfo {pages} {124033} (\bibinfo {year} {2013})},\ \Eprint
  {http://arxiv.org/abs/1310.3632} {arXiv:1310.3632 [gr-qc]} \BibitemShut
  {NoStop}%
\bibitem [{\citenamefont {Arodz}\ \emph {et~al.}(2008)\citenamefont {Arodz},
  \citenamefont {Klimas},\ and\ \citenamefont {Tyranowski}}]{Arodz:2007jh}%
  \BibitemOpen
  \bibfield  {author} {\bibinfo {author} {\bibfnamefont {H.}~\bibnamefont
  {Arodz}}, \bibinfo {author} {\bibfnamefont {P.}~\bibnamefont {Klimas}}, \
  and\ \bibinfo {author} {\bibfnamefont {T.}~\bibnamefont {Tyranowski}},\
  }\bibfield  {title} {\enquote {\bibinfo {title} {{Compact oscillons in the
  signum-Gordon model}},}\ }\href {\doibase 10.1103/PhysRevD.77.047701}
  {\bibfield  {journal} {\bibinfo  {journal} {Phys. Rev. D}\ }\textbf {\bibinfo
  {volume} {77}},\ \bibinfo {pages} {047701} (\bibinfo {year} {2008})},\
  \Eprint {http://arxiv.org/abs/0710.2244} {arXiv:0710.2244 [hep-th]}
  \BibitemShut {NoStop}%
\bibitem [{\citenamefont {Adam}\ \emph {et~al.}(2018)\citenamefont {Adam},
  \citenamefont {Foster}, \citenamefont {Krusch},\ and\ \citenamefont
  {Wereszczynski}}]{Adam:2017srx}%
  \BibitemOpen
  \bibfield  {author} {\bibinfo {author} {\bibfnamefont {C.}~\bibnamefont
  {Adam}}, \bibinfo {author} {\bibfnamefont {D.}~\bibnamefont {Foster}},
  \bibinfo {author} {\bibfnamefont {S.}~\bibnamefont {Krusch}}, \ and\ \bibinfo
  {author} {\bibfnamefont {A.}~\bibnamefont {Wereszczynski}},\ }\bibfield
  {title} {\enquote {\bibinfo {title} {{BPS sectors of the Skyrme model and
  their non-BPS extensions}},}\ }\href {\doibase 10.1103/PhysRevD.97.036002}
  {\bibfield  {journal} {\bibinfo  {journal} {Phys. Rev. D}\ }\textbf {\bibinfo
  {volume} {97}},\ \bibinfo {pages} {036002} (\bibinfo {year} {2018})},\
  \Eprint {http://arxiv.org/abs/1709.06583} {arXiv:1709.06583 [hep-th]}
  \BibitemShut {NoStop}%
\bibitem [{\citenamefont {Klimas}\ \emph {et~al.}(2018)\citenamefont {Klimas},
  \citenamefont {Streibel}, \citenamefont {Wereszczynski},\ and\ \citenamefont
  {Zakrzewski}}]{Klimas:2018woi}%
  \BibitemOpen
  \bibfield  {author} {\bibinfo {author} {\bibfnamefont {P.}~\bibnamefont
  {Klimas}}, \bibinfo {author} {\bibfnamefont {J.~S.}\ \bibnamefont
  {Streibel}}, \bibinfo {author} {\bibfnamefont {A.}~\bibnamefont
  {Wereszczynski}}, \ and\ \bibinfo {author} {\bibfnamefont {W.~J.}\
  \bibnamefont {Zakrzewski}},\ }\bibfield  {title} {\enquote {\bibinfo {title}
  {{Oscillons in a perturbed signum-Gordon model}},}\ }\href {\doibase
  10.1007/JHEP04(2018)102} {\bibfield  {journal} {\bibinfo  {journal} {JHEP}\
  }\textbf {\bibinfo {volume} {04}},\ \bibinfo {pages} {102} (\bibinfo {year}
  {2018})},\ \Eprint {http://arxiv.org/abs/1801.05454} {arXiv:1801.05454
  [hep-th]} \BibitemShut {NoStop}%
\bibitem [{\citenamefont {Klimas}\ and\ \citenamefont
  {Livramento}(2017)}]{Klimas:2017eft}%
  \BibitemOpen
  \bibfield  {author} {\bibinfo {author} {\bibfnamefont {P.}~\bibnamefont
  {Klimas}}\ and\ \bibinfo {author} {\bibfnamefont {L.~R.}\ \bibnamefont
  {Livramento}},\ }\bibfield  {title} {\enquote {\bibinfo {title} {{Compact
  Q-balls and Q-shells in CPN type models}},}\ }\href {\doibase
  10.1103/PhysRevD.96.016001} {\bibfield  {journal} {\bibinfo  {journal} {Phys.
  Rev. D}\ }\textbf {\bibinfo {volume} {96}},\ \bibinfo {pages} {016001}
  (\bibinfo {year} {2017})},\ \Eprint {http://arxiv.org/abs/1704.01132}
  {arXiv:1704.01132 [hep-th]} \BibitemShut {NoStop}%
\bibitem [{\citenamefont {Ferreira}\ and\ \citenamefont
  {Leite}(1999)}]{Ferreira:1998zx}%
  \BibitemOpen
  \bibfield  {author} {\bibinfo {author} {\bibfnamefont {Luiz~A.}\ \bibnamefont
  {Ferreira}}\ and\ \bibinfo {author} {\bibfnamefont {Erica~E.}\ \bibnamefont
  {Leite}},\ }\bibfield  {title} {\enquote {\bibinfo {title} {{Integrable
  theories in any dimension and homogeneous spaces}},}\ }\href {\doibase
  10.1016/S0550-3213(99)00090-5} {\bibfield  {journal} {\bibinfo  {journal}
  {Nucl. Phys. B}\ }\textbf {\bibinfo {volume} {547}},\ \bibinfo {pages}
  {471--500} (\bibinfo {year} {1999})},\ \Eprint
  {http://arxiv.org/abs/hep-th/9810067} {arXiv:hep-th/9810067} \BibitemShut
  {NoStop}%
\bibitem [{\citenamefont {Eichenherr}\ and\ \citenamefont
  {Forger}(1980)}]{EICHENHERR1980528}%
  \BibitemOpen
  \bibfield  {author} {\bibinfo {author} {\bibfnamefont {H.}~\bibnamefont
  {Eichenherr}}\ and\ \bibinfo {author} {\bibfnamefont {M.}~\bibnamefont
  {Forger}},\ }\bibfield  {title} {\enquote {\bibinfo {title} {More about
  non-linear sigma models on symmetric spaces},}\ }\href {\doibase
  https://doi.org/10.1016/0550-3213(80)90525-8} {\bibfield  {journal} {\bibinfo
   {journal} {Nuclear Physics B}\ }\textbf {\bibinfo {volume} {164}},\ \bibinfo
  {pages} {528--535} (\bibinfo {year} {1980})}\BibitemShut {NoStop}%
\bibitem [{\citenamefont {Ferreira}\ and\ \citenamefont
  {Olive}(1985)}]{Ferreira:1984bi}%
  \BibitemOpen
  \bibfield  {author} {\bibinfo {author} {\bibfnamefont {L.~A.}\ \bibnamefont
  {Ferreira}}\ and\ \bibinfo {author} {\bibfnamefont {David~I.}\ \bibnamefont
  {Olive}},\ }\bibfield  {title} {\enquote {\bibinfo {title} {{Noncompact
  Symmetric Spaces and the Toda Molecule Equations}},}\ }\href {\doibase
  10.1007/BF01240353} {\bibfield  {journal} {\bibinfo  {journal} {Commun. Math.
  Phys.}\ }\textbf {\bibinfo {volume} {99}},\ \bibinfo {pages} {365} (\bibinfo
  {year} {1985})}\BibitemShut {NoStop}%
\end{thebibliography}%

\end{document}